\documentclass[12pt,preprint]{aastex}

\newcommand{\ionhy}{H{\sc ii}}
\newcommand{\glim}{{\em GLIMPSE}}

\newcommand{\twofields}[2] {
  \begin{center}
    \begin{minipage}[t]{0.45\textwidth}
      \includegraphics[scale=0.40]{#1}
    \end{minipage}
    \hfill
    \begin{minipage}[t]{0.45\textwidth}
      \includegraphics[scale=0.40]{#2}
    \end{minipage}
  \end{center}
}

\shorttitle{Tracers of early high-mass star formation}
\shortauthors{Ellingsen}

\begin{document}

\title{Methanol masers : Reliable tracers of the early stages of
  high-mass star formation}

\author{S.P. Ellingsen}
\affil{School of Mathematics and Physics, University of Tasmania, 
  Private Bag 37, Hobart, TAS 7001, Australia}
\email{Simon.Ellingsen@utas.edu.au}

\begin{abstract}
  The \glim\/ and {\em MSX} surveys have been used to examine the
  mid-infrared properties of a statistically complete sample of
  6.7~GHz methanol masers.  The \glim\/ point sources associated with
  methanol masers are clearly distinguished from the majority,
  typically having extremely red mid-infrared colors, similar to those
  expected of low-mass class~0 young stellar objects.  The intensity
  of the \glim\/ sources associated with methanol masers is typically
  4 magnitudes brighter at 8.0~\micron\/ than at 3.6~\micron.
  Targeted searches towards \glim\/ point sources with [3.6]-[4.5] $>$
  1.3 and an 8.0~\micron\ magnitude less than 10 will detect more than
  80\% of class~II methanol masers.  Many of the methanol masers are
  associated with sources within infrared dark clouds (IRDC) which are
  believed to mark regions where high-mass star formation is in its
  very early stages.  The presence of class~II methanol masers in a
  significant fraction of IRDC suggests that high-mass star formation
  is common in these regions.  Different maser species are thought to
  trace different evolutionary phases of the high-mass star formation
  process.  Comparison of the properties of the \glim\/ sources
  associated with class~II methanol masers and other maser species
  shows interesting trends, consistent with class~I methanol masers
  tracing a generally earlier evolutionary phase and OH masers tracing
  a later evolutionary phase.
\end{abstract}


\section{Introduction}

Interstellar masers from a number of molecular species (in particular
main-line OH, water and methanol) are known to be associated with
high-mass star formation regions.  Class~II methanol masers are unique
amongst these in that they are only observed towards high-mass star
formation regions \citep{MENB03}, in contrast to OH and water masers
which are also observed towards other astrophysical objects, such as
late-type stars.  The strongest methanol masers often show emission in
multiple transitions and are projected against ultra-compact \ionhy\ 
regions \citep[e.g. W3(OH),][]{MRP+92,SSE+01}.  However, the majority
of methanol masers only show emission in the 6.7 (and sometimes
12.1~GHz) transition and are not associated with detectable \ionhy\ 
regions \citep{PNEM98,WBHR98}.  Recent observations have shown that
those sources which do not show centimeter radio continuum emission
are associated with millimeter \citep{PHB02} and sub-millimeter
\citep{WMA+03} sources.  They have a spectral energy distribution
consistent with them being deeply embedded \citep{MBH+05} and being
associated with a very early phase of the star formation process.
Masers are potentially powerful probes of the high-mass star formation
process, but an understanding of the environment in which they arise
is required before they can be used as such.

Theoretical models of class~II methanol masers invoke excitation to
the second torsional state from the radiative field of warm dust at
sub-millimeter and far-infrared (FIR) wavelengths to explain the high
brightness temperatures observed \citep{SD94,SCG97}.  Many methanol
masers have been discovered in searches towards {\em IRAS} sources
which meet the \citet{WC89} criteria for being ultra-compact \ionhy\ 
regions \citep[e.g.][]{SVGM93,VGM95,WHRB97,SVK+99,SHK00}.  However, in
many cases the masers are not directly associated with the {\em IRAS}
sources, but rather with the larger star formation region.  The {\em
  IRAS} satellite had relatively poor angular resolution and suffered
well documented problems in the Galactic Plane with saturation and
confusion, due to the high density of strong sources.  This does not
mean that the masers are not associated with FIR sources, but rather
that they are not necessarily associated with the strongest source
within a given star formation region.  So {\em IRAS} observations are
of little use in determining the properties of the sources responsible
for pumping the methanol masers.

There have been a number of subsequent infrared surveys, such as the
{\em Midcourse Space Experiment} ({\em MSX}) and the Two Micron
All-Sky Survey (2MASS). {\em MSX} undertook a survey of the Galactic
Plane at mid-infrared wavelengths, with an angular resolution of
18\arcsec.  However, this is insufficient angular resolution to be
able to resolve individual sources in high-mass star formation
regions, which are typically at distances greater than 1~kpc.  Hence
the {\em MSX} sources closest to methanol masers usually have colors
consistent with massive young stellar objects \citep{LHOR02}, but are
typically not directly associated with the masers \citep{E05}.  It is
well established that many stars are in binary or higher order
multiple systems, and resolving the individual stars in such systems
requires angular resolutions significantly better than 1\arcsec\/.
However, the overall infrared characteristics of a multiple system
will be dominated by the most massive star and it is the properties of
this star that will have the primary influence on the masers.  Where
reference is made to an infrared ``source'' throughout the paper this
includes multiple star systems unresolved at arcsecond resolutions.
2MASS has sufficient angular resolution to be able to resolve the
source associated with the maser from others within the cluster star
forming environment.  However, the longest observing wavelength for
2MASS is only 2.2~\micron\ and the majority of the maser sources are
optically thick at near-infrared wavelengths.  As for {\em IRAS} and
{\em MSX}, typically the closest 2MASS sources are significantly
offset from the maser location and are likely unrelated objects within
the larger star formation region \citep{E05}.  There have been a
number of near-infrared observations targeted at methanol maser sites
\citep{TFPR98,WBHR99,GVG02}, many of the masers have no NIR
counterpart, and in other cases the NIR source is offset from the
maser location.  Typically these nearby sources are extremely red,
but, as for the 2MASS associations, it is likely that in most cases
they are sources unrelated to the masers within the larger star
formation region.

Targeted mid-infrared observations have been made toward a number of
methanol masers at wavelengths as long as 20~\micron\ 
\citep{DPT00,WBBN01}.  These find mid-infrared sources
associated with many of the masers that do not have near-infrared
counterparts (67\% in the case of \citeauthor{DPT00}), but also show
that sometimes the sources associated with the methanol masers are
optically thick to wavelengths longer than 20~\micron, suggesting that
they are deeply embedded and very young.  The most significant
problems with existing ground-based mid-infrared observations is
limited sensitivity \citep[e.g. of the order of 15~mJy at 10~\micron\ 
for][and higher in the other cases]{DPT00} and large astrometric
uncertainty, due to the lack of guide stars at these wavelengths.
This makes it impossible to accurately determine the relative location
of the masers and mid-infrared sources, which is particularly
important in situations where there are extended, or multiple
mid-infrared emission regions.  \citet{DPT00} use a method of
cross-correlating the centimeter radio continuum and mid-infrared
emission to determine the best registration.  However, there are
significant differences between the emission at the two wavelengths,
which means that the relative registration remains uncertain.  The
astrometry is highly dependent on the telescope and more recent
ground-based mid-infrared observations targeted towards water maser
sites have been able to achieve sub-arcsecond precision
\citep{DRTP05}.  Further sensitive mid-infrared observations with good
astrometric accuracy are highly desirable, as they will yield
important information on the nature of the mid-infrared sources
associated with methanol masers.

\section{GLIMPSE observations}

\glim\/\ (Galactic Legacy Infrared Mid-Plane Survey Extraordinare) is
a legacy science program of the {\em Spitzer Space Telescope} that
surveyed the inner Galactic plane ($|\ell|$ = 10\degr--65\degr) in a
number of mid-infrared wavelength bands at 1.4--1.9\arcsec\
resolution, using the {\em IRAC}\ (Infrared Array Camera) instrument
\citep{BCB+03}.  The observations were made in the 3.6, 4.5, 5.8 and
8.0~\micron\ bands of {\em IRAC}, and the quoted positional accuracy
in the point source catalog is approximately 0\farcs4.  As such it
provides the best opportunity to date to investigate the mid-infrared
properties of a large number of methanol masers, and compare them with
other sources.  I have used the 15 April 2005 release of the \glim\/
point source catalog and mosaicked images from the longest wavelength
band (8.0~\micron) to investigate the mid-infrared properties of the
fifty-six methanol masers in the Galactic Longitude range $\ell$ =
325\degr--335\degr.  The positions of the methanol masers are known to
an accuracy of 0\farcs5 from Australia Telescope Compact Array (ATCA)
observations \citep{E05}.  The methanol masers under investigation
include a statistically complete sample detected in the Mt Pleasant
survey \citep{EVM+96}.  In contrast, previous ground based
mid-infrared observations have been primarily targeted either at {\em
IRAS}-associated, or strong maser peak flux density based samples
\citep{DPT00,WBBN01}.

\section{Results} \label{sec:results}

The fifty-six class~II methanol masers in the region $\ell$ =
325\degr--335\degr\ are listed in Table~\ref{results}, along with
any \glim\/ point sources within 2\arcsec\ of the maser location.
Forty-one of these maser sources are part of a statistically complete
sample.  Those which are not, are indicated in column 1 of
Table~\ref{results}.  Three of the sources not within the
statistically complete sample (G$\,331.120\!-\!0.118$ ,
G$\,333.029\!-\!0.063$, G$\,333.646\!+\!0.058$) were not detected in the Mt
Pleasant survey, but rather in the survey of \citet{C96}, the
remainder were detected at Mt Pleasant, but lie outside the spatial or
velocity range for which the search is complete \citep{EVM+96}.
Twenty-nine of the methanol masers in Table~\ref{results} have a
\glim\/ point source within 2\arcsec\ (52\%), considering only the
statistically complete sources this becomes twenty-three of forty-one
masers (56\%).  The magnitudes and flux densities observed in each of
the 4 {\em IRAC} bands for each of the \glim\/ sources associated with
methanol masers are listed in Table~\ref{glimprop}.  

\begin{figure}
\plotone{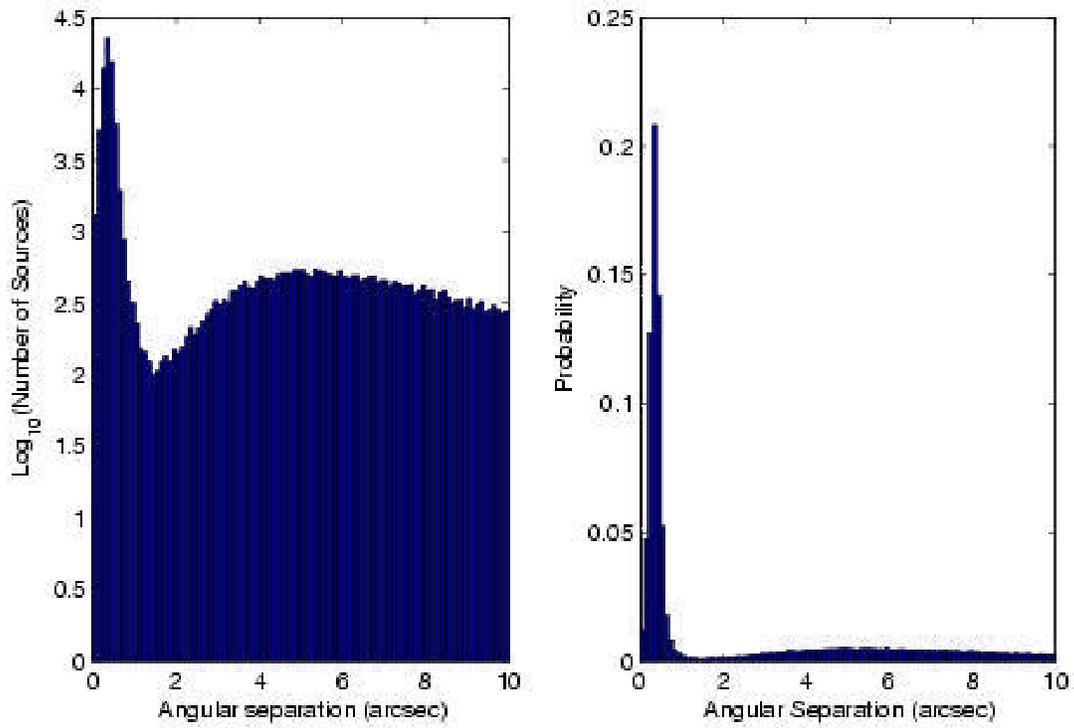}
\caption{Left : A histogram showing the distribution of the separation between
  a \glim\/ point source and the nearest 2MASS point source, for
  sources within a 30\arcmin\ radius of $\ell$=326.5\degr,
  $b$=0.0\degr.  Right : The same distribution plotted on a linear
  scale in terms of probability.}
\label{fig:angsep}
\end{figure}

To cross-check the positional accuracy of the \glim\/ catalog and
estimate the probablity of confusion, I measured the angular
separation between each \glim\/ point source and the nearest 2MASS
point source for sources within a 30\arcmin\ radius of $\ell$ =
326.5\degr, $b$ = 0.0\degr.  In total there are approximately 109,000
\glim\/ and 83,500 2MASS point sources in this region.
Figure~\ref{fig:angsep} shows the distribution of angular separations
plotted as a histogram with 0.1\arcsec\ bins for the range
1-10\arcsec.  The distribution peaks very strongly at 0.3-0.4\arcsec\/
as predicted with a much smaller secondary peak at around 5\arcsec\
which can be used to estimate the probability of chance coincidence as
it relates to the source density in the two catalogs.  Approximately
34,000 \glim\/ sources have a 2MASS point source within 1\arcsec\ and
extrapolation of the tail of the coincidence distribution implies
(conservatively) that more than 97\% of these are valid associations.
This implies that the probability of any of the associations between
\glim\/ point sources and methanol masers listed in
Table~\ref{results} being chance coincidences is low, particularly for
the sources where the separation is less than 1\arcsec.

The minimum requirement for a source to be included in the \glim\/
point source catalogue was that it had to be detected with a
signal-to-noise ratio greater than 5, twice in one {\em IRAC} band and
once in an adjacent band \footnote{GLIMPSE Data Products Document v1.5, \\
http://data.spitzer.caltech.edu/popular/glimpse/20050415\_enhanced\_v1/Documents/\\ glimpse\_dataprod\_v1.5.pdf}.
The point source catalog is believed to be 99\% reliable.  The \glim\/
archive is a more complete second catalog with less stringent
conditions, requiring two detections with a signal-to-noise ratio
greater than 5 in any bands.  The archive contains approximately 50\%
more sources than the point source catalog.  A search of the \glim\/
archive finds sources within 2\arcsec\ of an additional 9 methanol
masers that do not have an associated \glim\/ point source.  The names
of the sources from the \glim\/ archive (where there is no associated
\glim\/ point source) are also listed in Table~\ref{results}.  \glim\
sources from the point source catalog have the prefix GLMC and those
from the archive GLMA, the archive sources are also listed in italics
in Tables~\ref{results} \& \ref{glimprop} so that they stand out.
Considering both \glim\/ point sources and archive sources,
thirty-eight of the fifty-six methanol masers (68\%) have an associated
\glim\/ source within 2\arcsec.  For the forty-one methanol masers in
the statistically complete sample there are thirty-two with associated
\glim\ sources (78\%)

For each of the eighteen methanol masers that have no associated
\glim\/ point or archive source there are two possible explanations,
one is that the associated mid-infrared emission is too faint, the
other is that the emission in the region is extended and hence not
present in the point source catalog.  It is possible to distinguish
between these two possibilities by examining the \glim\/ mosaics in
the region surrounding each maser source.
Figures~\ref{fig:images1}-\ref{fig:images13} show the 8.0~\micron\ 
\glim\/ observations (grayscale) and {\em MSX} E-band (21~\micron)
emission (contours) in a 2\arcmin\ square field centered on the
location of each maser (in some cases there are multiple maser sites
within one field).  The locations of the methanol masers, nearest
\glim\/ point source (within 2\arcsec) and any {\em MSX} point source
within 5\arcsec\ are also shown in each figure.  Comments on selected
individual sources are given in section~\ref{sec:indiv}.  There are
image artifacts present in some of the fields, wherever there are
strong 8.0~\micron\ sources they are accompanied by linear features
running from a position angle of approximately 250\degr\ through to
70\degr\, (the scanning direction of the observations).  For some
point sources (e.g.  G\,$329.610\!+\!0.114$) the point spread function
of the {\em IRAC} instrument is evident, and a few fields (e.g.
G\,$333.029\!-\!0.015$) have blanked pixels due to saturation.
However, these and other minor image artifacts do not have a
significant impact on our ability to investigate the mid-infrared
emission associated with the methanol masers.  It is clear from
Fig.~\ref{fig:images1}-\ref{fig:images13} that in the majority of
cases where there is no \glim\/ point source near the masers, they are
projected against regions of extended 8.0~\micron\ emission.  So at
the sensitivity of the \glim\/ survey only a small percentage
(somewhere in the range of 5-10\%) of methanol masers are not
associated with mid-infrared emission at wavelengths of 8.0~\micron\ 
or shorter.

Information on associated water, OH and class~I methanol masers is
available for the 6.7~GHz methanol masers from the Mt Pleasant sample
in the region $\ell$=325-335\degr\ region, and these are listed in the
last column of Table~\ref{results}, along with the relevant
references.  The 6.7~GHz methanol and the OH masers have both been
observed with the ATCA and only sources where the two species are
coincident to within 1\arcsec\ are listed.  However, the water and
class~I methanol maser observations were both made with single dish
radio telescopes and were targeted towards the 6.7~GHz maser position,
so a definite association between the two different maser species has
not been made for these transitions.  The locations of these
associated masers are not shown in
Fig.~\ref{fig:images1}-\ref{fig:images13}, as the OH masers lie at
essentially the same location as the methanol masers and for the water
and class~I methanol masers the position relative to the class~II
methanol masers has not been determined.

\subsection{Comments on individual sources of interest} \label{sec:indiv}

\begin{figure}
  \twofields{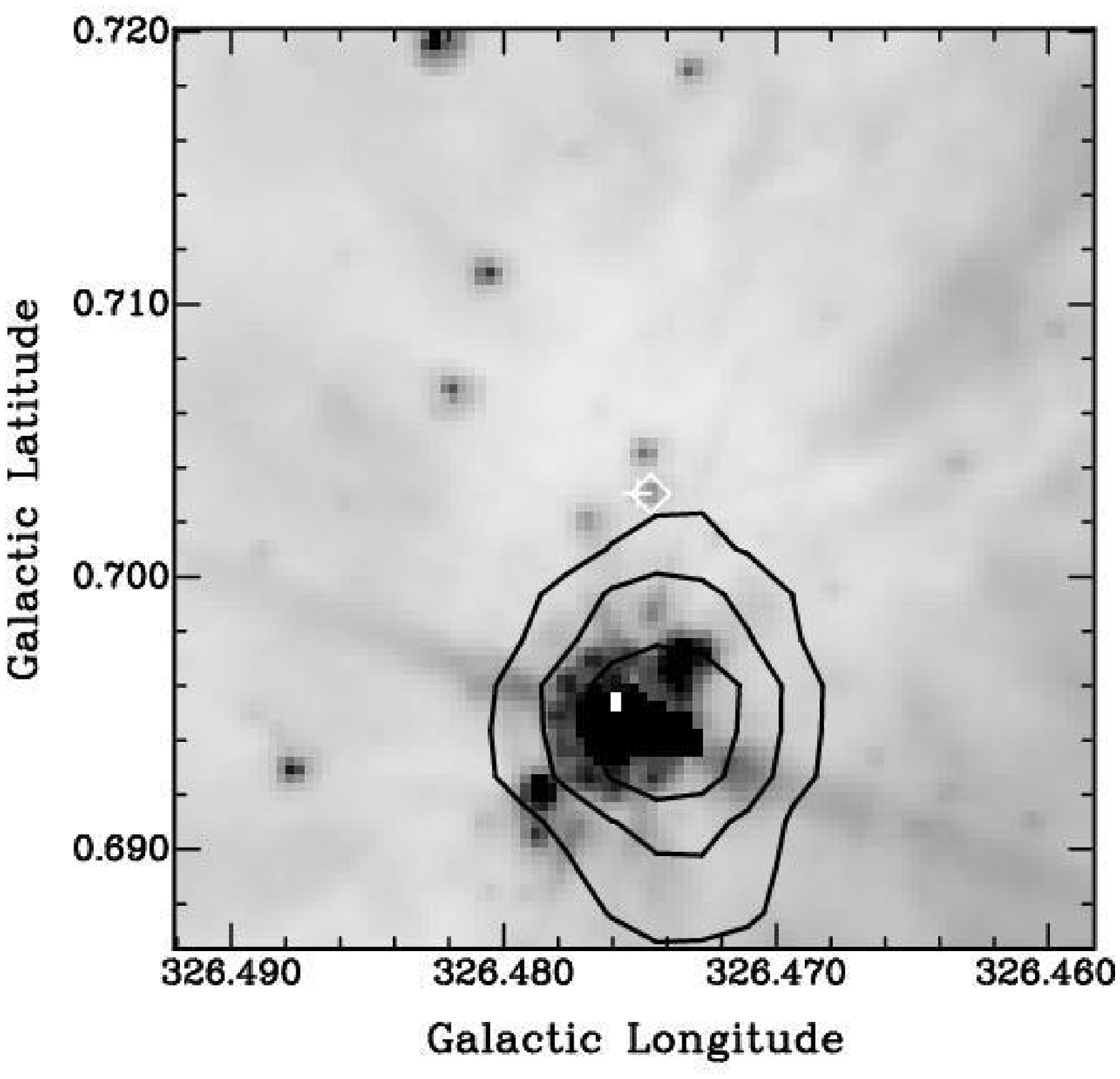}{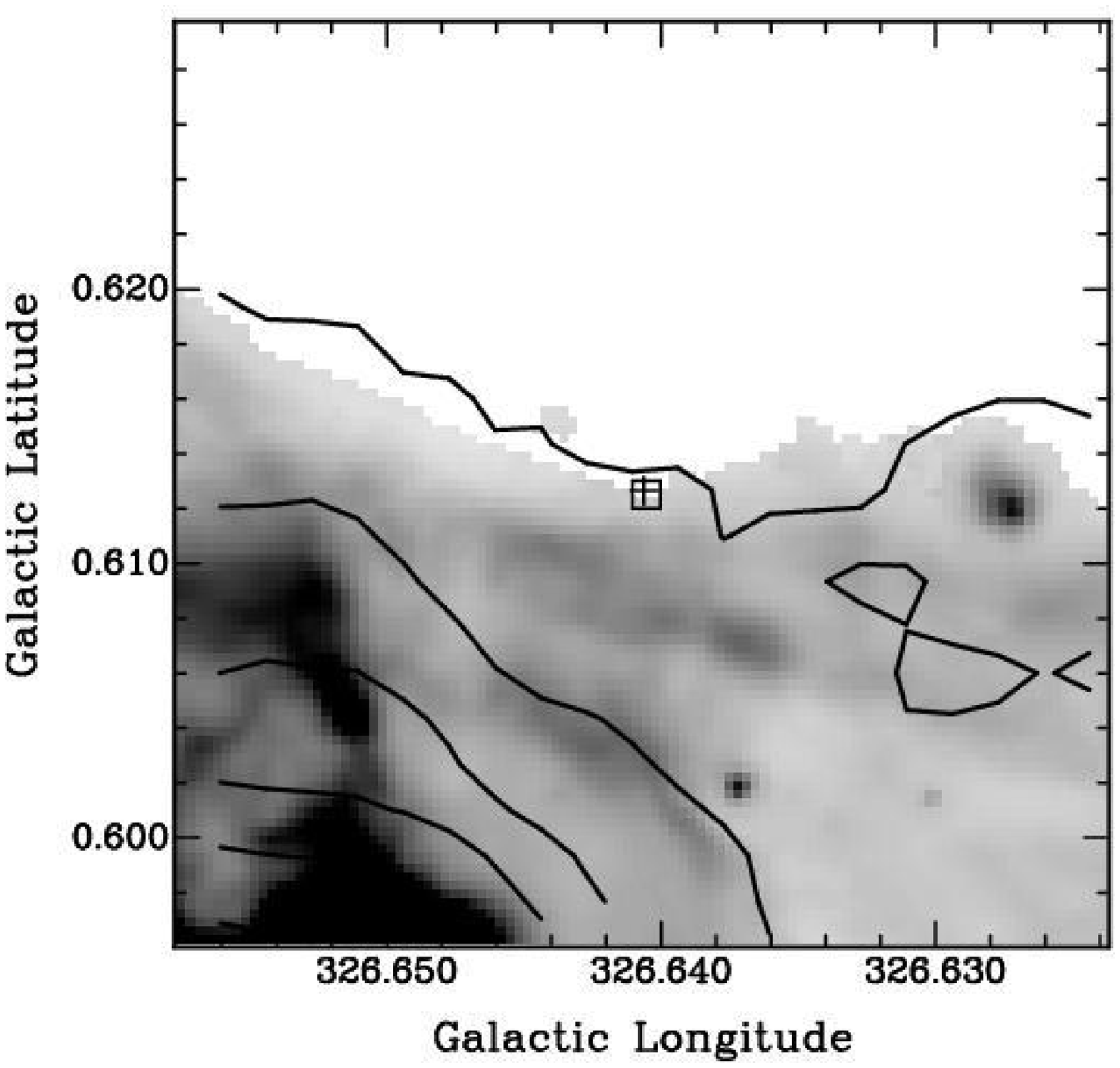}
  \twofields{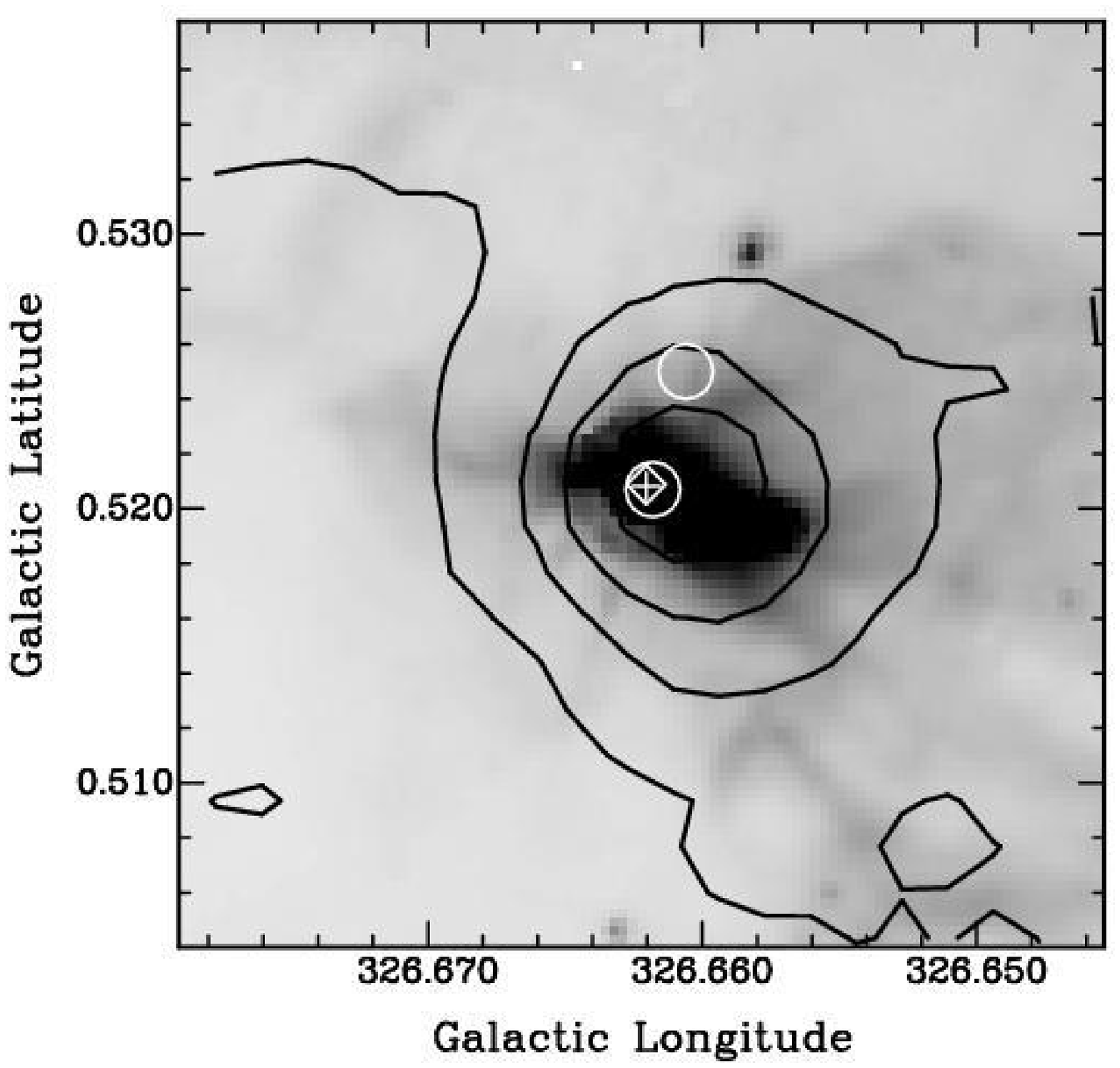}{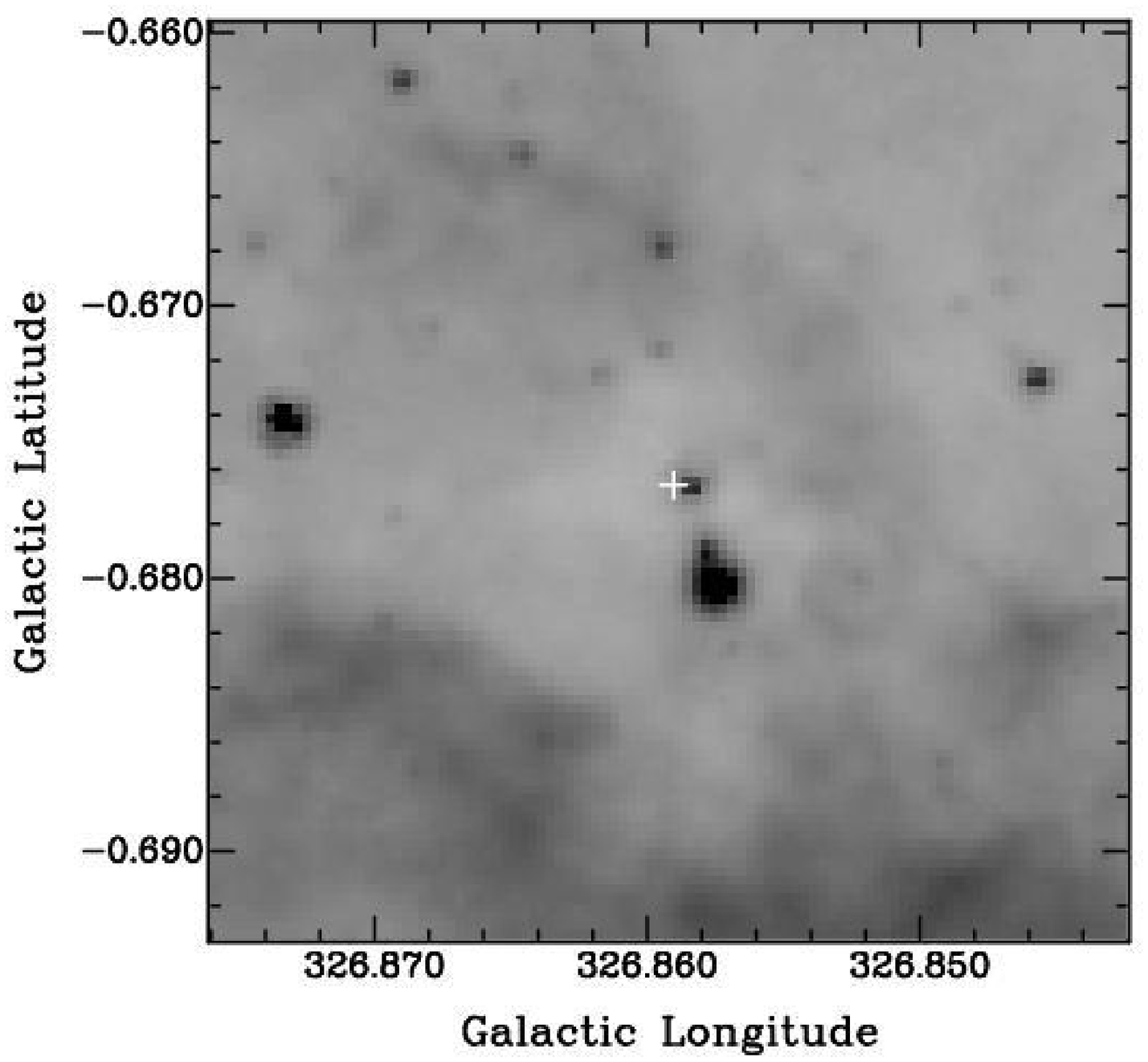}
 \caption{The grey scale is the GLIMPSE band 4 emission (8.0~\micron), the 
   contours are the {\em MSX} E-band emission (21~\micron), the
   crosses mark the positions of 6.7~GHz methanol masers, the squares
   the positions of \glim\ point sources, the diamonds the positions of
   \glim\ archive sources and the circles the positions of {\em MSX} point 
   sources.  Moving from top left to bottom right, the sources are 
   G\,$326.475\!+\!0.703$, G\,$326.641\!+\!0.613$, G\,$326.662\!+\!0.521$,
   G\,$326.859\!-\!0.677$}
  \label{fig:images1}
\end{figure}

\begin{figure}
  \twofields{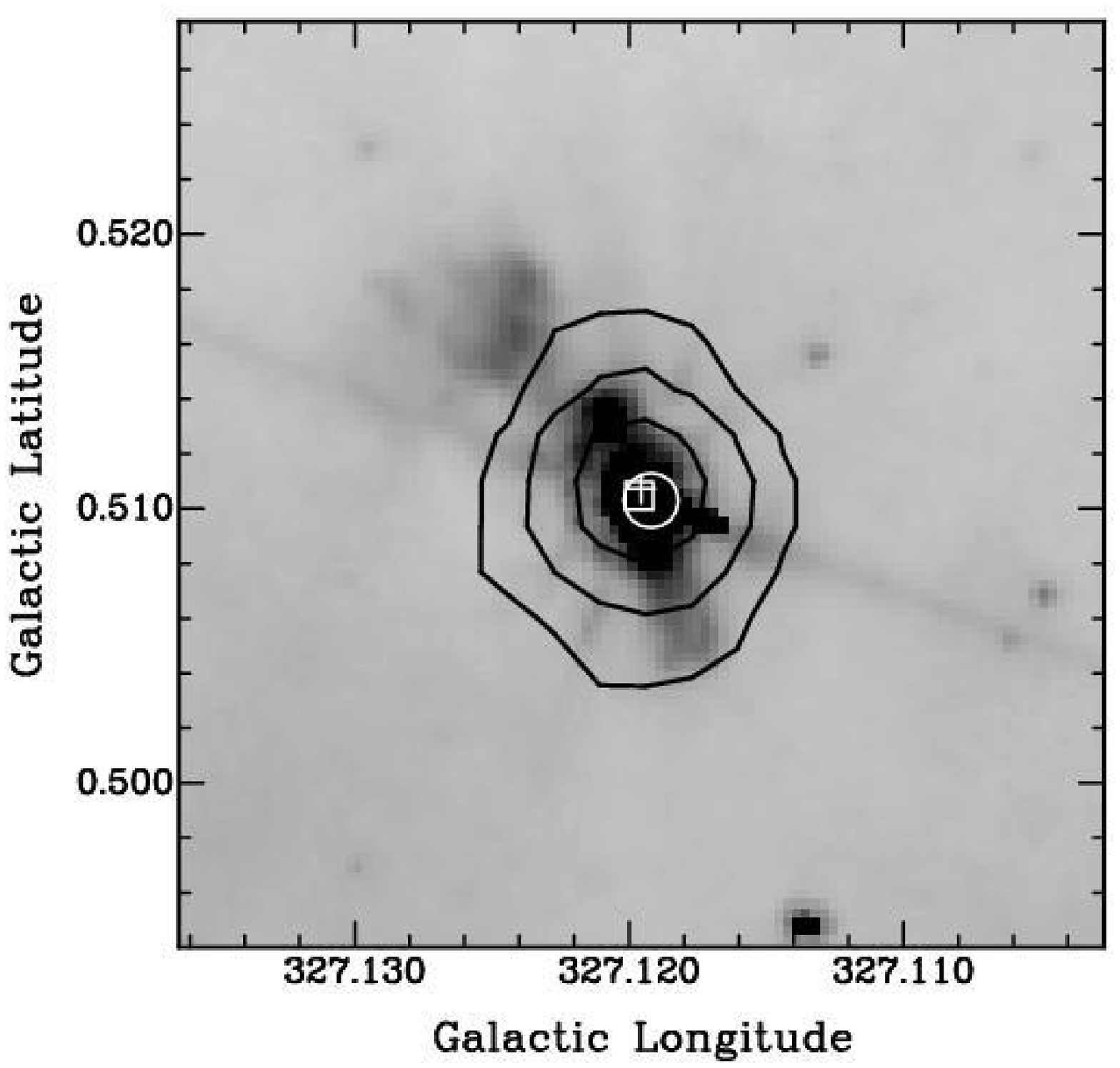}{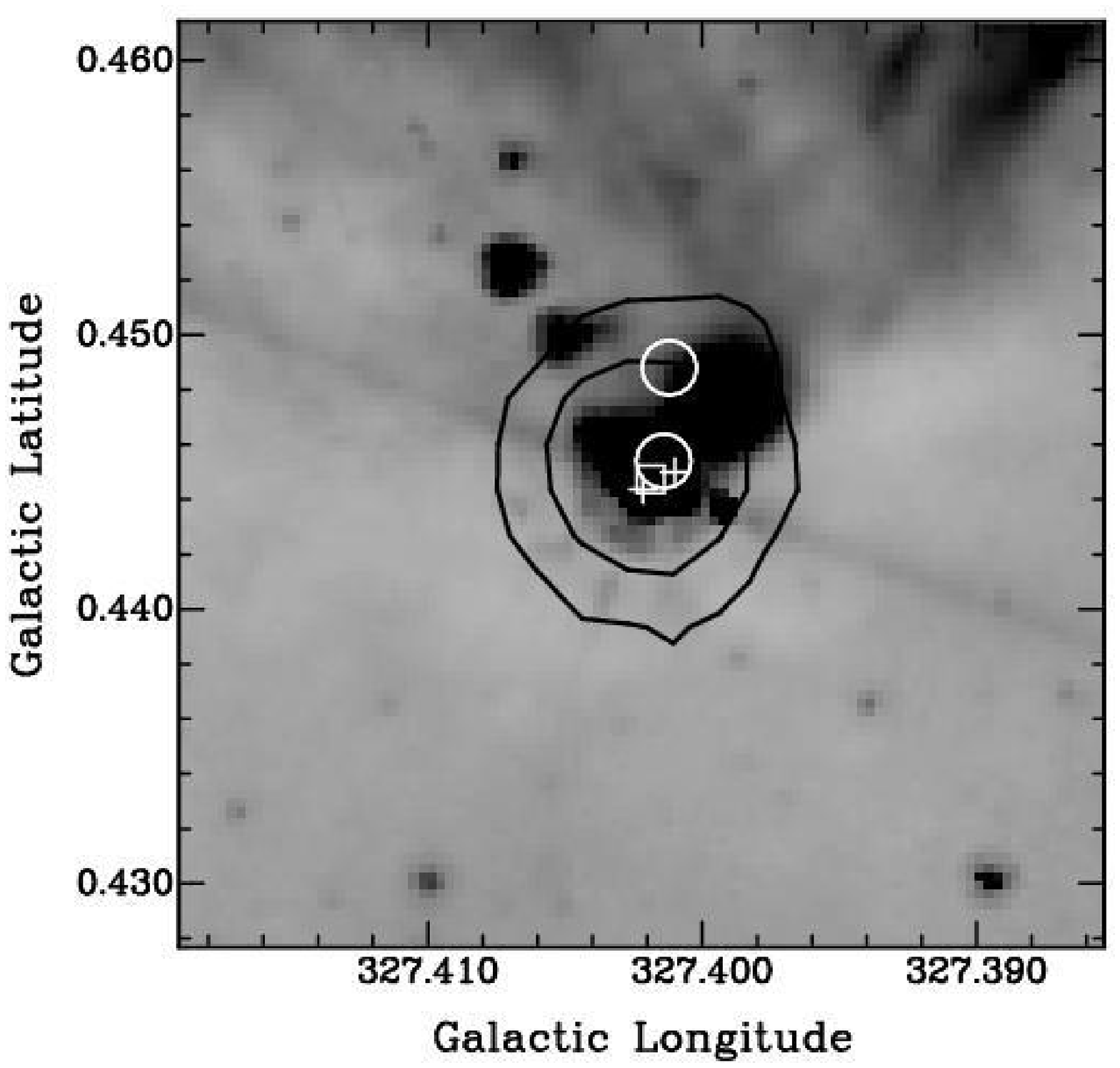}
  \twofields{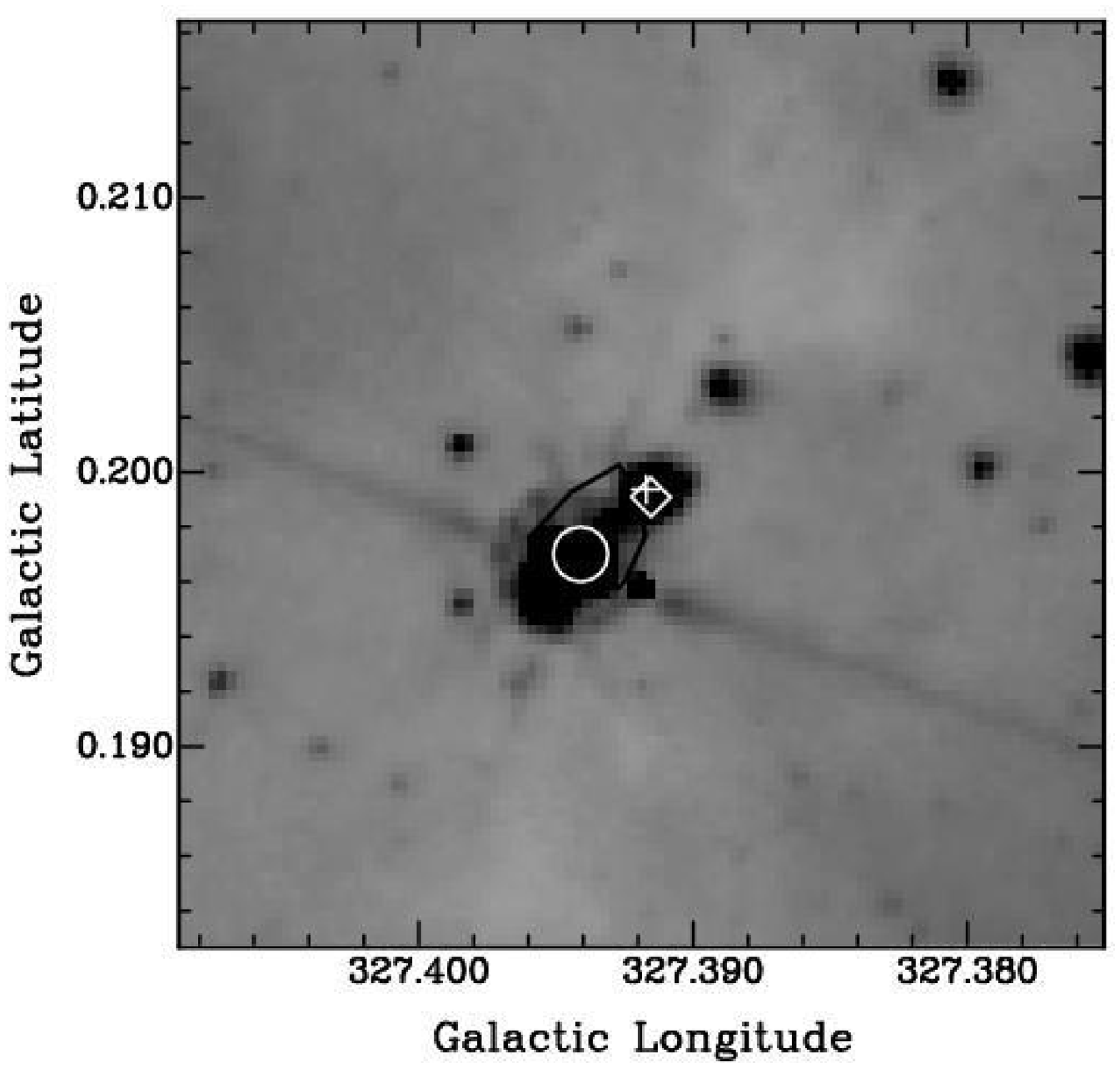}{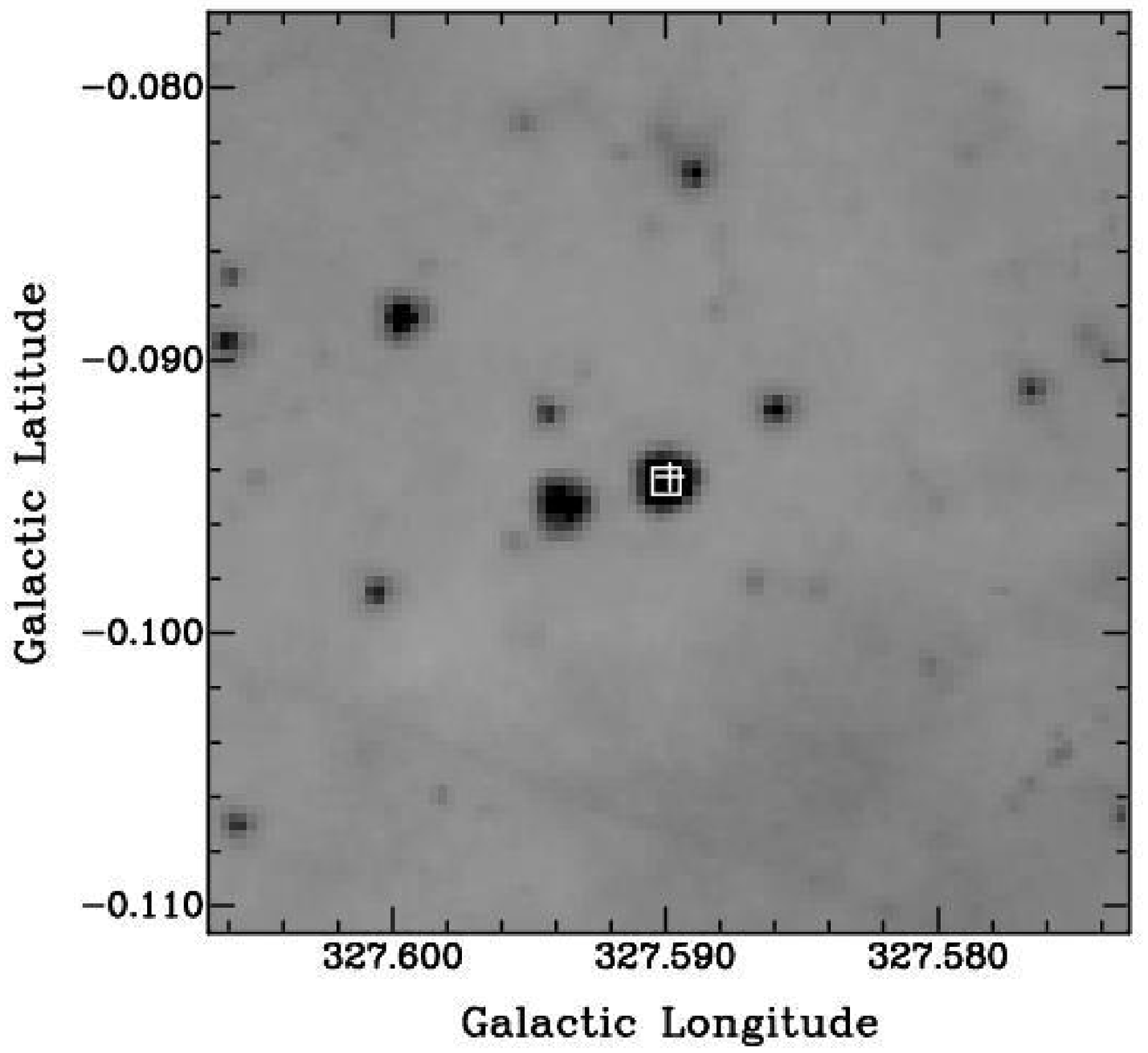}
  \caption{As for Fig.~\ref{fig:images1} for the sources  
    G\,$327.120\!+\!0.511$, G\,$327.402\!+\!0.444$,
    G\,$327.392\!+\!0.199$, G\,$327.590\!-\!0.094$}
  \label{fig:images2}
\end{figure}

\begin{figure}
  \twofields{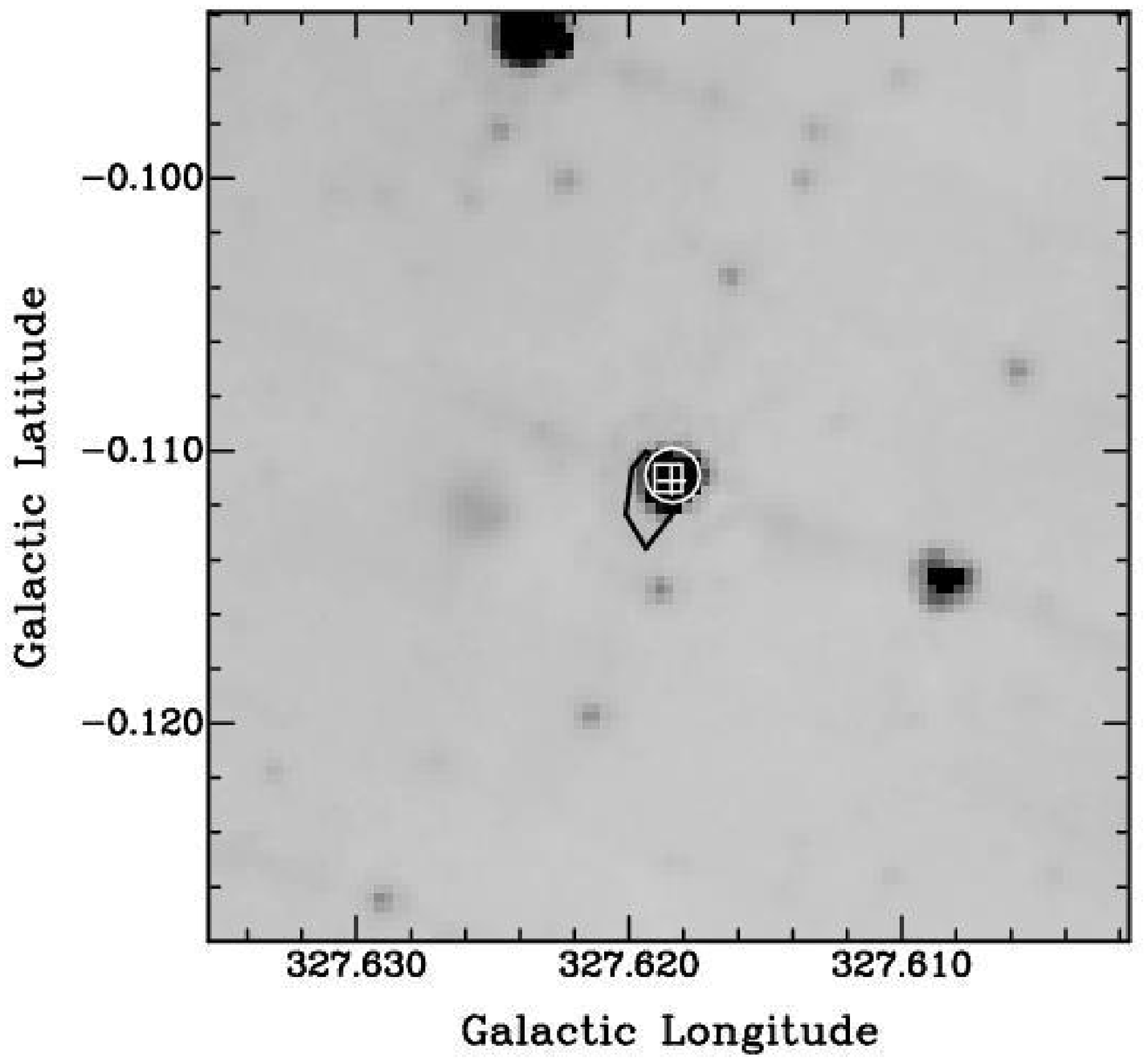}{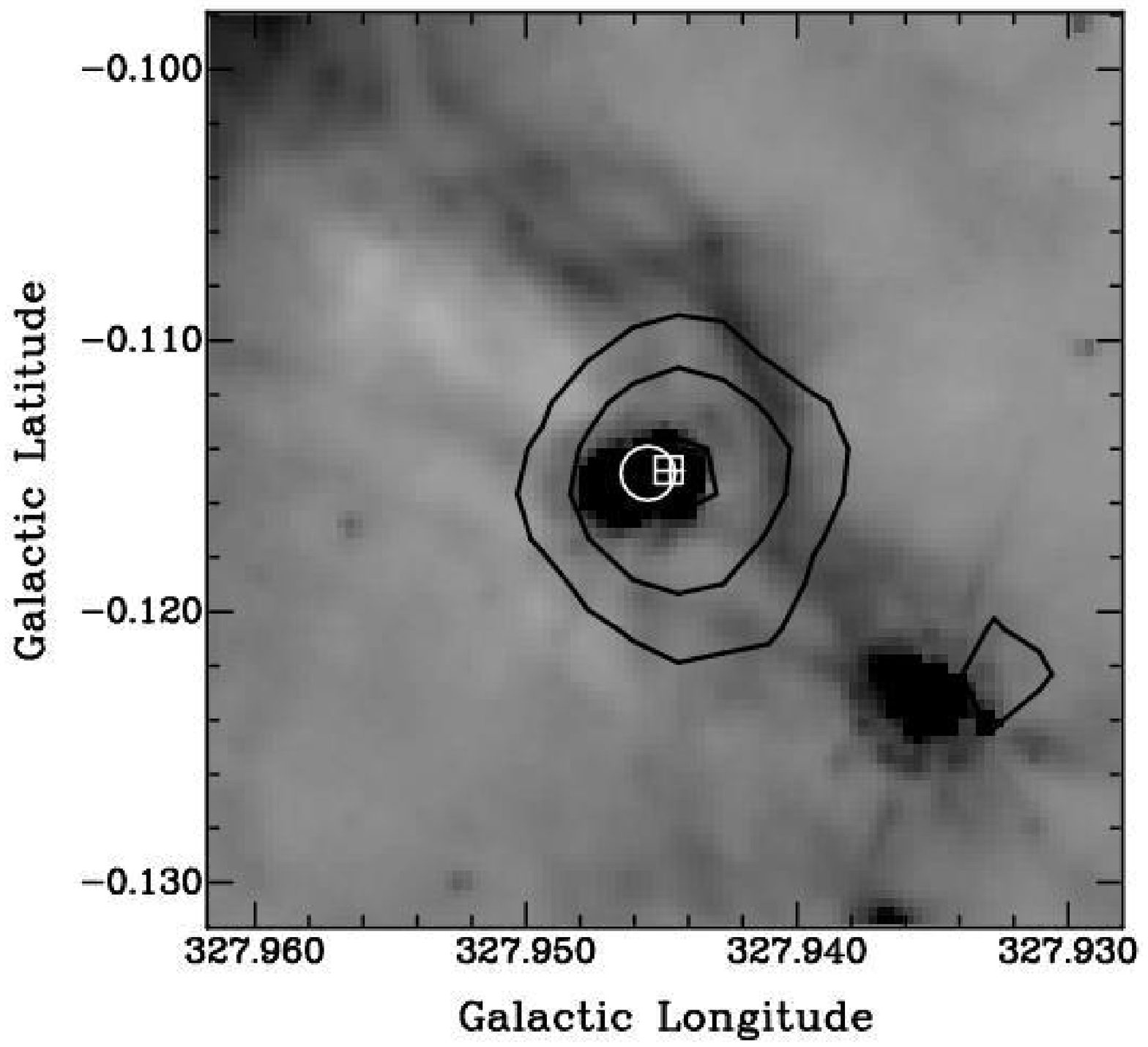}
  \twofields{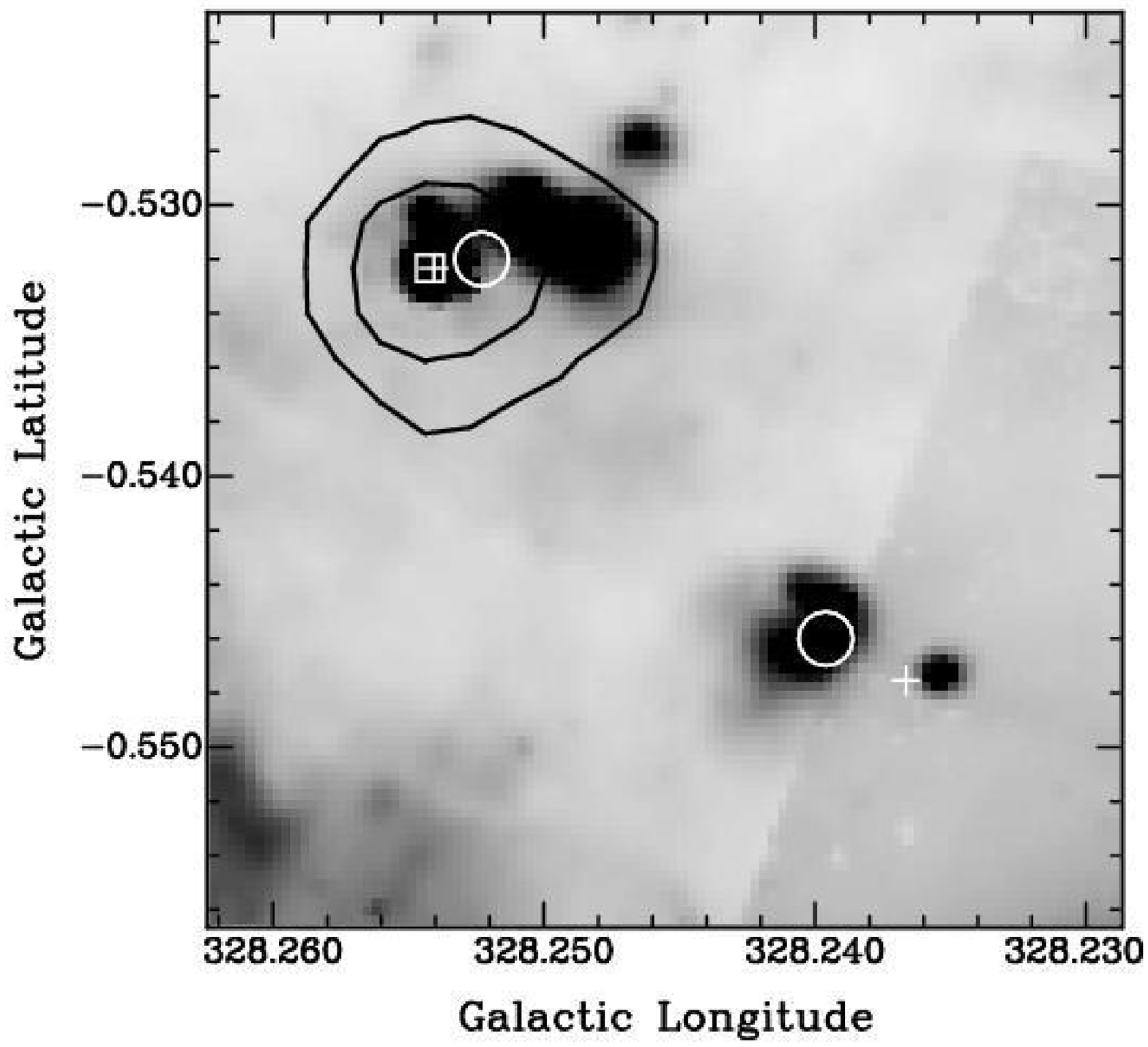}{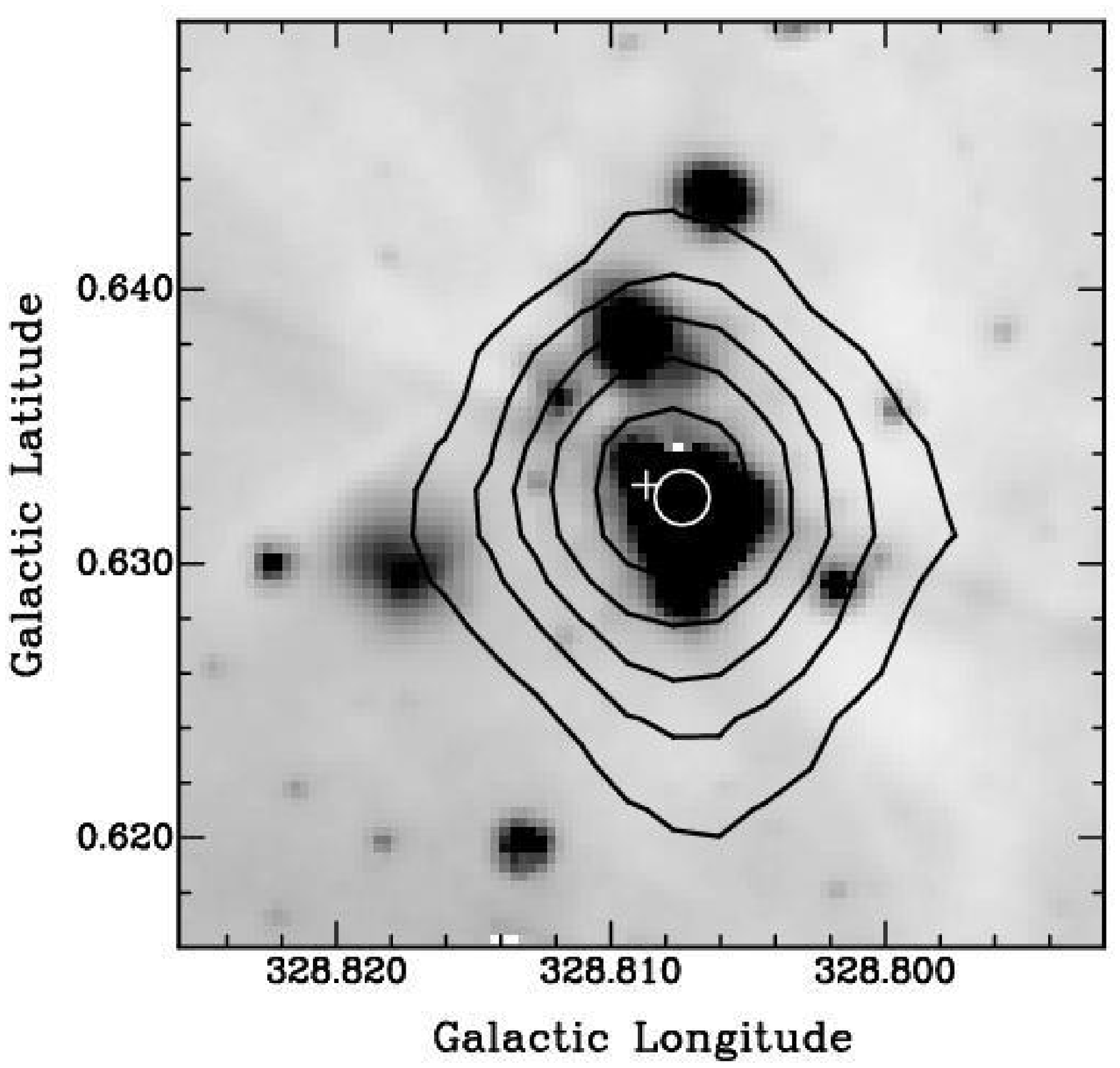}
  \caption{As for Fig.~\ref{fig:images1} for the sources
    G\,$327.618\!-\!0.111$, G\,$327.945\!-\!0.115$,
    G\,$328.237\!-\!0.548$ \& G\,$328.254\!-\!0.532$, G\,$328.809\!+\!0.633$}
  \label{fig:images3}
\end{figure}

G$\,326.475\!+\!0.703$ (Fig.~\ref{fig:images1}) : The methanol maser
lies to the north of a strong 8 and 21~\micron\/ source.  The maser is
projected against one of a number of faint, compact 8~\micron\/
sources and may be associated with the nearest \glim\/ archive source.

G$\,326.641\!+\!0.613$ (Fig.~\ref{fig:images1}) : The maser is at the
edge of a region of strong 21~\micron\/ emission; the (8~\micron)
\glim\/ image is completely saturated.  The maser is associated with a
\glim\/ point source, but it is one for which only the 3.6 \&
4.5~\micron\ measurements are given.

G$\,326.662\!+\!0.521$ (Fig.~\ref{fig:images1}) : This methanol maser
is associated with an {\em MSX} point source and a \glim\/ archive
source.  The 8~\micron\/ image shows the maser projected against
strong extended emission, presumably a moderately evolved star forming
complex.

G$\,326.859\!-\!0.677$ (Fig.~\ref{fig:images1}) : The maser is
associated with a faint 8~\micron\/ source that lies within an IRDC,
approximately 15\arcsec\ to the south there is a stronger source which
has a single narrow extension pointing towards the maser and its
associated \glim\/ point source.

G$\,327.120\!+\!0.511$ (Fig.~\ref{fig:images2}) : This is a moderately
strong methanol maser source that is associated with OH and water
masers and \glim\/ and {\em MSX} point sources.  It is projected
against extended 8 \& 21~\micron\/ emission and lies near the edge of
the large scale diffuse PAH emission from the Galactic Plane.

G$\,327.402\!+\!0.444$ \& G$\,327.401\!+\!0.445$
(Fig.~\ref{fig:images2}) : This complex region contains a number of
extended 8~\micron\/ sources with diffuse emission to the north and
low-contrast absorption to the south and east perhaps indicating the
remnants of an IRDC.  There are water, OH and two sites of methanol
masers in this region separated by 5\arcsec.  The methanol masers are
projected against extended 8~\micron\/ emission and their separation
is approximately perpendicular to the direction of extension of the
dust emission. The VLBI images of \citet{DOE04} show the maser
emission is elongated along approximately the same direction.

G$\,327.392\!+\!0.199$ (Fig.~\ref{fig:images2}) : This region shows
weak 21~\micron\/ emission, with a number of 8~\micron\/ sources at
its edges, all of this embedded in a low-contrast IRDC.  The methanol
maser is associated with a slightly extended 8~\micron\/ source, there
is a \glim\/ archive source with measurements at 5.8 \& 8.0~\micron\/ within
2~\arcsec.

G$\,328.237\!-\!0.548$ \& G$\,328.254\!-\!0.532$
(Fig.~\ref{fig:images3}) : These two strong methanol masers each lie
at the edge of extended 8~\micron\/ emission which is embedded within
a single large IRDC.  G$\,328.237\!-\!0.548$ has associated OH and
water masers, while G$\,328.254\!-\!0.532$ is coincident with the
strongest 8.0~\micron\/ emission in the vicinity and is also the centre
of {\em MSX} 21~\micron\/ emission.  Previous mid-infrared images by
\citet{DPT00} and \citet{WBBN01} detected the emission associated with
G$\,328.254\!-\!0.532$, but showed the masers offset.  The better
astrometry of the \glim\/ observations shows that the methanol masers
are in fact projected against the mid-infrared source.

G$\,328.810\!+\!0.633$ (Fig.~\ref{fig:images3}) : This is another
strong, well studied methanol maser associated with class I methanol
and OH masers (but not water).  The masers are projected against an
\ionhy\/ region which is extended in the same direction as the
mid-infrared emission, but to a much smaller degree \citep{ESK05} and
this region is seen in 8~\micron\/ \glim\/ emission.  Previous
observations by \citet{DPT00} and \citet{WBBN01} are consistent
with the \glim\ observations, but with the greater sensitivity and
positional accuracy of \glim, all the masers can be seen to be
projected against mid-infrared emission.

\begin{figure}
  \twofields{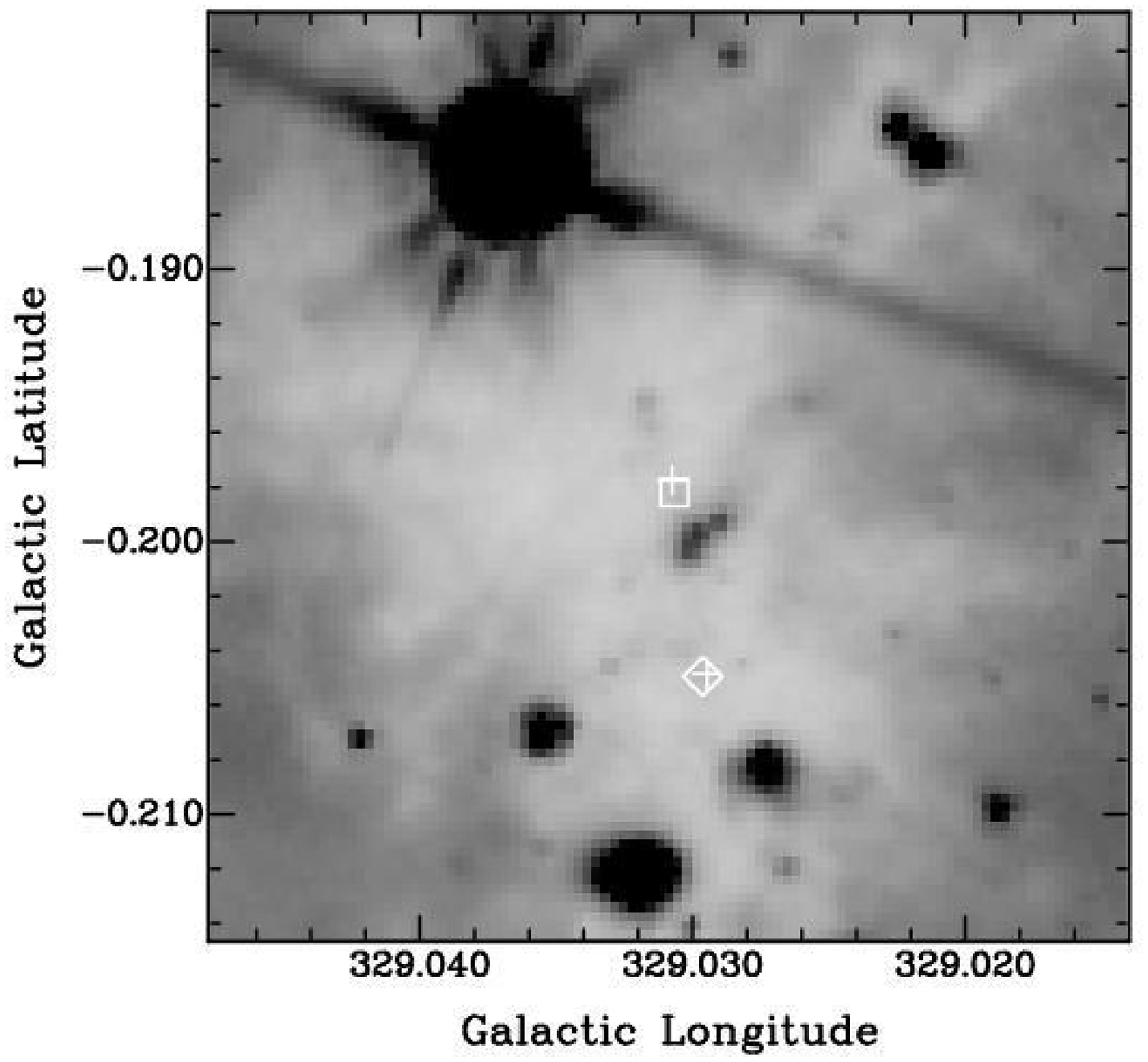}{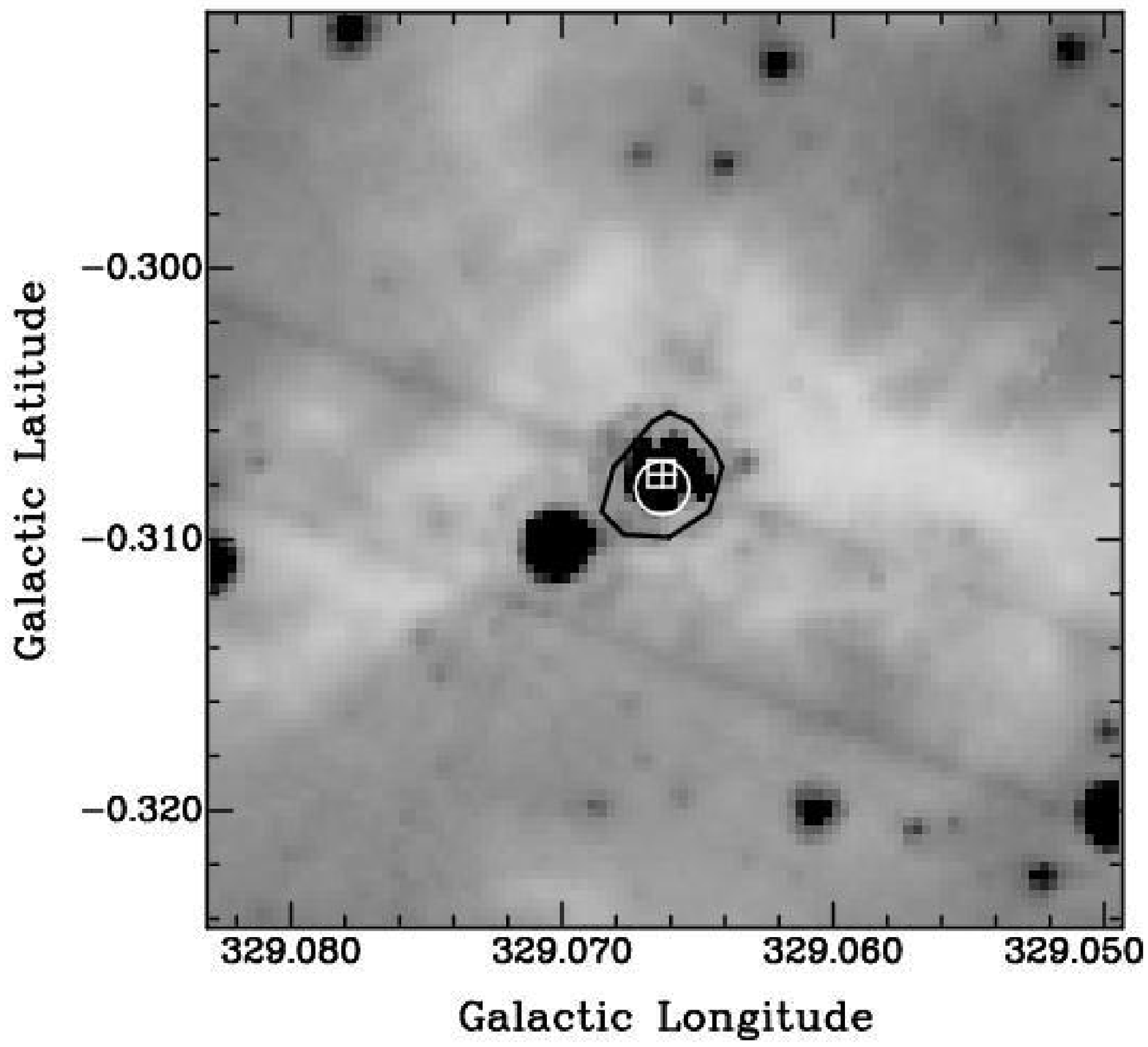}
  \twofields{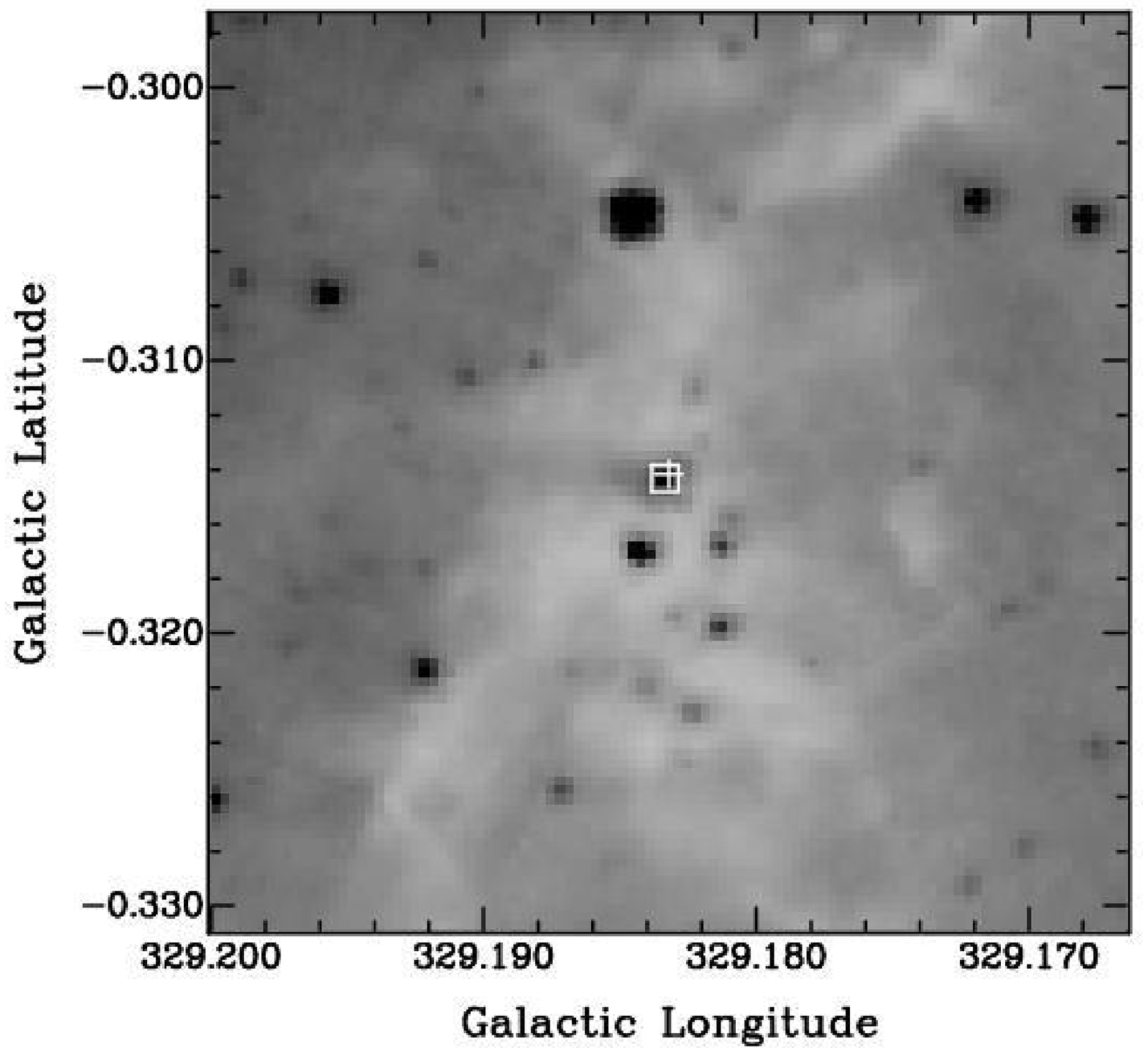}{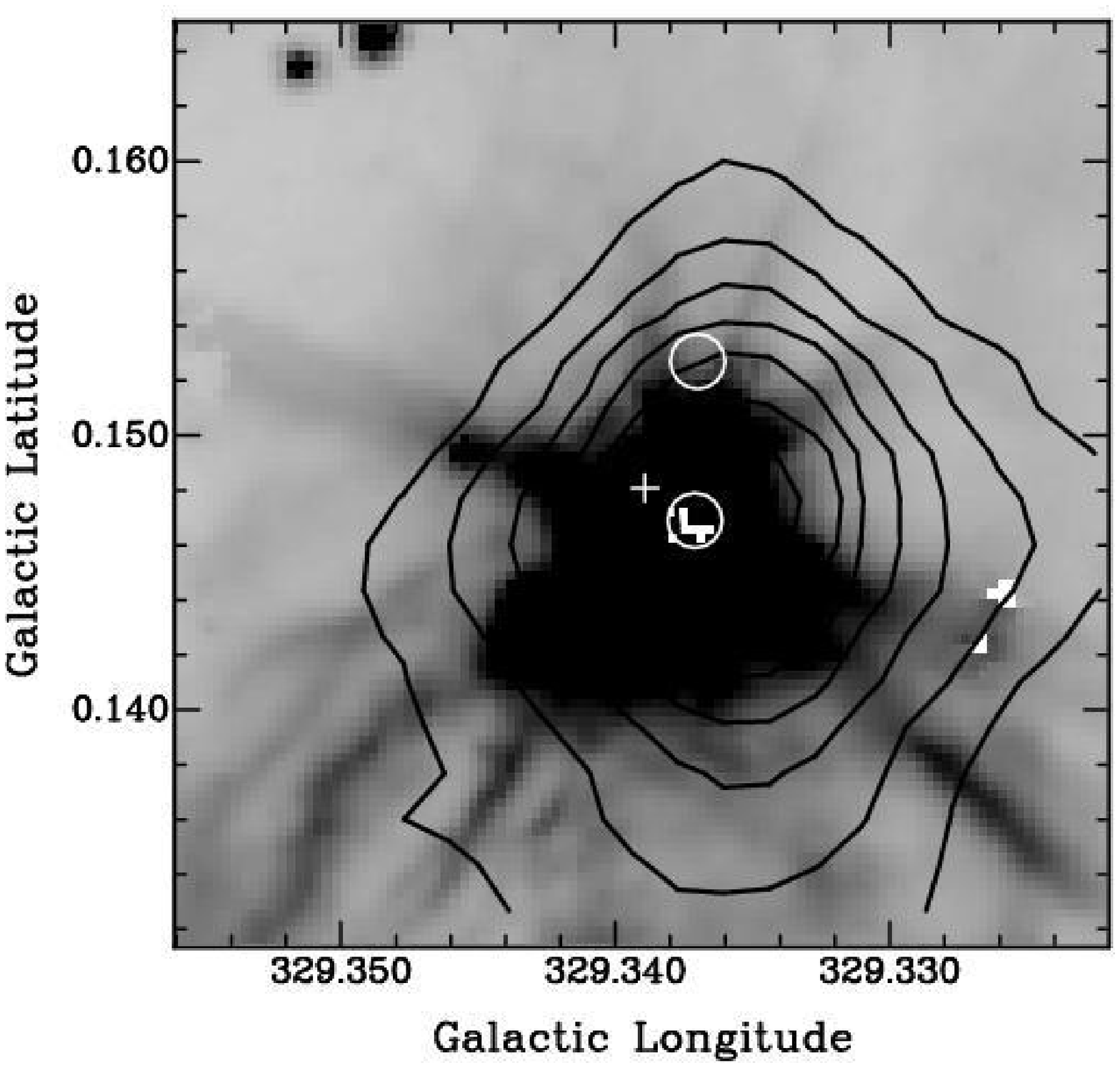}
  \caption{As for Fig.~\ref{fig:images1} for the sources
    G\,$329.031\!-\!0.198$ \&
    G\,$329.029\!-\!0.205$, G\,$329.066\!-\!0.308$,
    G\,$329.183\!-\!0.314$, G\,$329.339\!+\!0.148$}
  \label{fig:images4}
\end{figure}

\begin{figure}
  \twofields{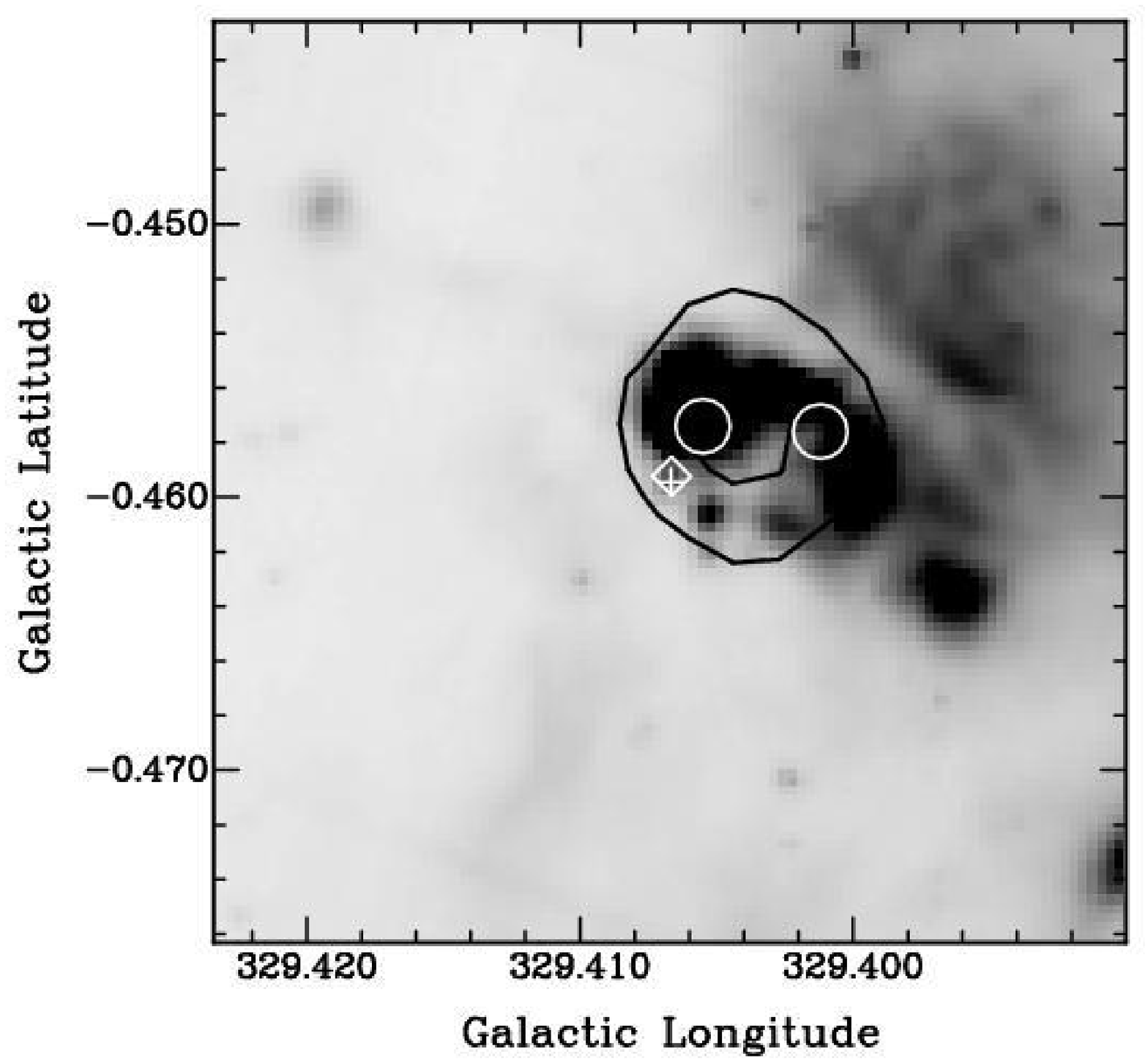}{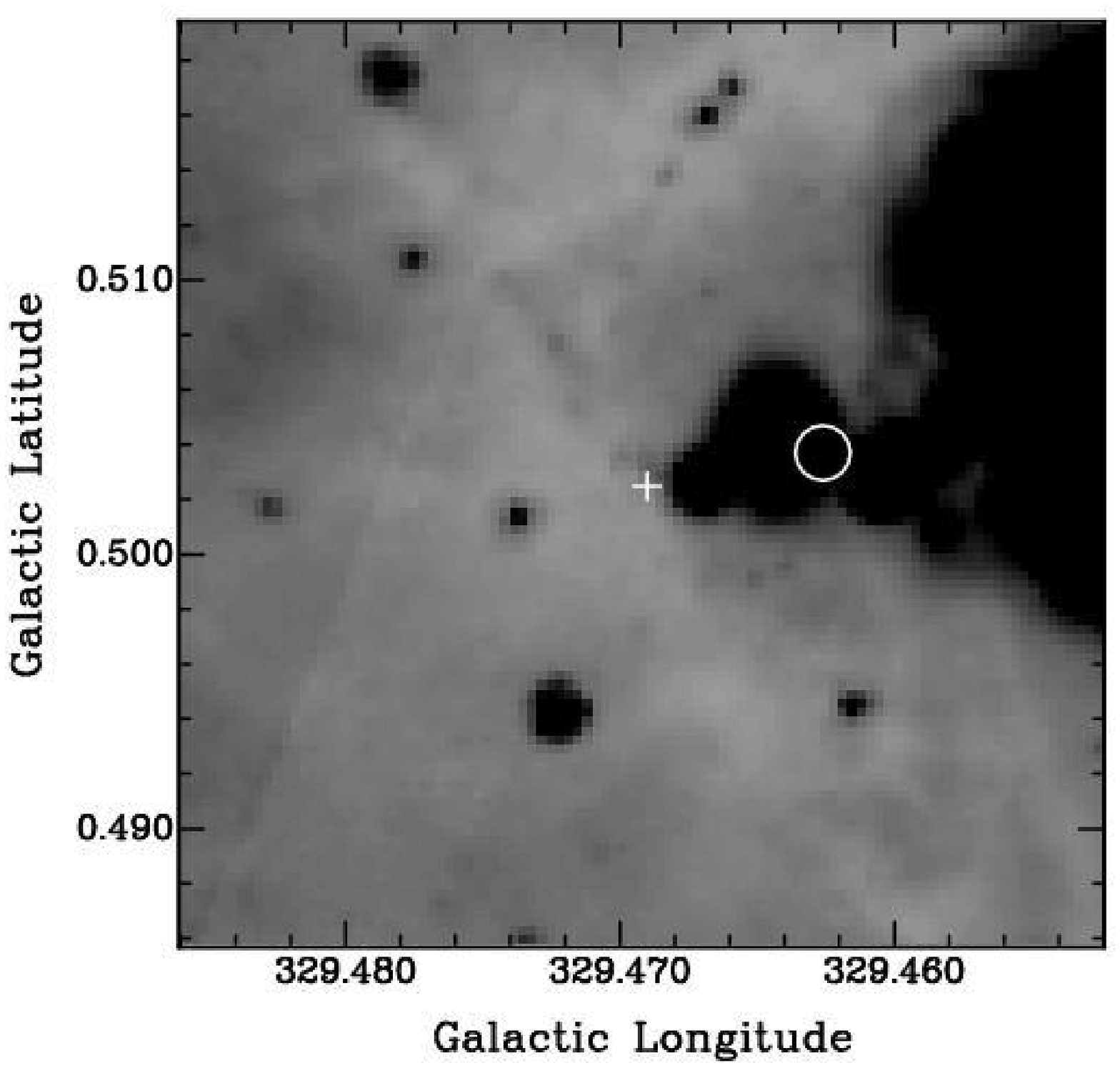}
  \twofields{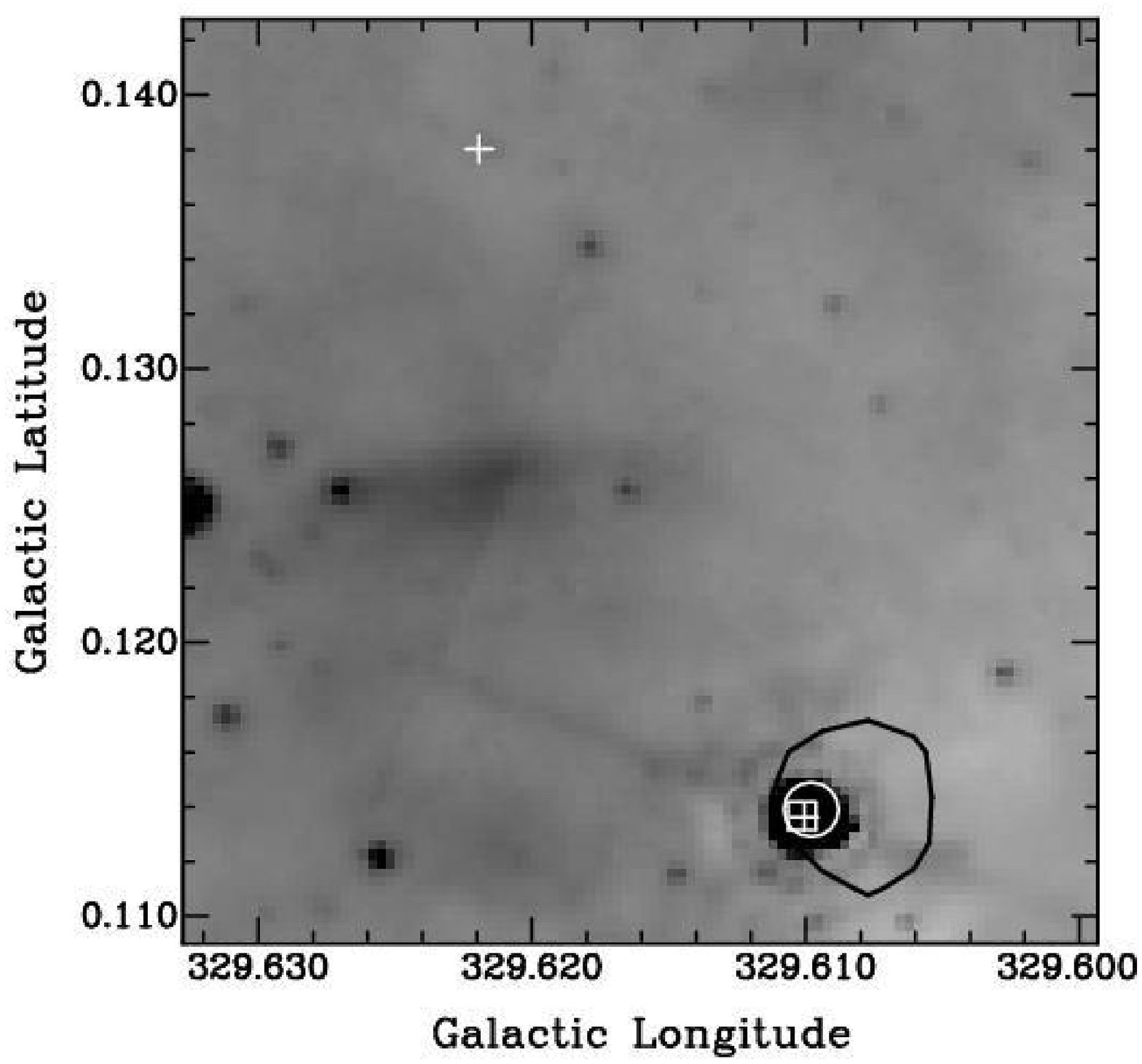}{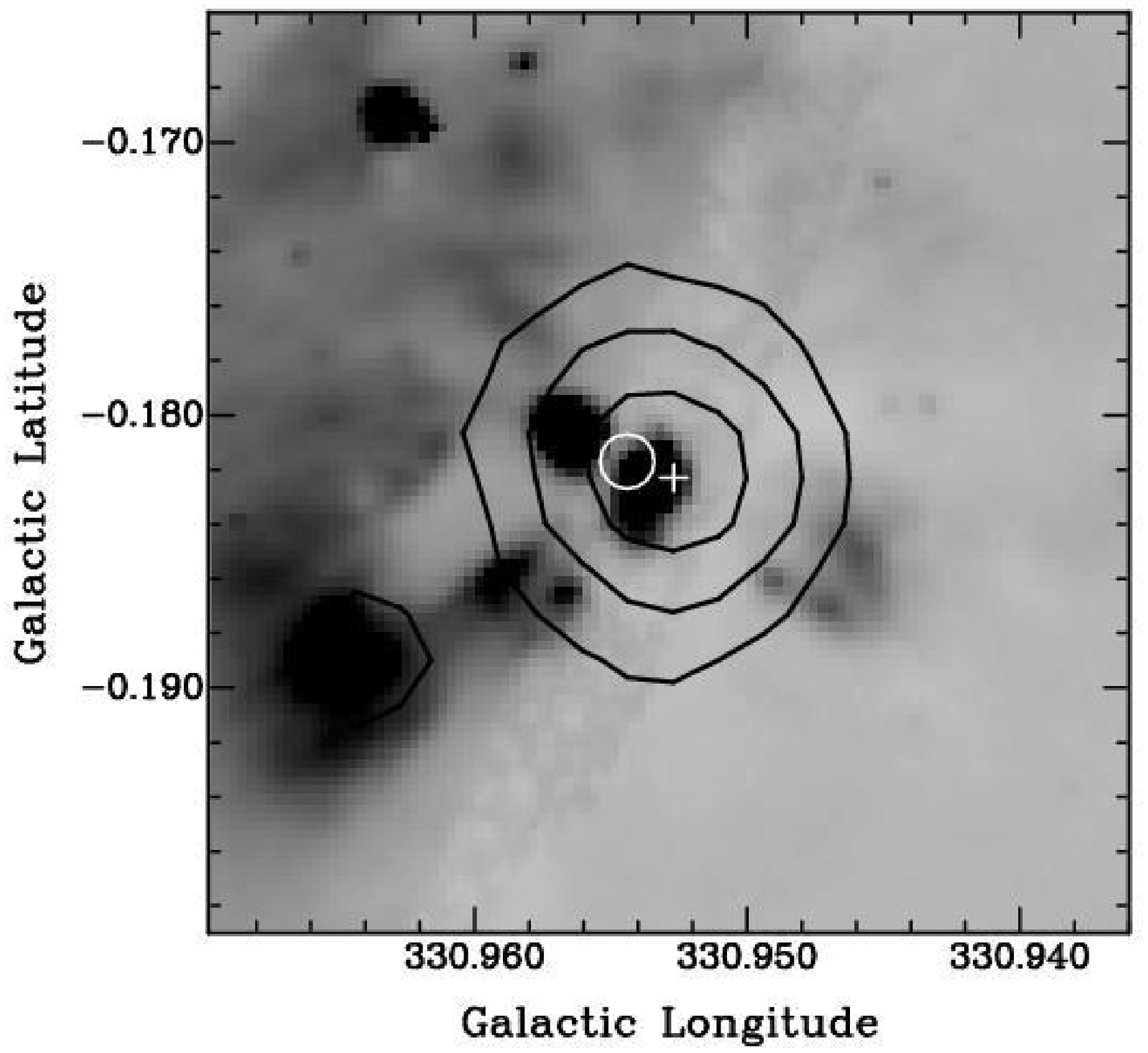}
  \caption{As for Fig.~\ref{fig:images1} for the sources
    G\,$329.407\!-\!0.459$, G\,$329.469\!+\!0.502$, 
    G\,$329.622\!+\!0.138$ \& G\,$329.610\!+\!0.114$,
    G\,$330.952\!-\!0.182$}
  \label{fig:images5}
\end{figure}

\begin{figure}
  \twofields{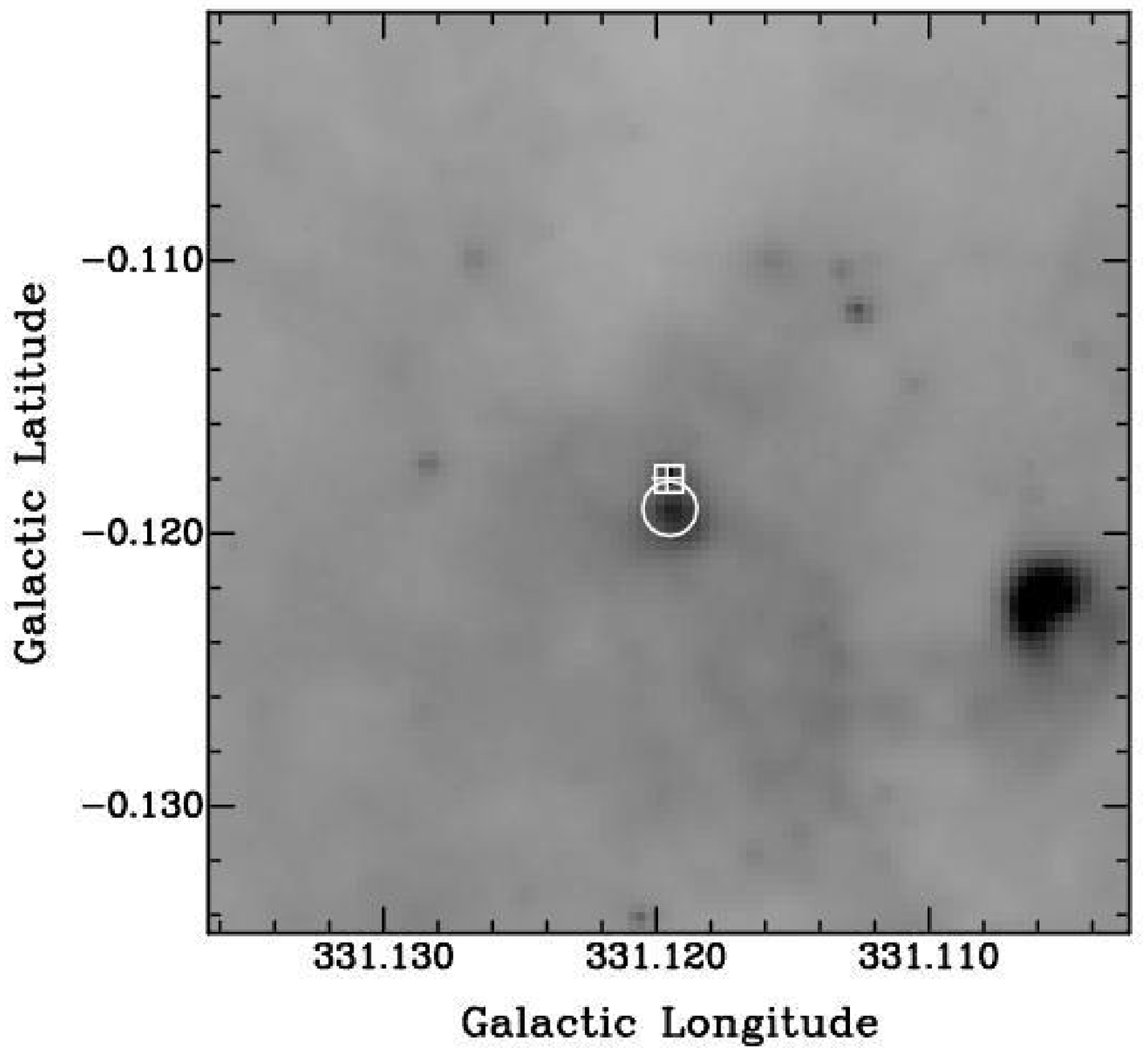}{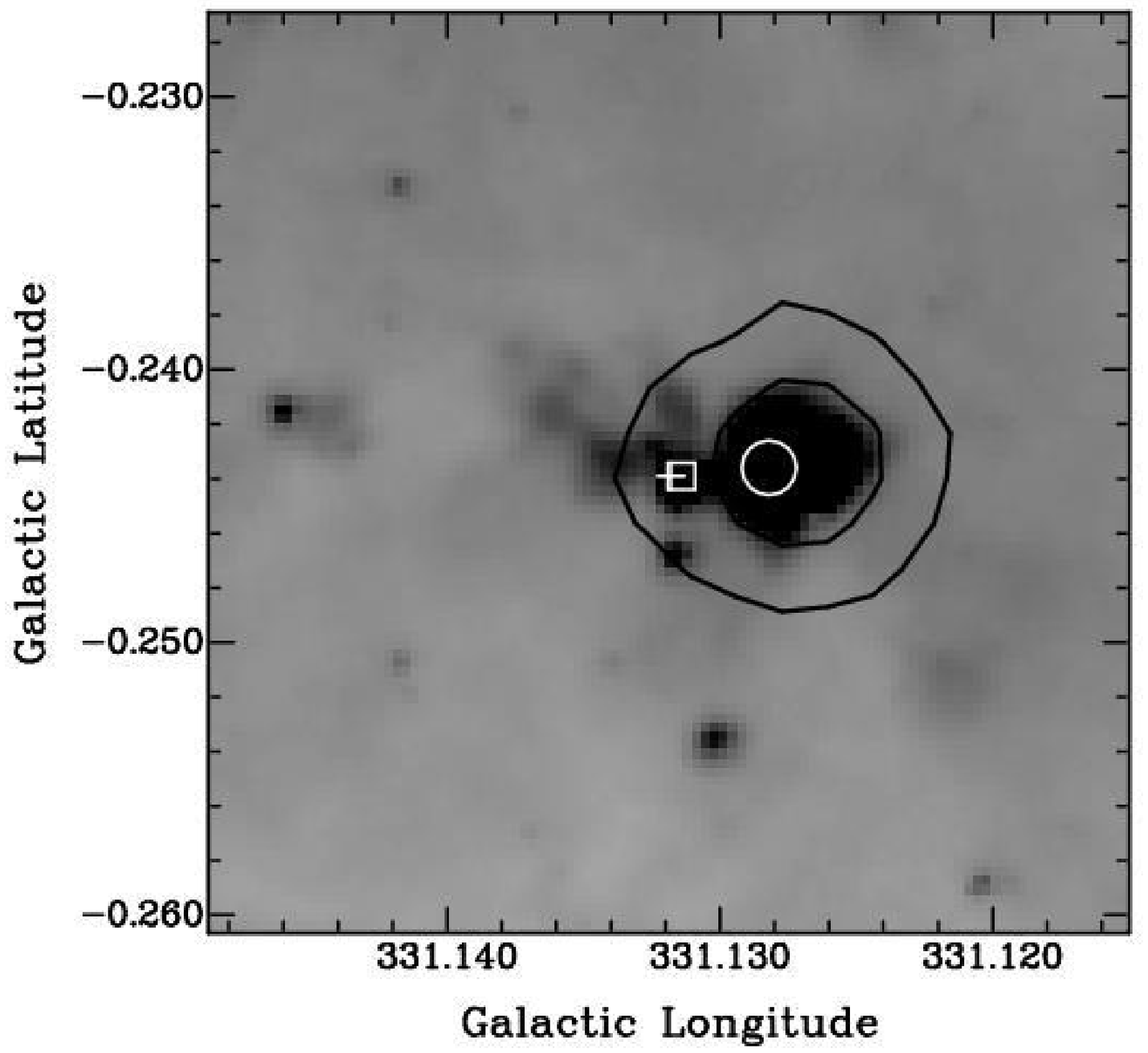}
  \twofields{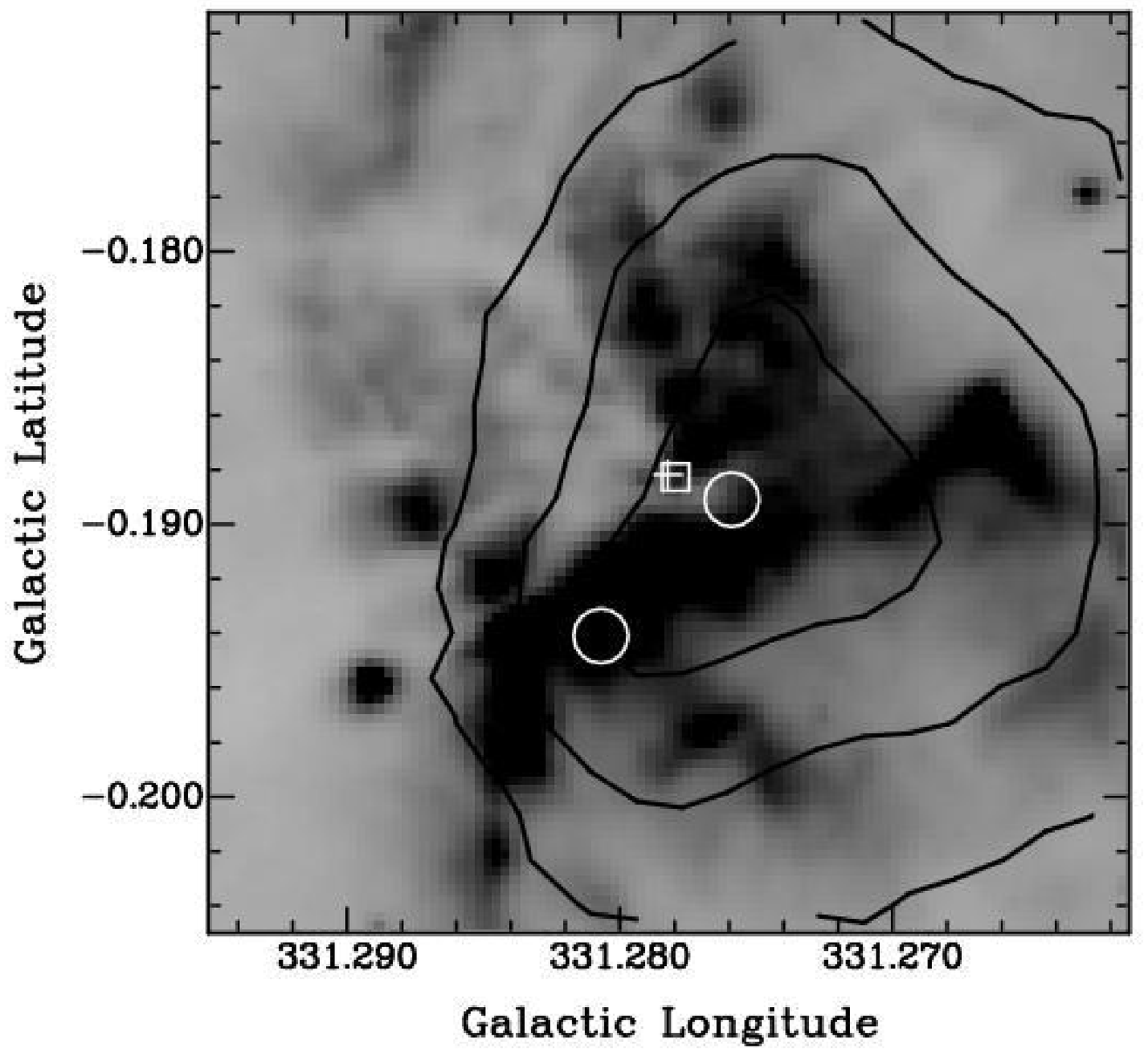}{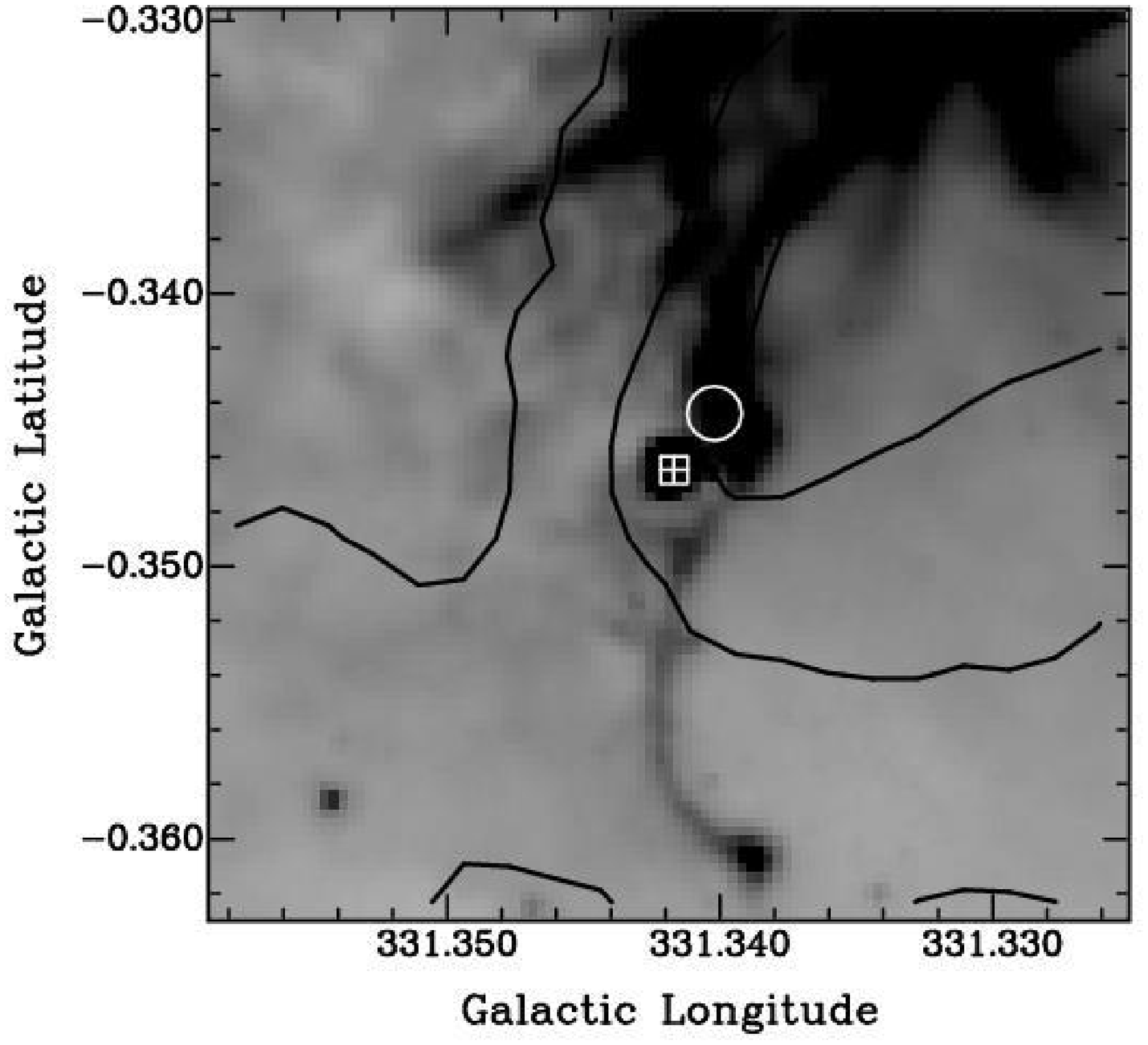}
  \caption{As for Fig.~\ref{fig:images1} for the sources    
    G\,$331.120\!-\!0.118$, G\,$331.132\!-\!0.244$,
    G\,$331.278\!-\!0.188$, G\,$331.342\!-\!0.346$}
  \label{fig:images6}
\end{figure}

\begin{figure}
  \twofields{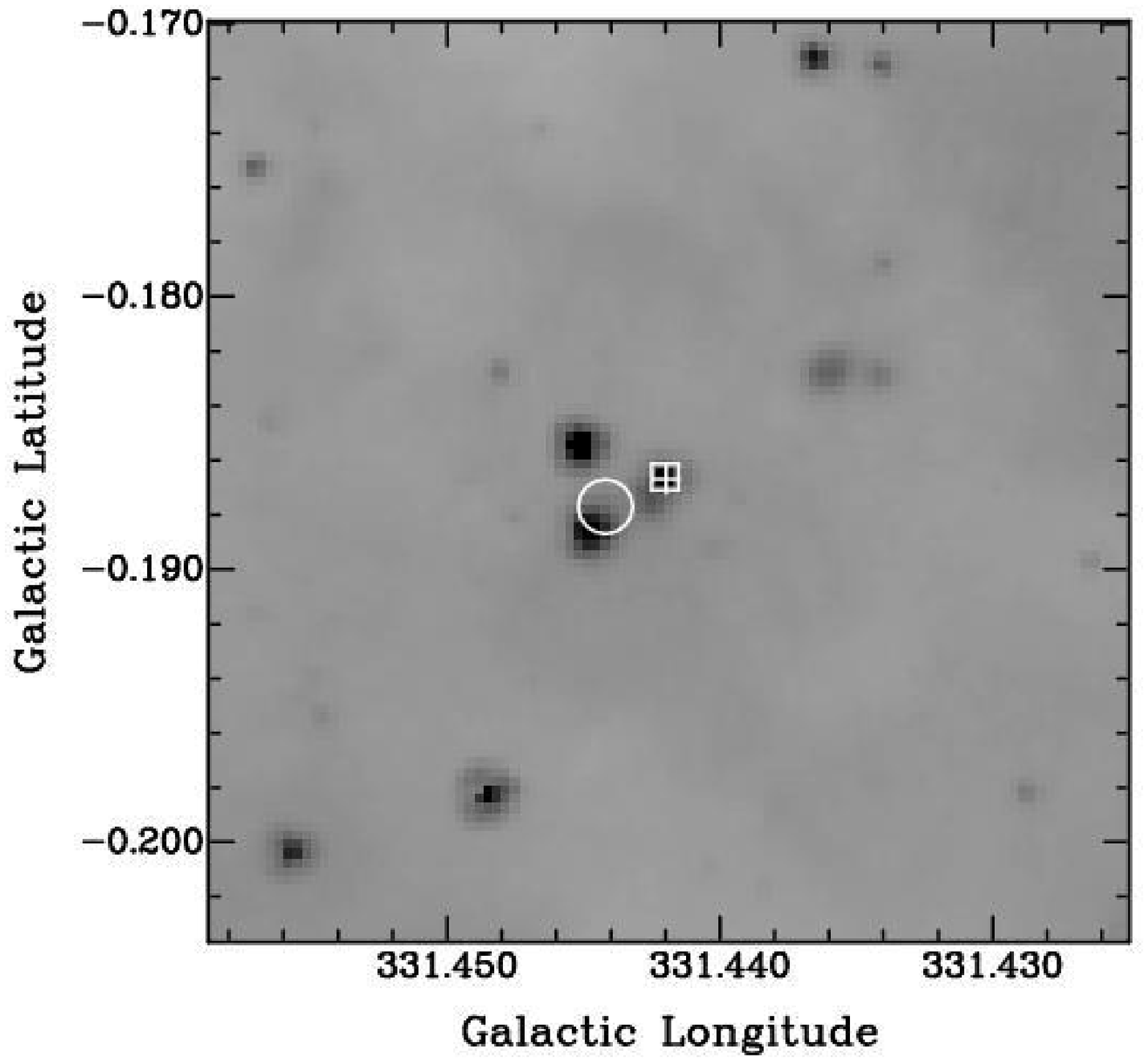}{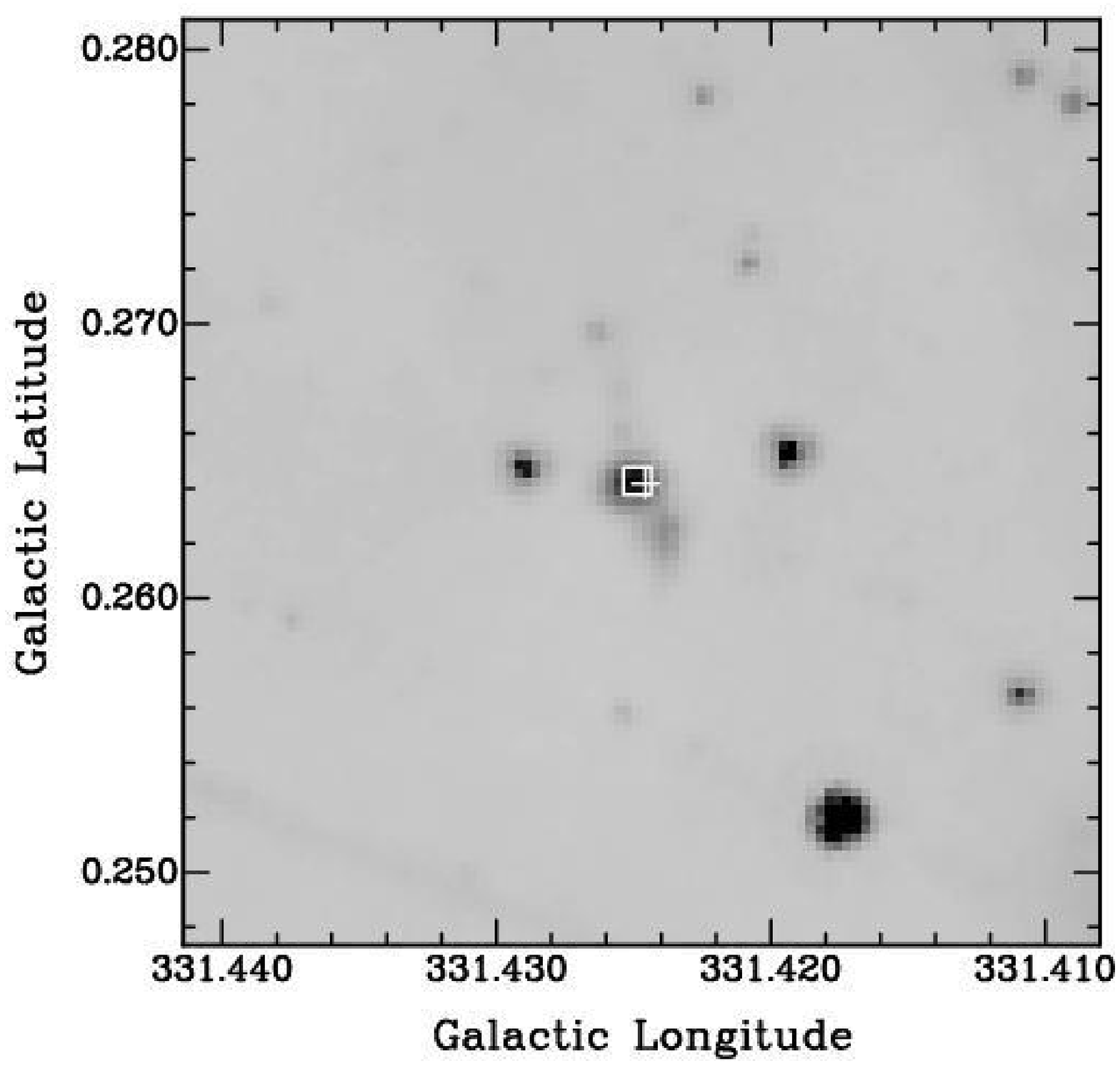}
  \twofields{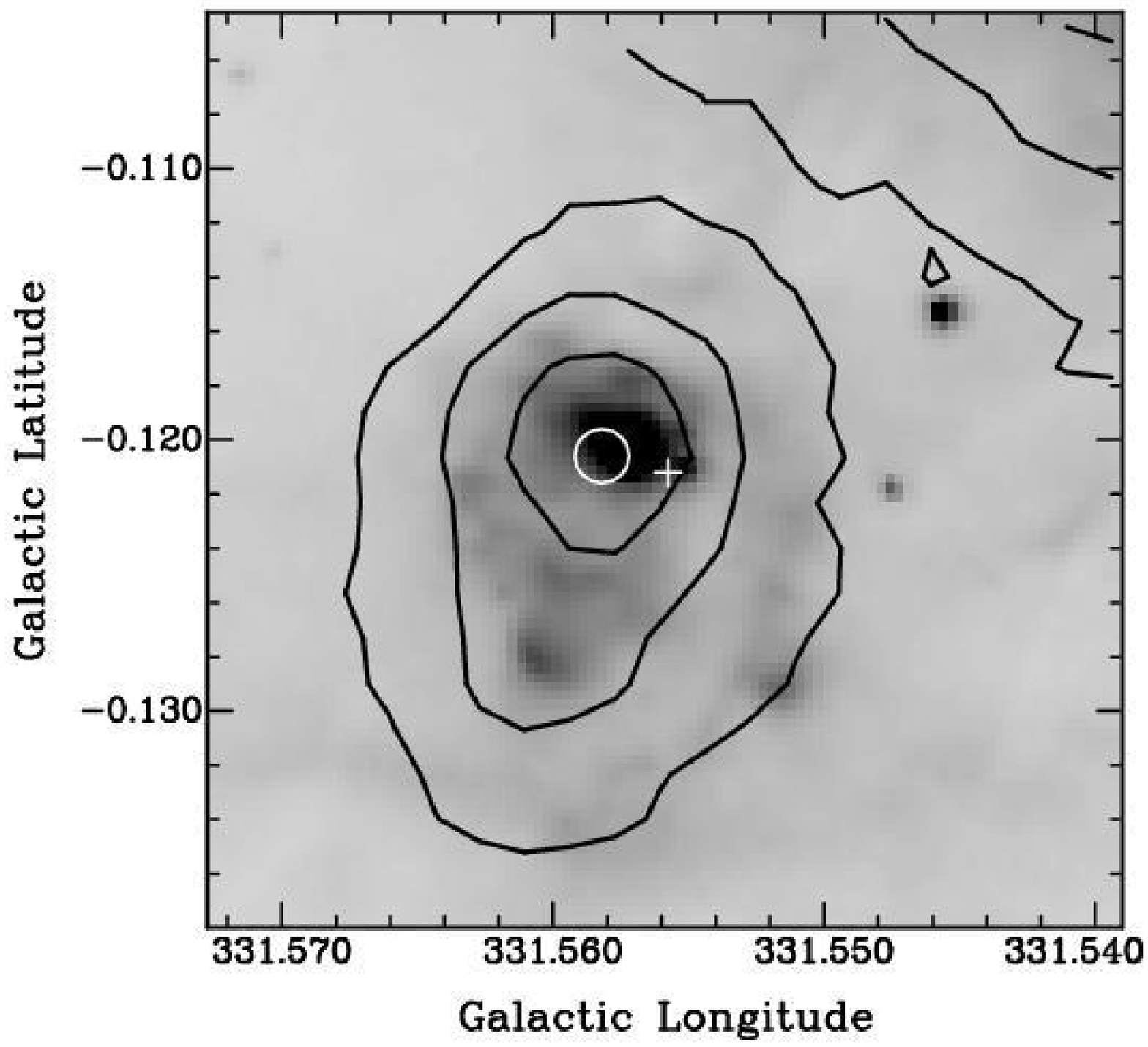}{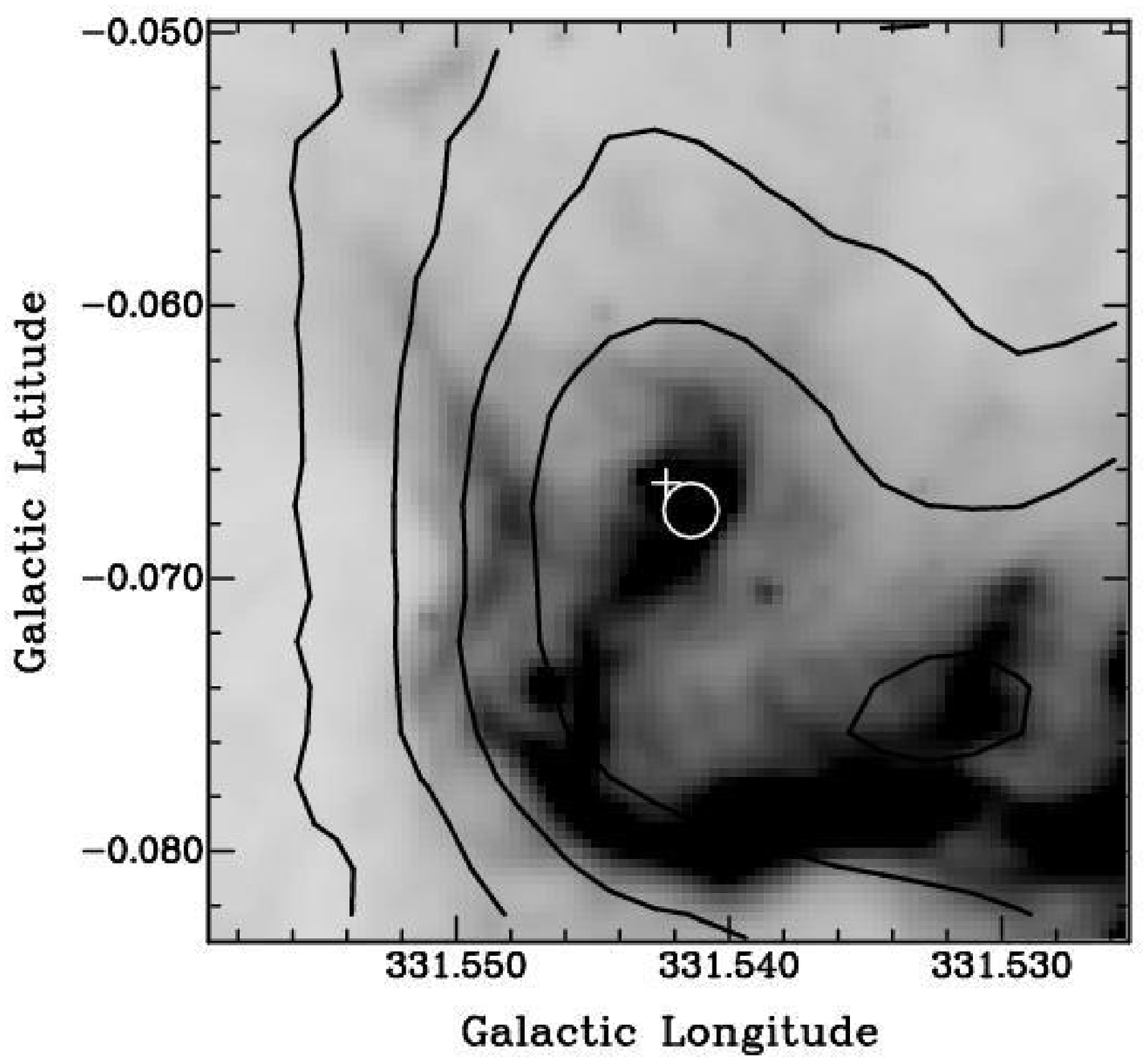}
  \caption{As for Fig.~\ref{fig:images1} for the sources
    G\,$331.442\!-\!0.187$, G\,$331.425\!+\!0.264$
    G\,$331.542\!-\!0.066$, G\,$331.556\!-\!0.121$}
  \label{fig:images7}
\end{figure}

G\,$329.031\!-\!0.198$ \& G\,$329.029\!-\!0.205$
(Fig.~\ref{fig:images4}) : These two methanol maser sources lie at the
northern end of a very long dark filament (G\,$329.05\!-\!0.30$),
which also contains G$\,329.066\!-\!0.308$ and G$\,329.183\!-\!0.314$
and is discussed in more detail in section~\ref{sec:filament}.
Figure~\ref{fig:images4} appears to show that neither of the masers
are associated with 8.0~\micron\/ emission, however, each is within
2~\arcsec\ of a \glim\/ source that Table~\ref{glimprop} shows to have
a detection in this band.  Comparison with the other sources listed in
Table~\ref{glimprop} shows that in each case the emission is several
magnitudes weaker than any of the other maser associated sources and
this emission is only visible when the image is scaled in such a way
that the majority of the region is saturated.  The unusual weakness of
the 8.0~\micron\/ emission in these sources suggests that either the
extinction in this IRDC is very high, or these methanol masers are
associated with a particularly young high-mass star formation region.
There are class~I methanol, water and OH masers associated with both
6.7~GHz maser sites.

G\,$329.066\!-\!0.308$ \& G$\,329.183\!-\!0.314$
(Fig.~\ref{fig:images4}) : These masers are both projected against
8.0~\micron\ emission near the middle and opposite extremity
respectively of the G\,$329.05\!-\!0.30$ dark filament to
G\,$329.029\!-\!0.205$.  Each of the masers is associated with a
\glim\/ point source with significantly stronger 8.0~\micron\/
emission than G\,$329.031\!-\!0.198$ or G\,$329.029\!-\!0.205$.

G$\,329.339\!+\!0.148$ (Fig.~\ref{fig:images4}) : The maser is
projected against an extended region of mid-infrared emission, some of
which has saturated at 8~\micron.  \citet{WBHR98} did not detect any
radio continuum emission associated with the masers.

G$\,329.407\!-\!0.459$ (Fig.~\ref{fig:images5}) : The maser emission
in this region lies projected against a small extension of a large
extended mid-infrared source visible at both 8 \& 21~\micron.  The
maser lies at the edge of the extended mid-infrared emission.

G\,$329.469\!+\!0.502$ (Fig.~\ref{fig:images5}) : The methanol masers
lie at the edge of a large region of 8.0~\micron\/ emission which is
relatively bright in comparison to the majority of those with
associated masers and is projected against an IRDC which has lower
contrast than most, perhaps indicating that it is at the far kinematic
distance.  The maser emission is not associated with a \glim\/ point
or archive source and \citet{WBHR98} did not detect any radio
continuum emission associated with the masers.

G\,$330.952\!-\!0.182$ (Fig.~\ref{fig:images5}) : The methanol masers
are associated with both an OH and water maser and lie near the
interface between a large complex region of mid-infrared emission and
an IRDC.  \citet{WBHR98} found that the methanol masers lie at the
edge of a larger \ionhy\/ region which is likely associated with the
\glim\/ and {\em MSX} emission.

G\,$331.132\!-\!0.244$ (Fig.~\ref{fig:images6}) : The methanol masers
in this source are projected against an \ionhy\/ region \citep{PNEM98}
which also has associated OH, water and a class~I methanol maser.  In
terms of the mid-infrared emission the masers lie near the edge of a
complex region which is surrounded by a low contrast dark cloud.

G\,$331.278\!-\!0.188$ (Fig.~\ref{fig:images6}) : This source has
previously been imaged in the mid-infrared by \citet{DPT00} and
\citet{WBBN01}, the latter finding the masers associated with a
compact 10~\micron\/ source and the former finding the masers
offset from the same source.  The \glim\/ images show that the masers
are projected against a finger of low level 8.0~\micron\/ emission, at
the edge of a large complex region.  The masers are projected against
weak, relatively diffuse radio continuum emission which is extended
along the same axis as the mid-infrared emission \citep{PNEM98}.
\citet{LWBA01} detected H$_{2}$ emission in the same region, perhaps
powered by an outflow from the source associated with the masers.

G\,$331.342\!-\!0.346$ (Fig.~\ref{fig:images6}) : The methanol masers
are projected against 8.0~\micron\ emission at the edge of a large
complex region.  There is a smooth, thin curved finger of mid-infrared
emission that commenences just south of the masers and extends for
approximately 1\arcmin.

G\,$331.542\!-\!0.066$ (Fig.~\ref{fig:images7}) : This is one of the
less common cases where the methanol masers lie near the centre of an
extended region of mid-infrared emission at both 8.0 and 21~\micron.

\begin{figure}
  \twofields{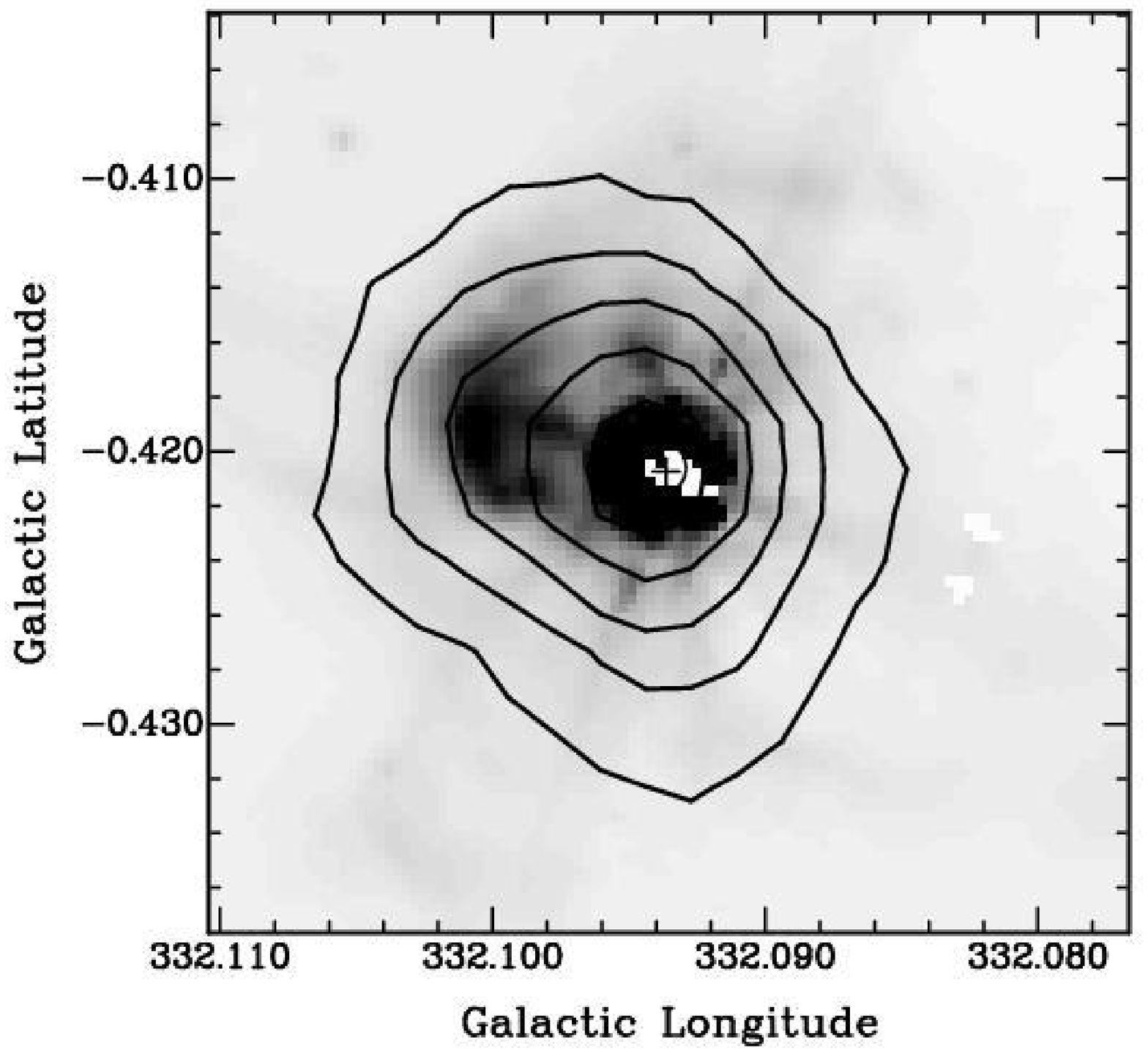}{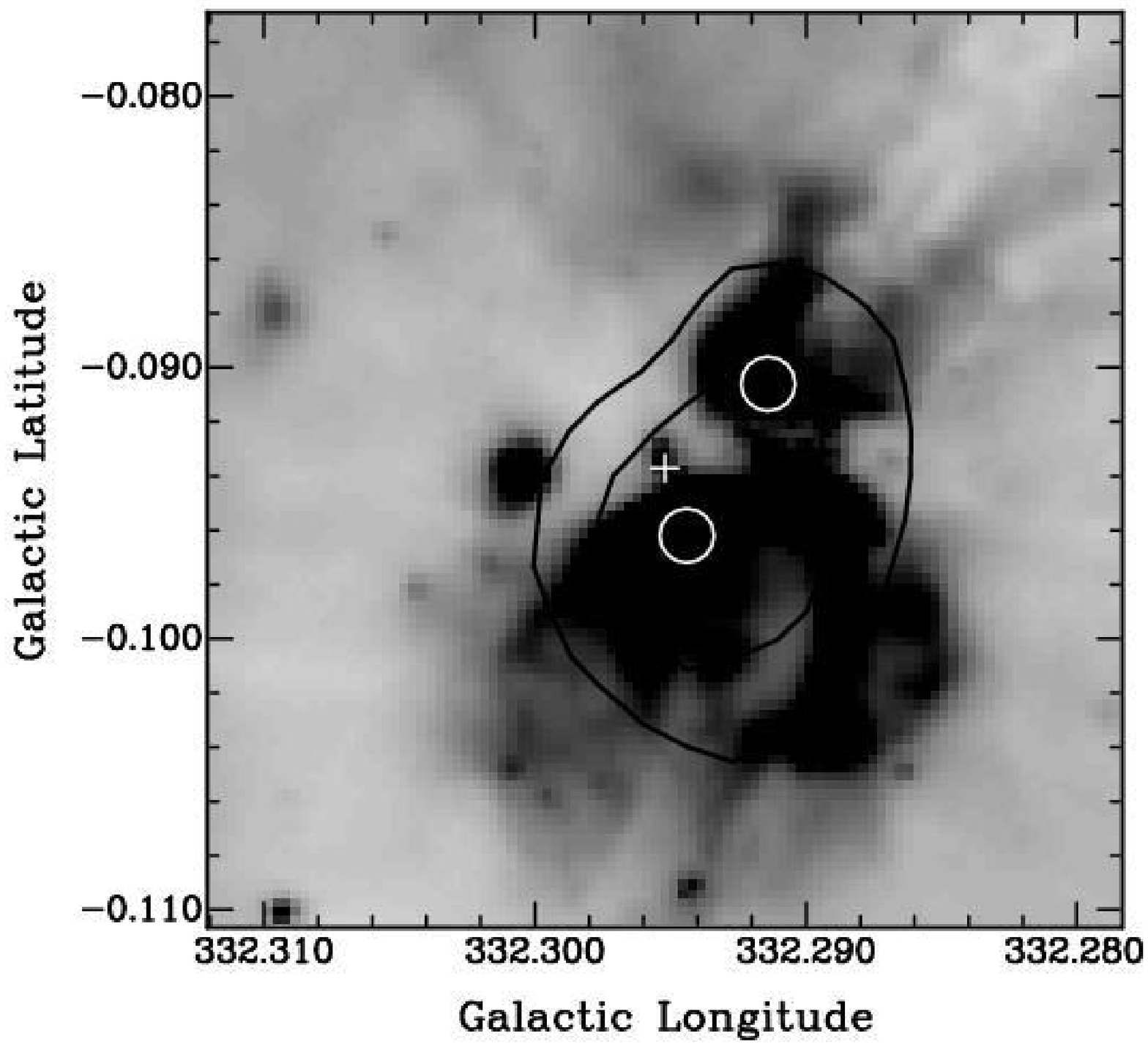}
  \twofields{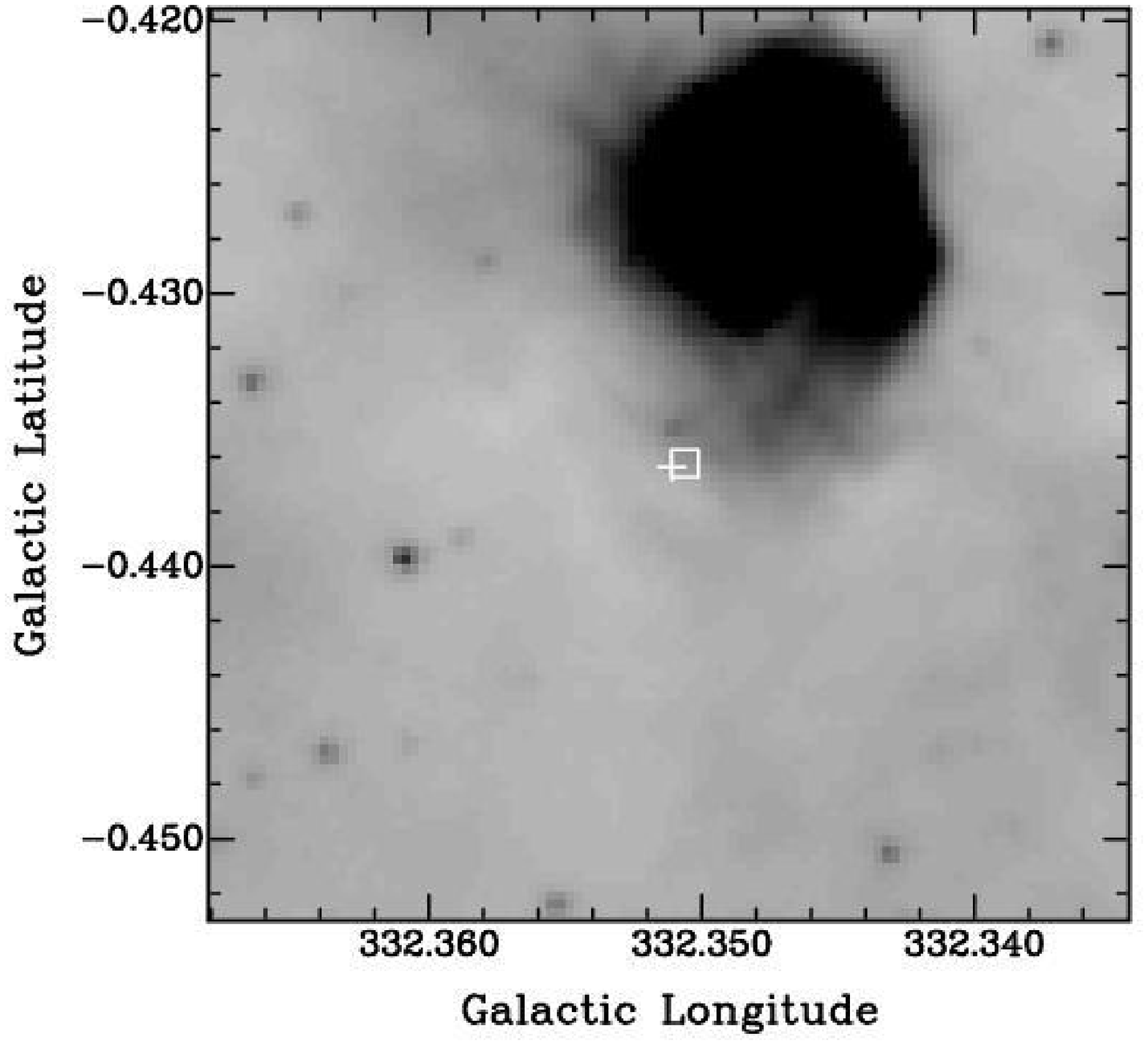}{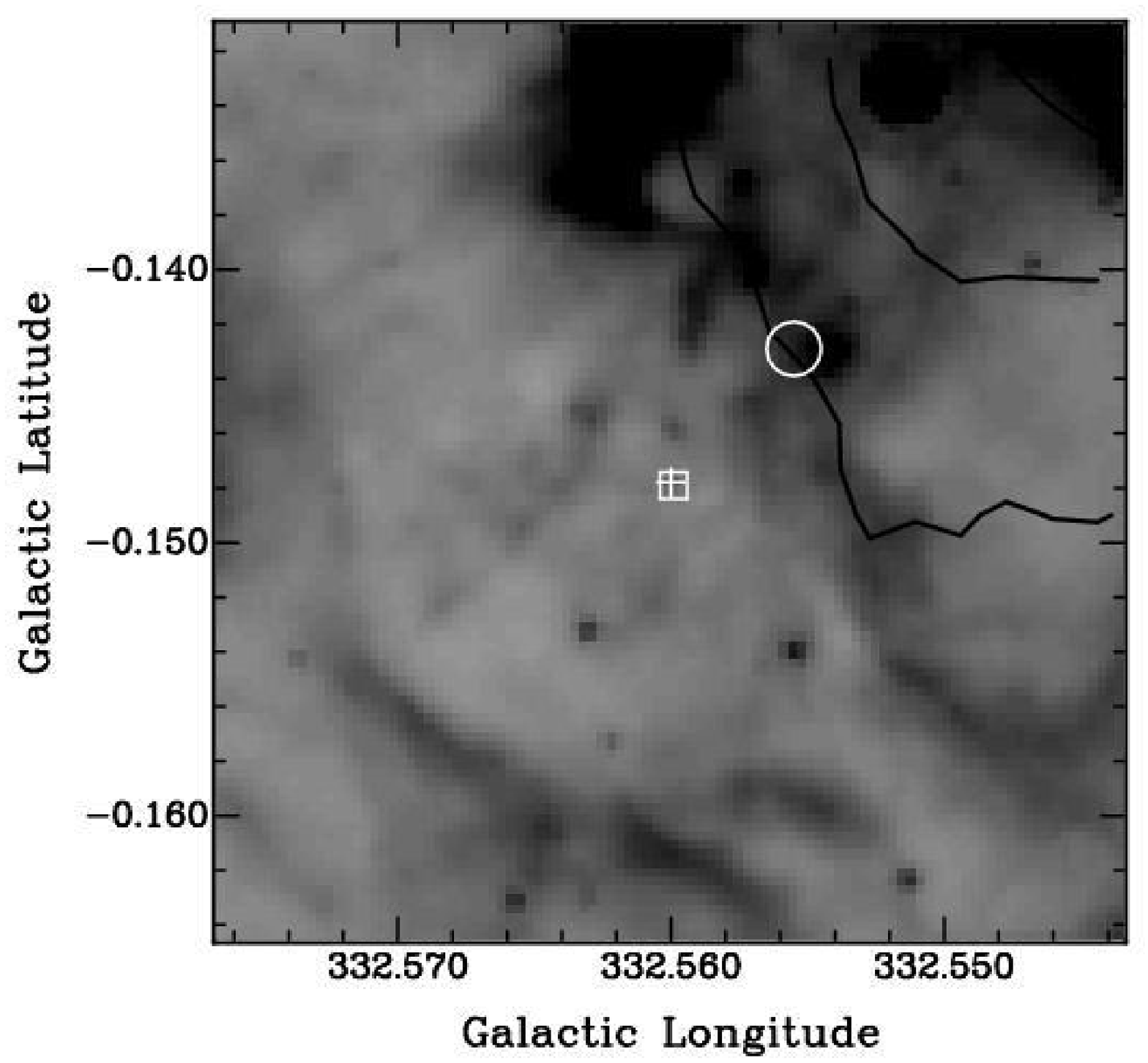}
  \caption{As for Fig.~\ref{fig:images1} for the sources
    G\,$332.094\!-\!0.421$, G\,$332.295\!-\!0.094$,
    G\,$332.351\!-\!0.436$, G\,$332.560\!-\!0.148$}
  \label{fig:images8}
\end{figure}

\begin{figure}
  \twofields{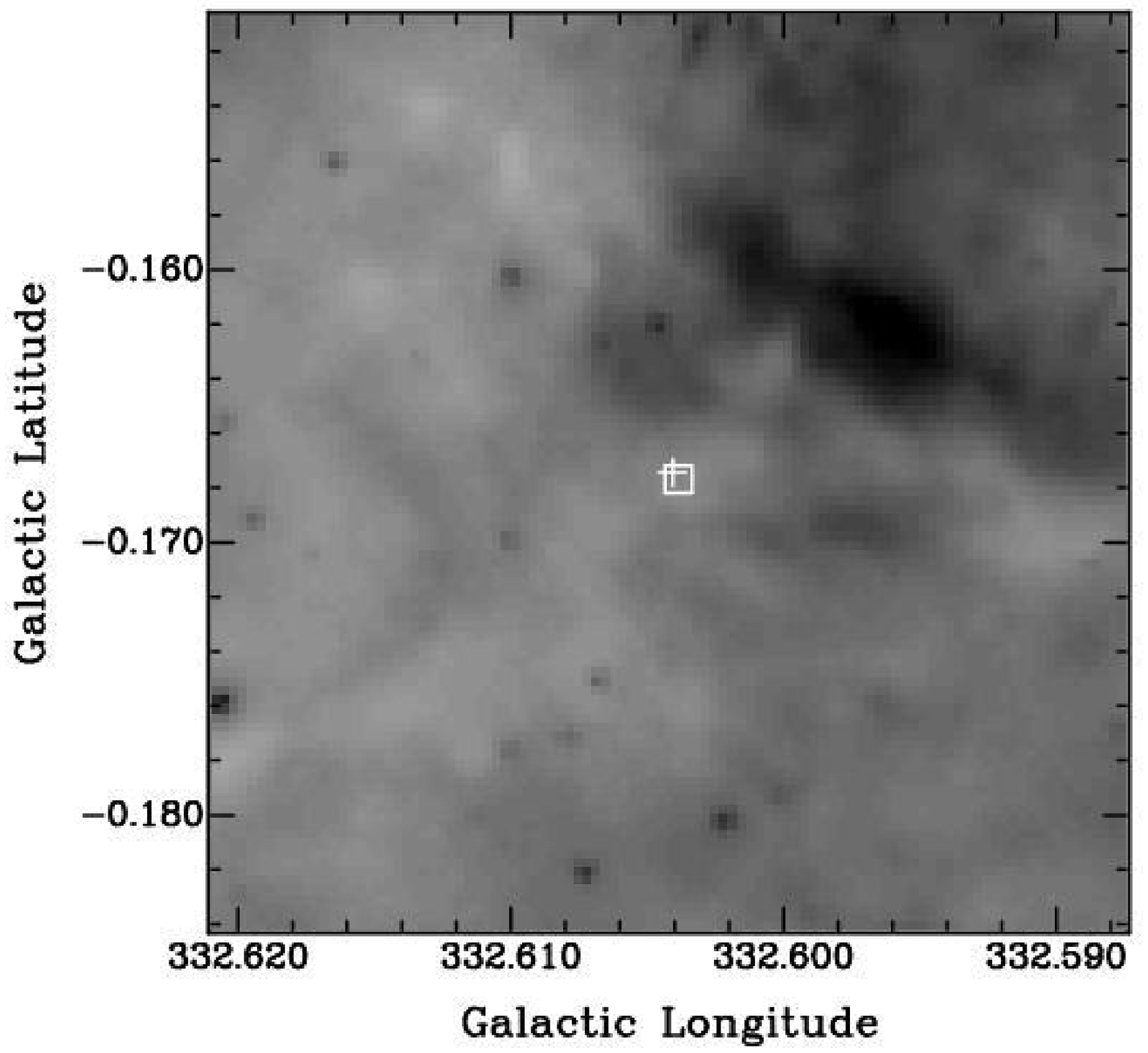}{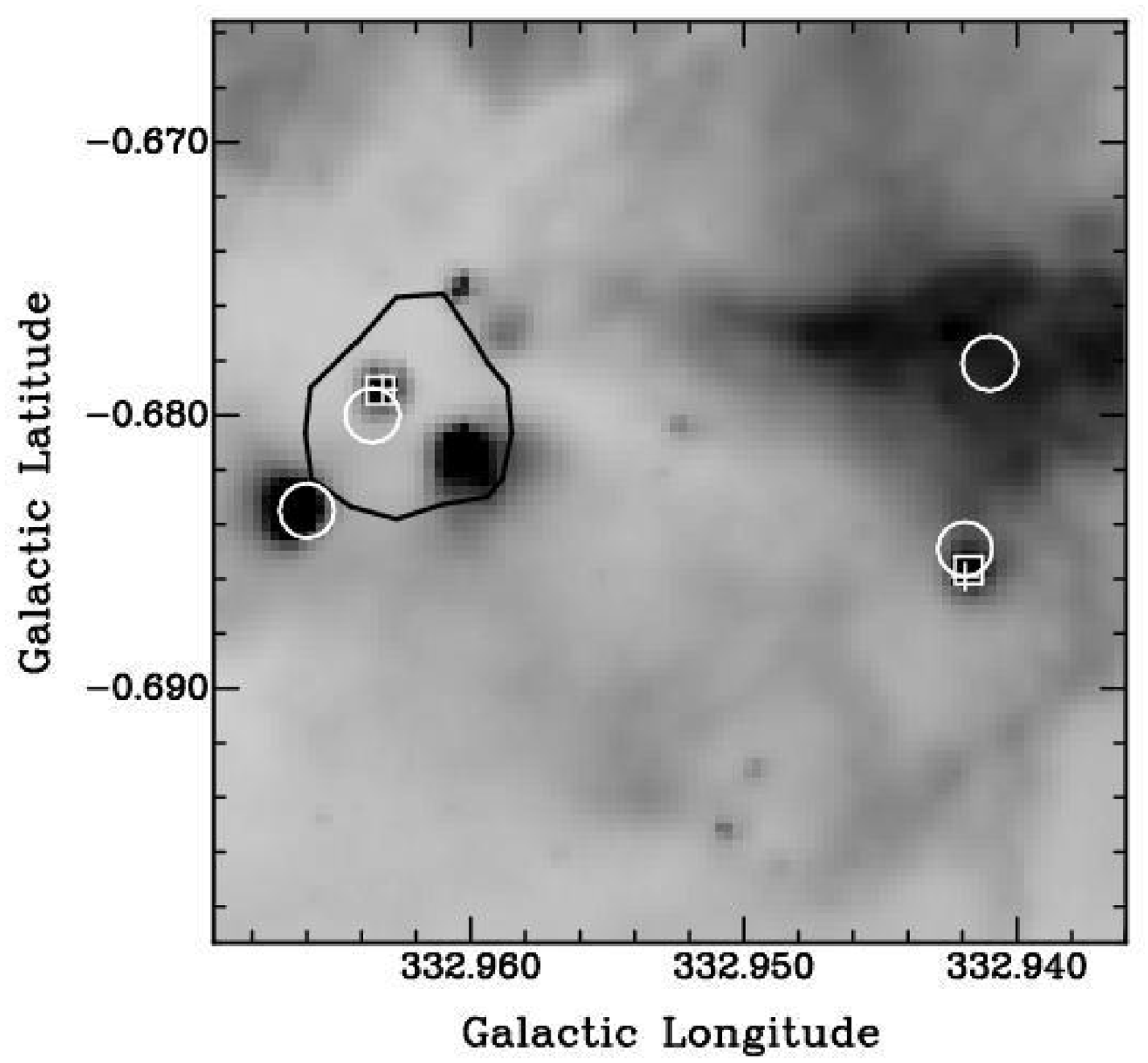}
  \twofields{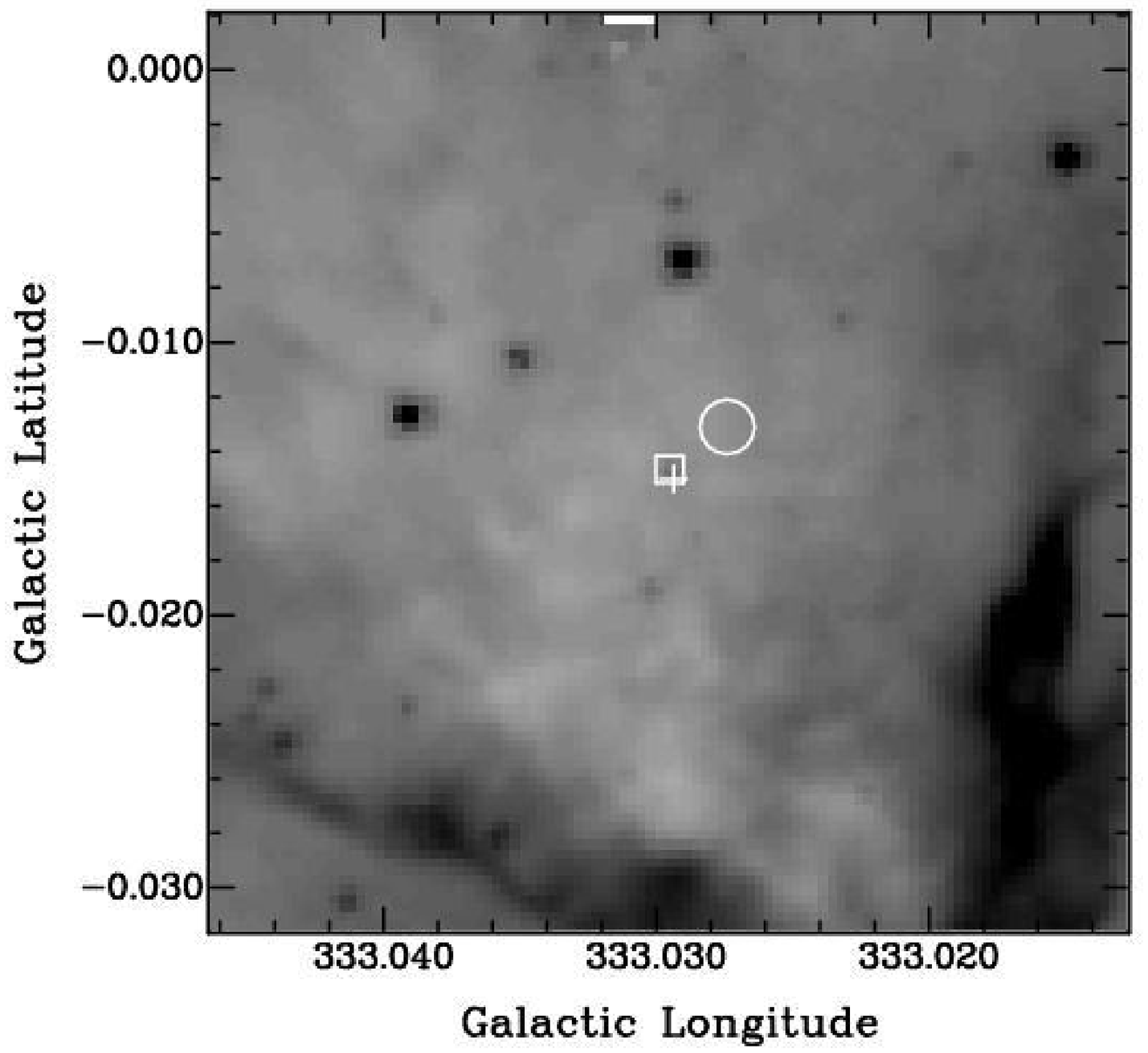}{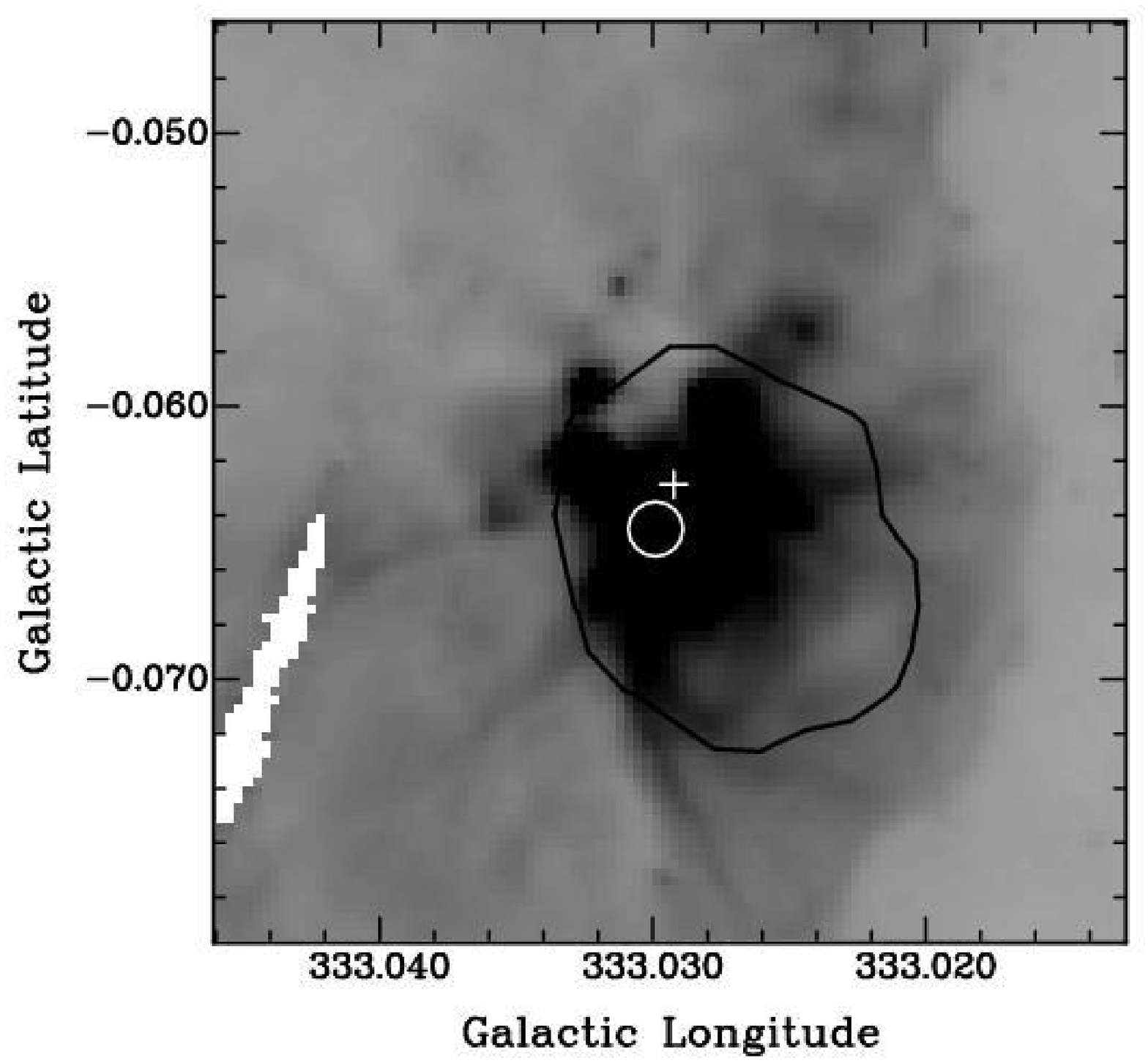}
  \caption{As for Fig.~\ref{fig:images1} for the sources  
    G\,$332.604\!-\!0.167$, G\,$332.942\!-\!0.686$ \&
    G\,$332.963\!-\!0.679$, G\,$333.029\!-\!0.015$, 
    G\,$333.029\!-\!0.063$}
  \label{fig:images9}
\end{figure}

\begin{figure}
  \twofields{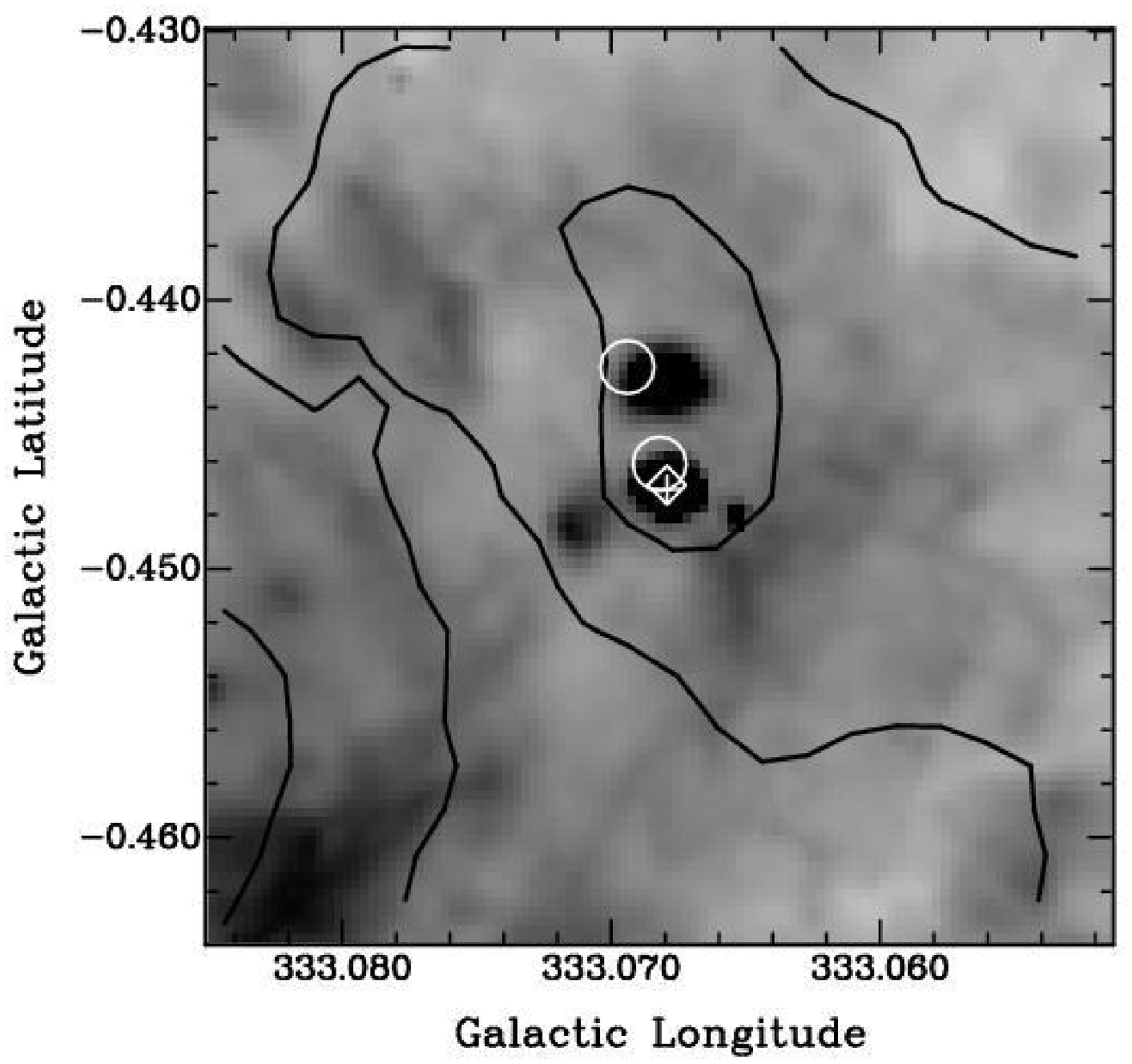}{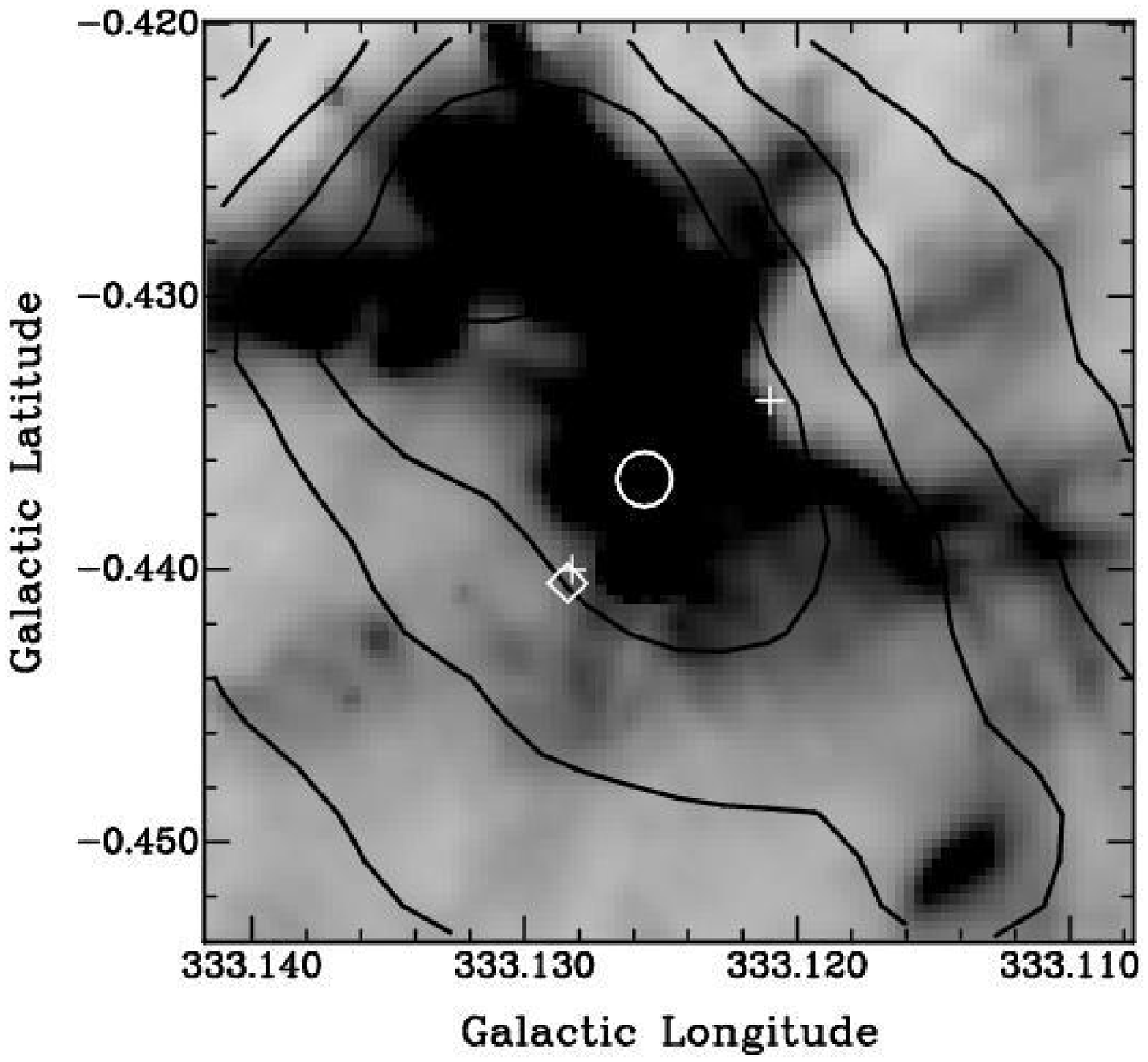}
  \twofields{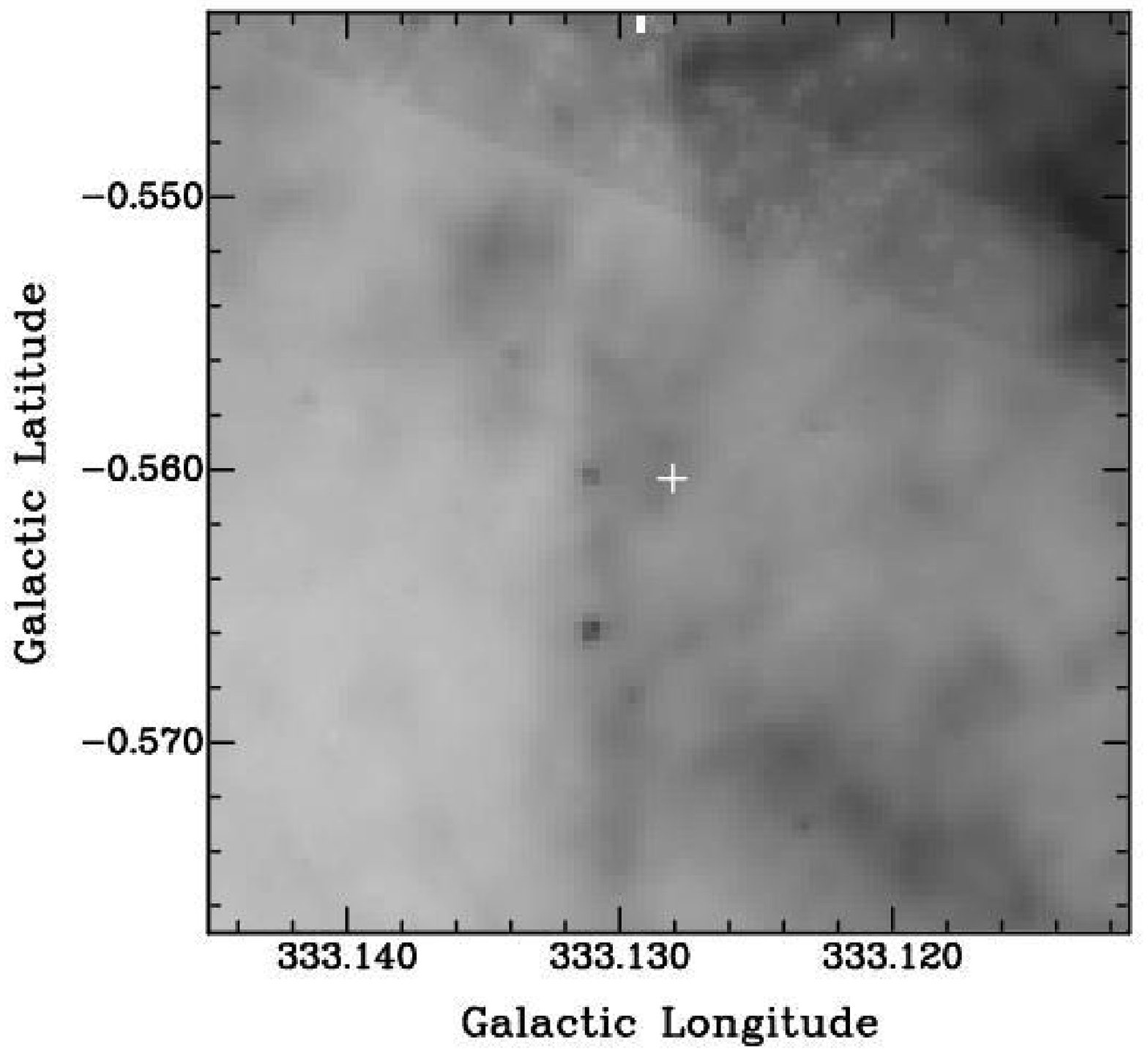}{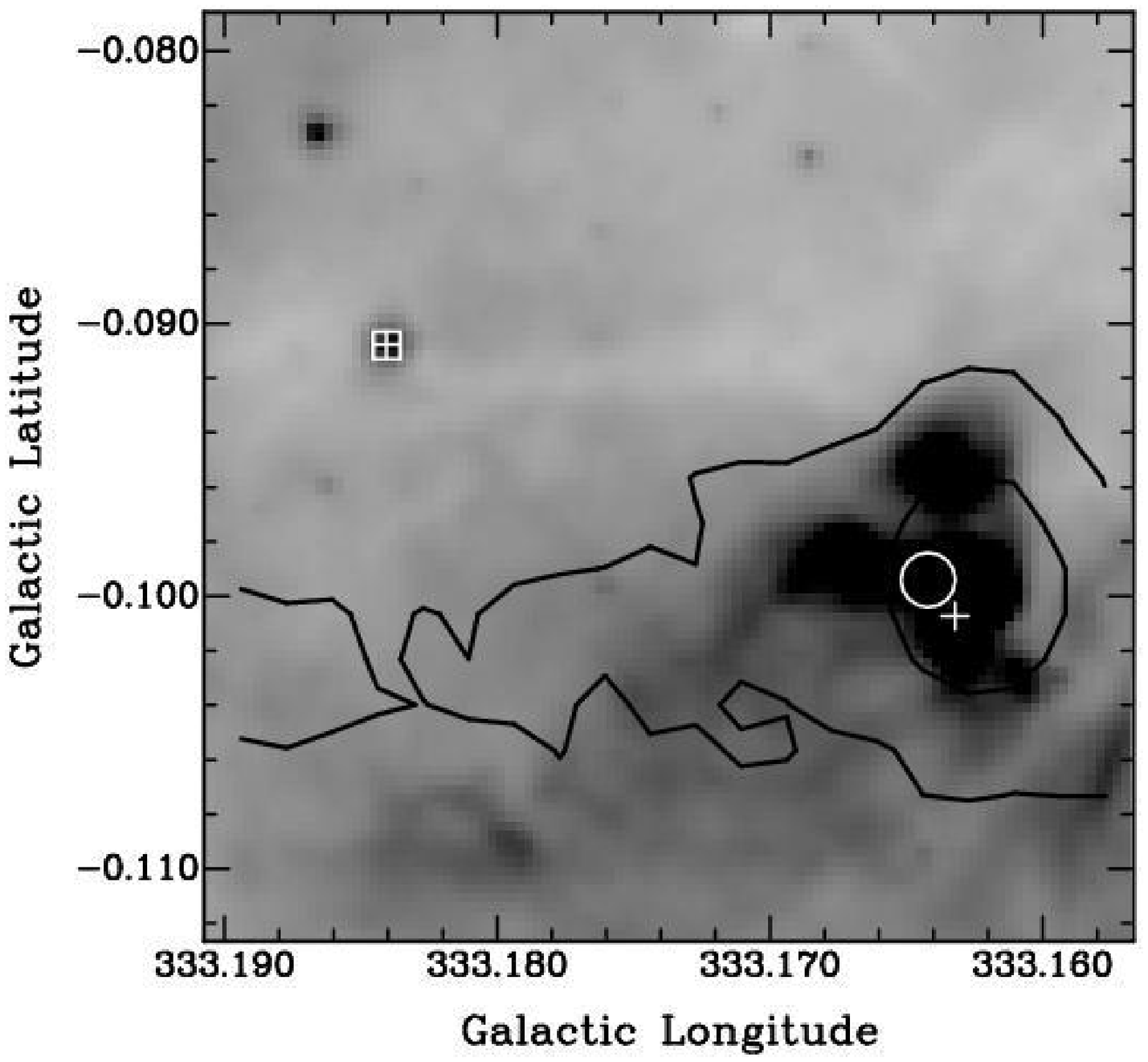}
  \caption{As for Fig.~\ref{fig:images1} for the sources  
    G\,$333.068\!-\!0.447$, G\,$333.121\!-\!0.434$ \& G\,$333.128\!-\!0.440$,
    G\,$333.130\!-\!0.560$, G\,$333.163\!-\!0.101$ \& G\,$333.184\!-\!0.091$} 
  \label{fig:images10}
\end{figure}

\begin{figure}
  \twofields{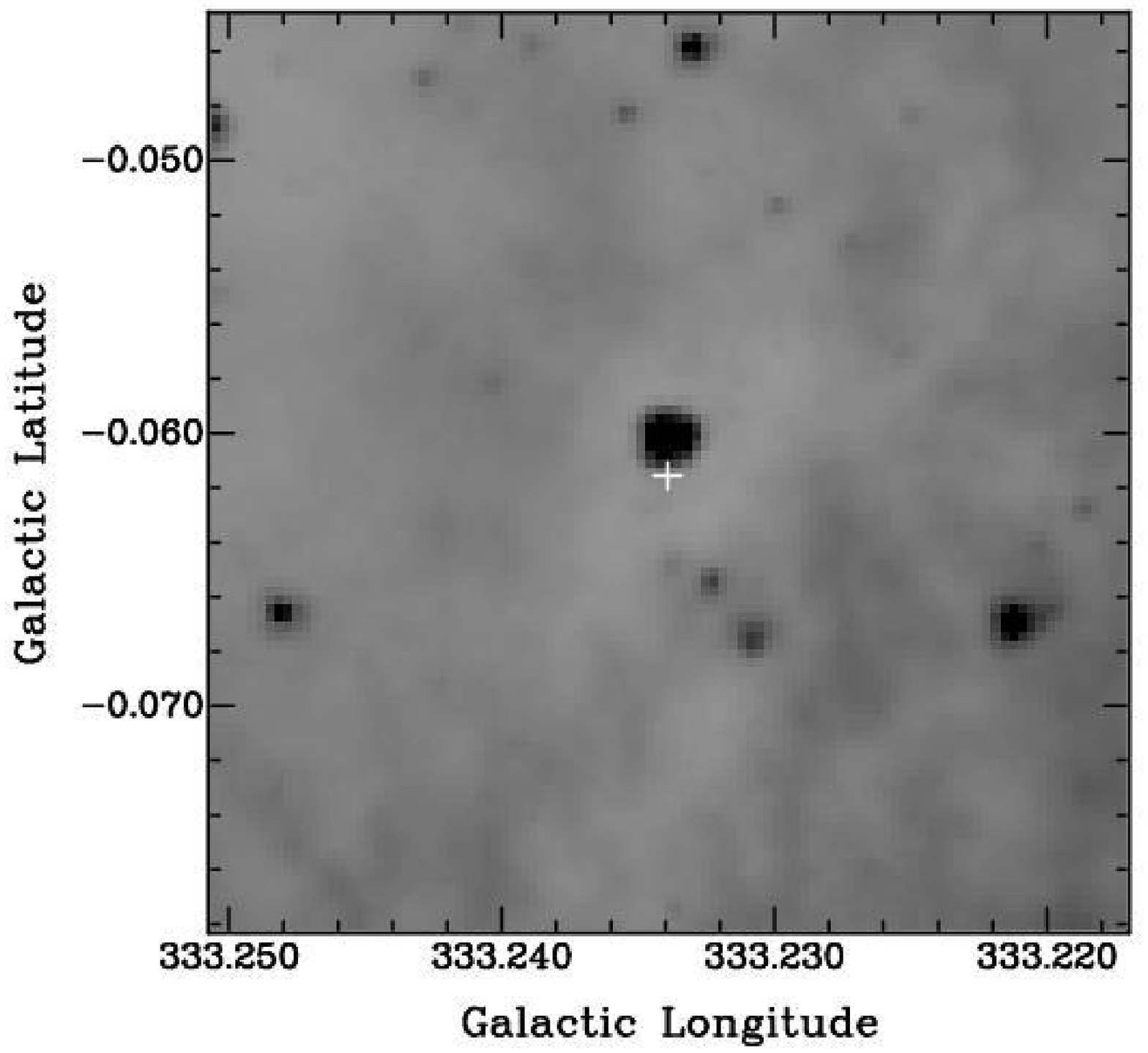}{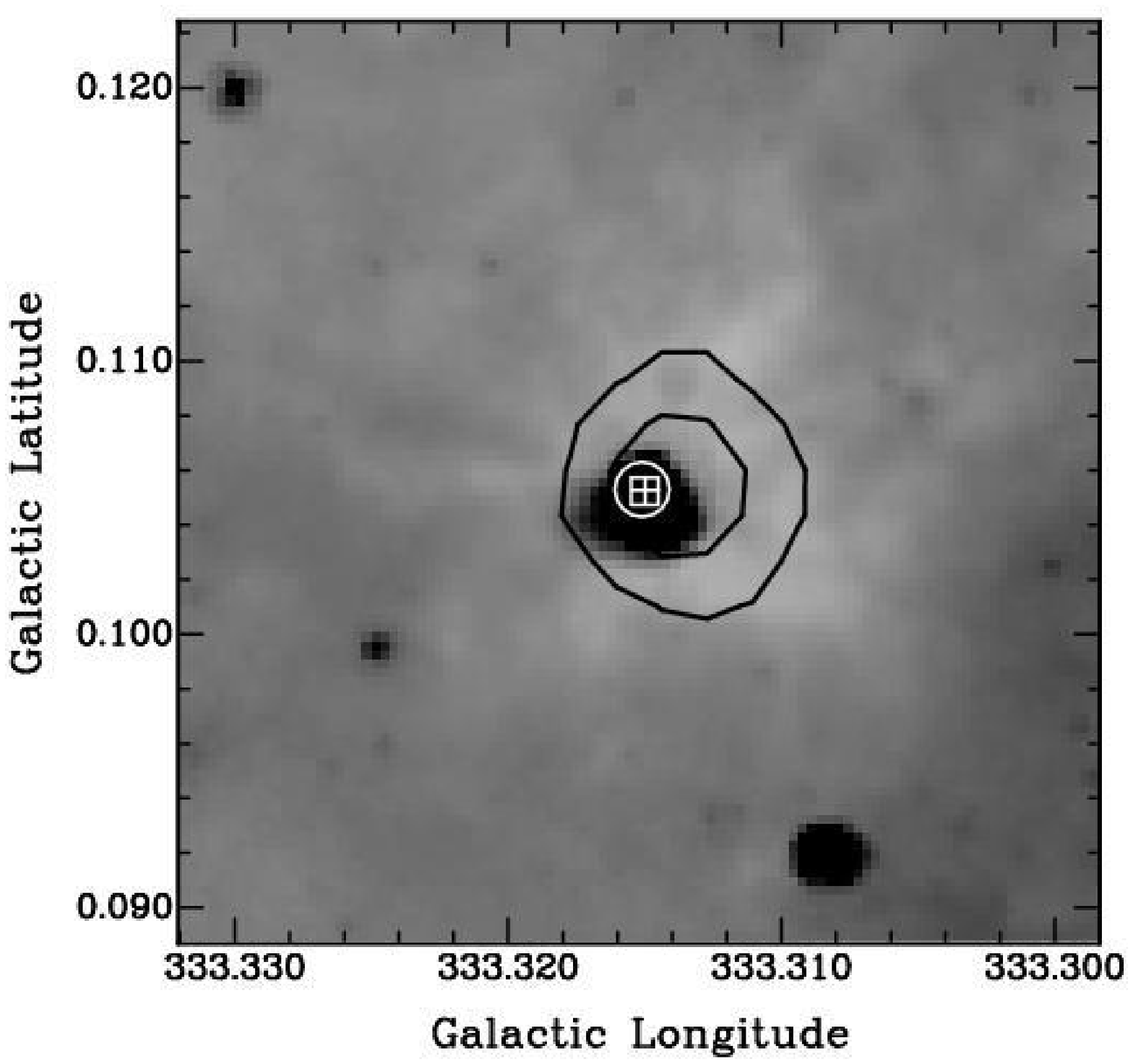}
  \twofields{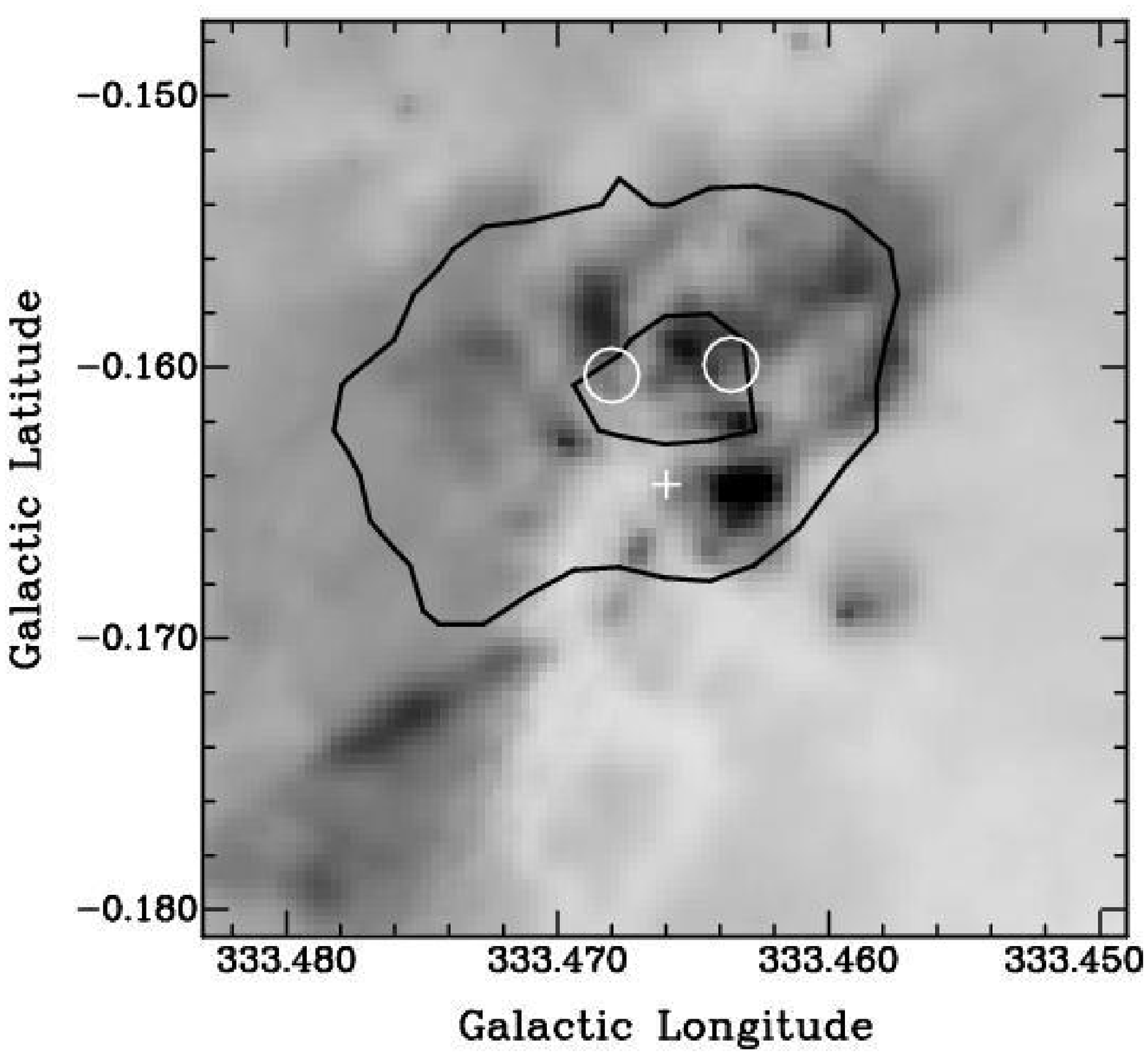}{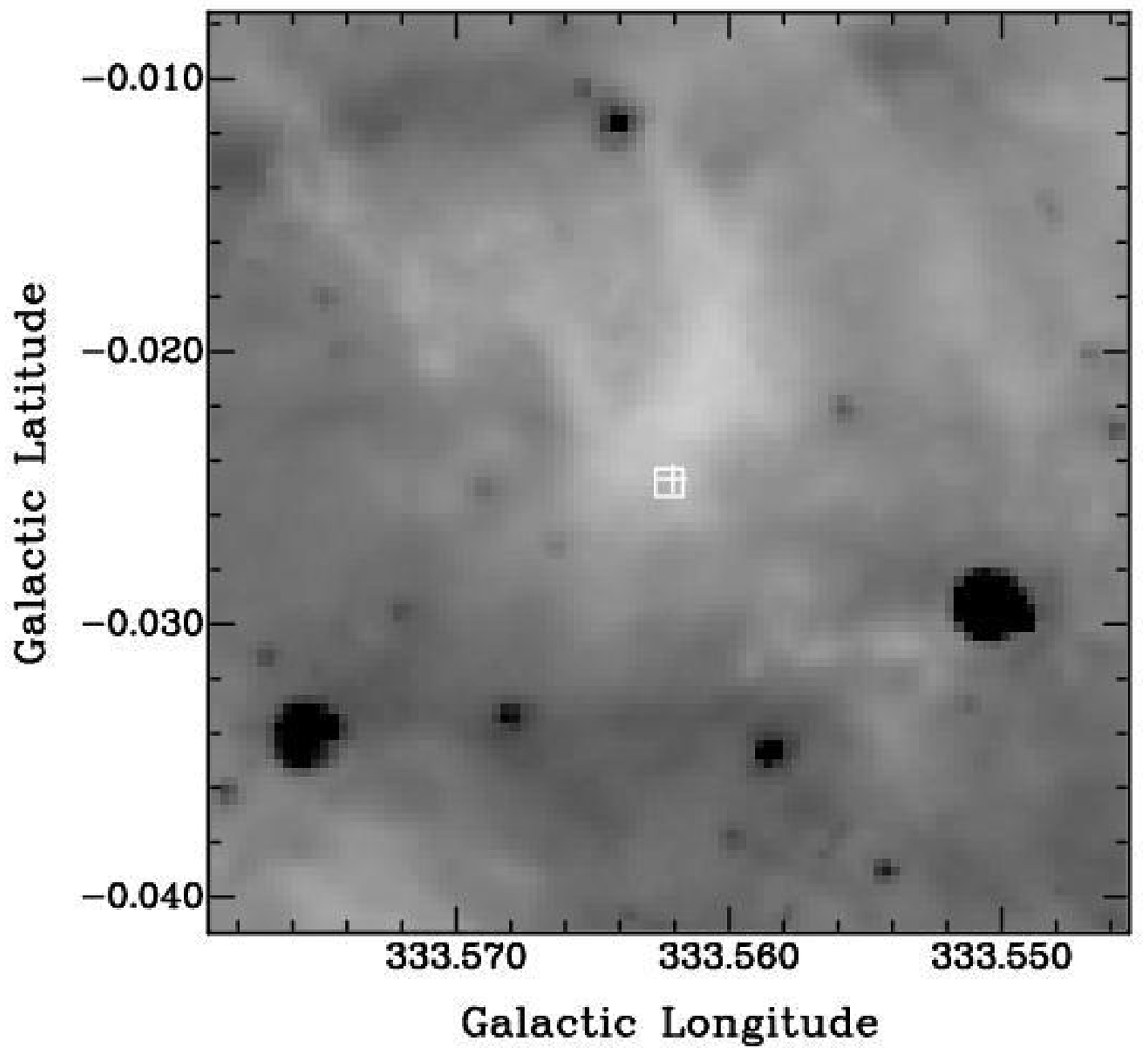}
  \caption{As for Fig.~\ref{fig:images1} for the sources  
    G\,$333.234\!-\!0.062$, G\,$333.315\!+\!0.105$,
    G\,$333.466\!-\!0.164$, G\,$333.562\!-\!0.025$}
  \label{fig:images11}
\end{figure}

\begin{figure}
  \twofields{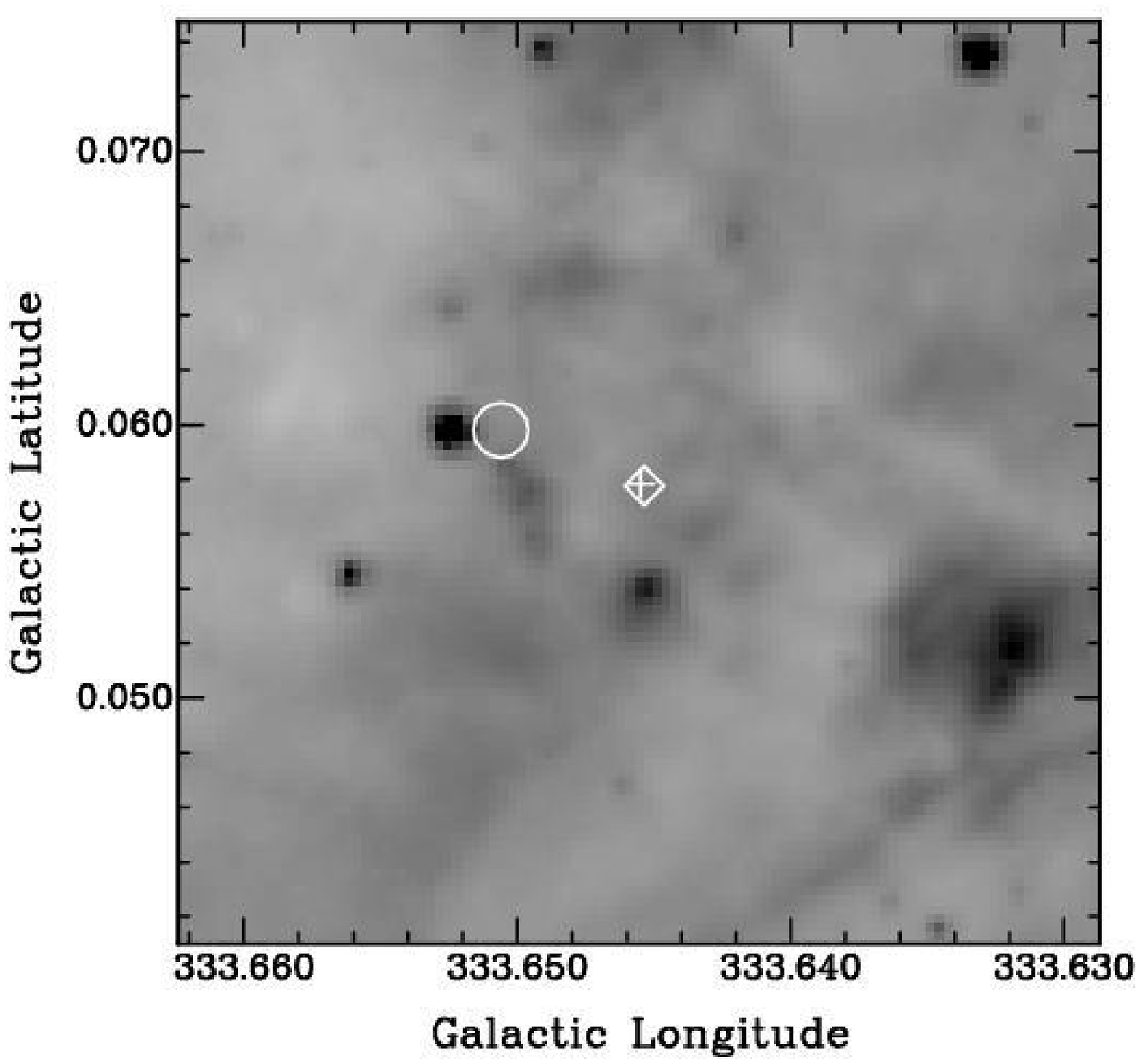}{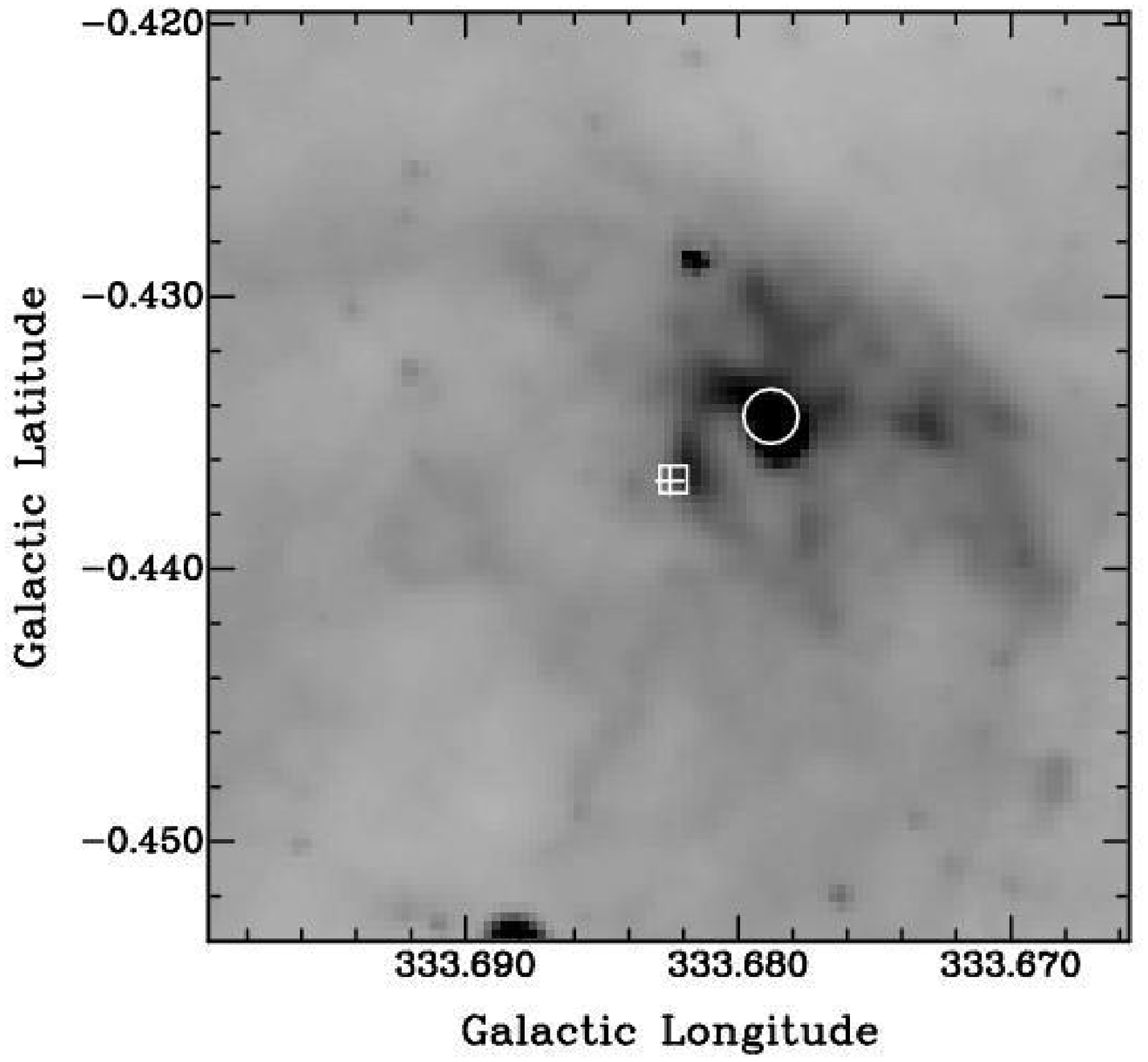}
  \twofields{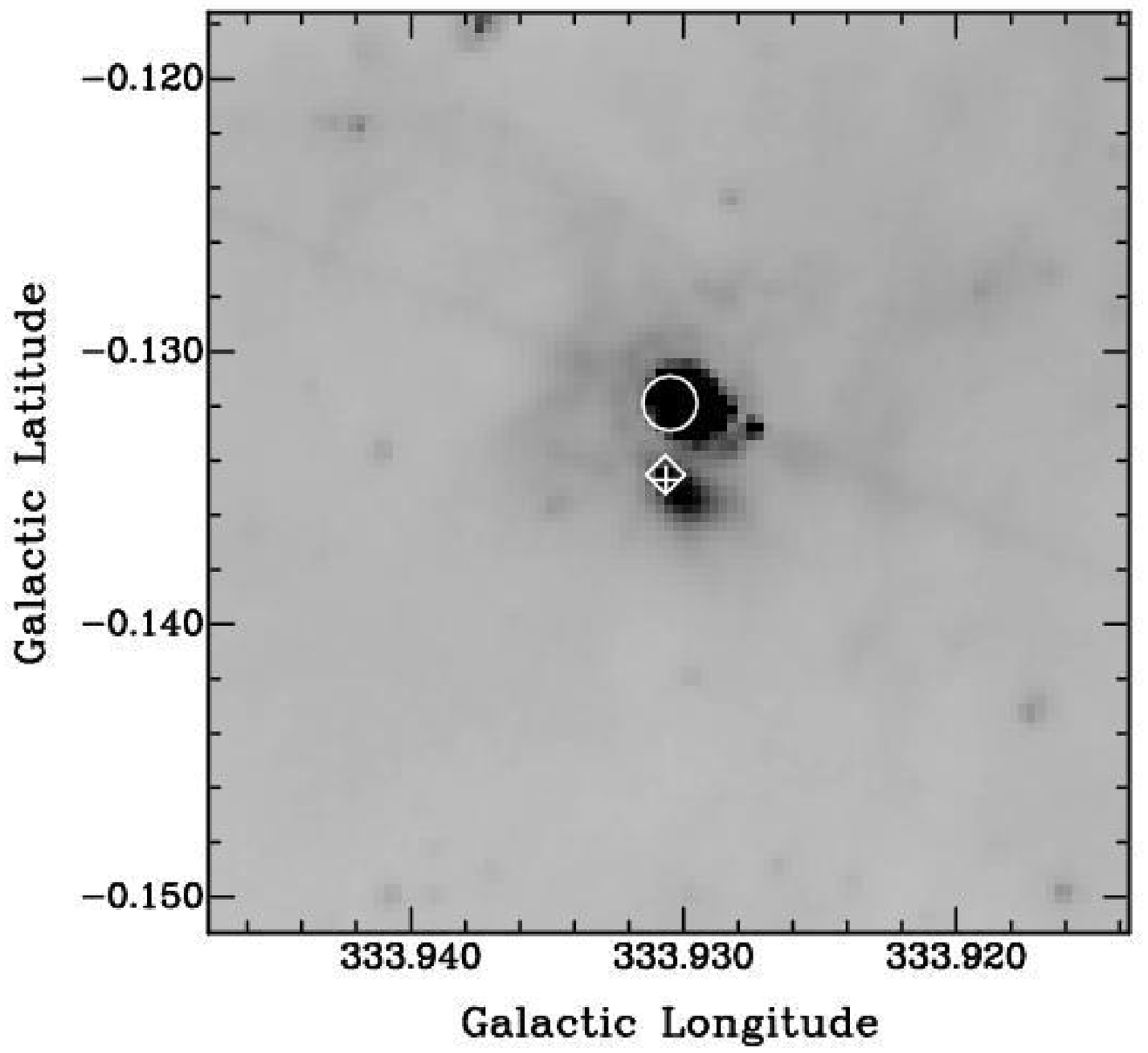}{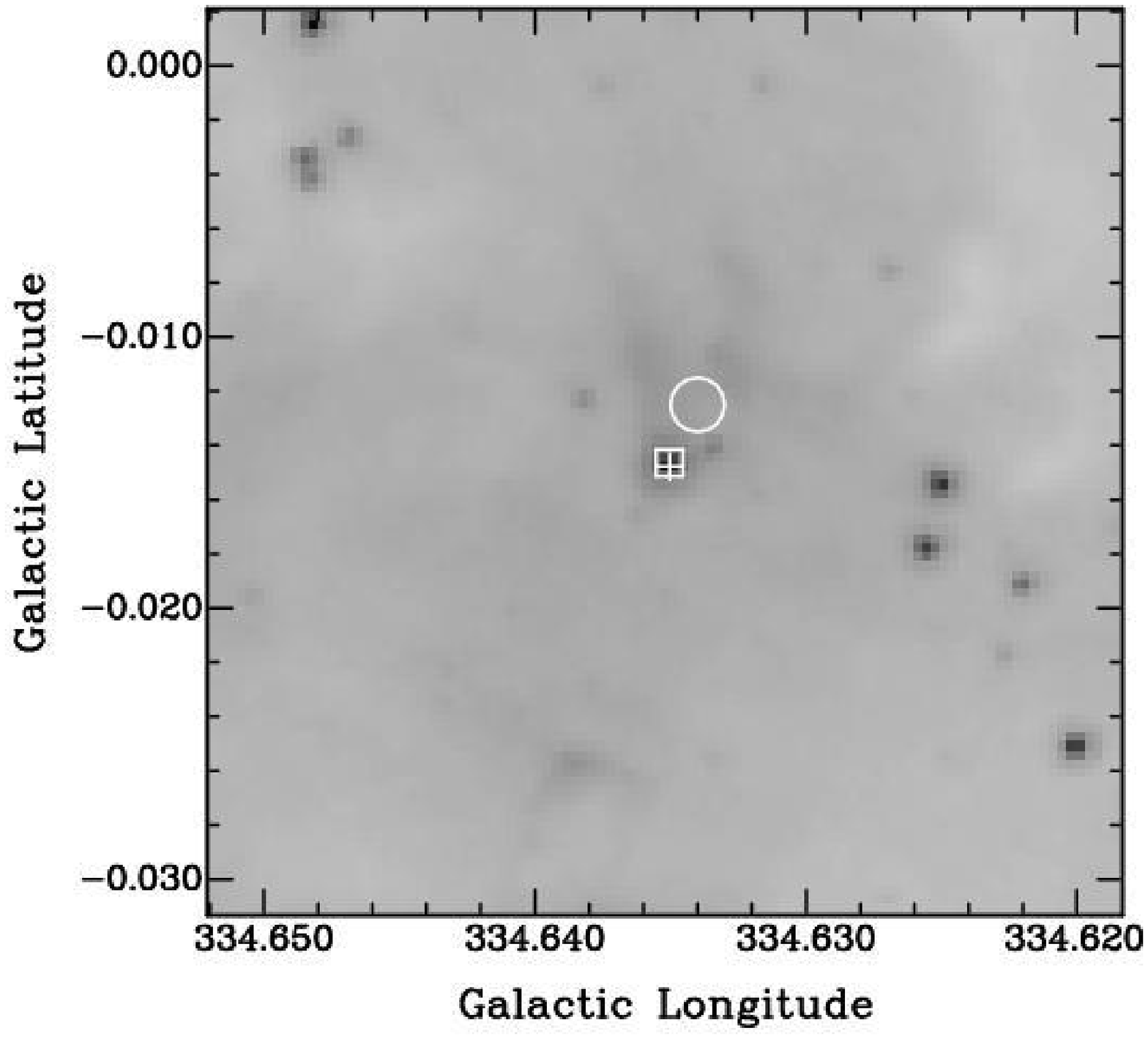}
  \caption{As for Fig.~\ref{fig:images1} for the sources  
    G\,$333.646\!+\!0.058$, G\,$333.683\!-\!0.437$,
    G\,$333.931\!-\!0.135$, G\,$334.635\!-\!0.015$}
  \label{fig:images12}
\end{figure}

\begin{figure}
  \twofields{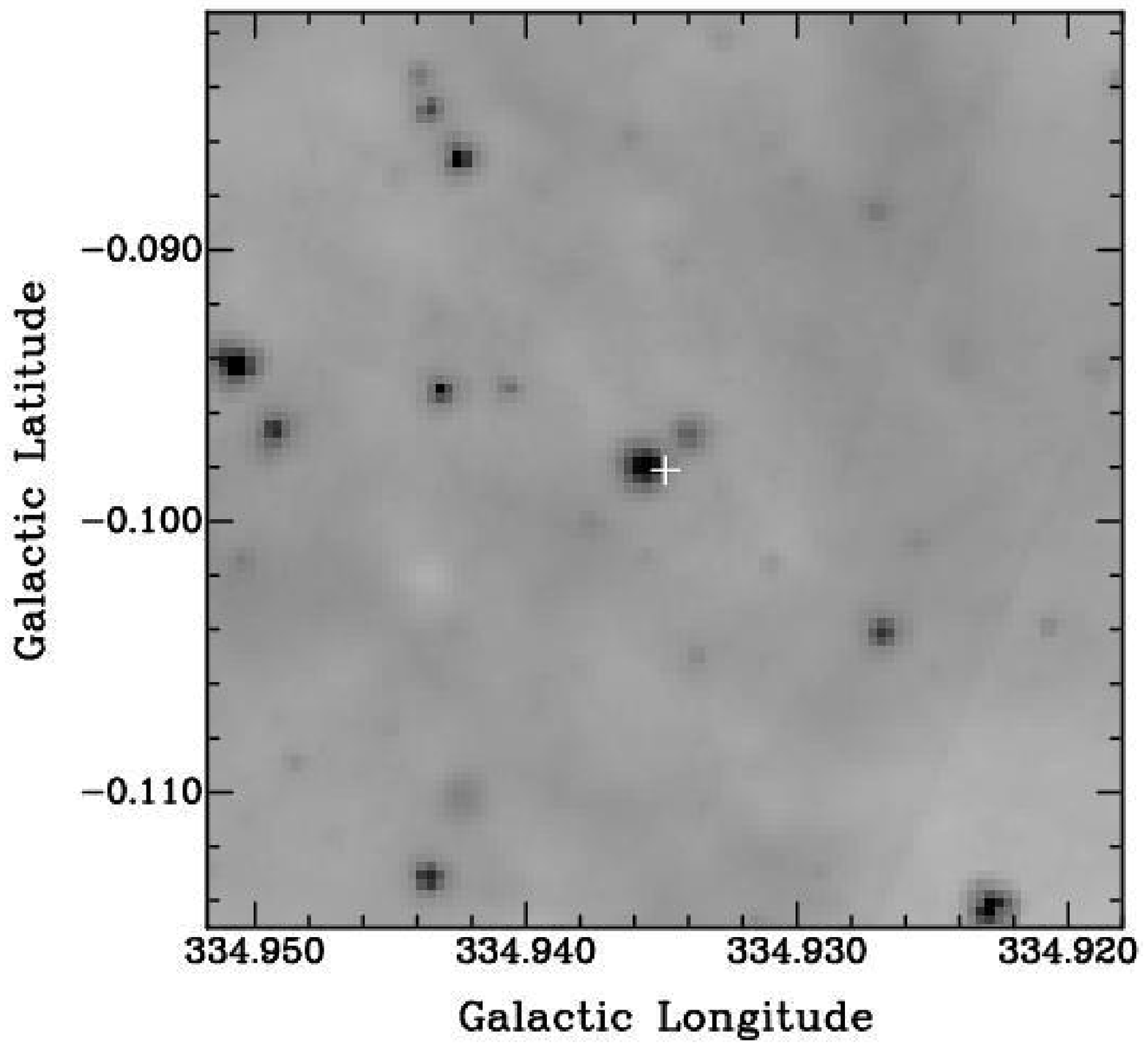}{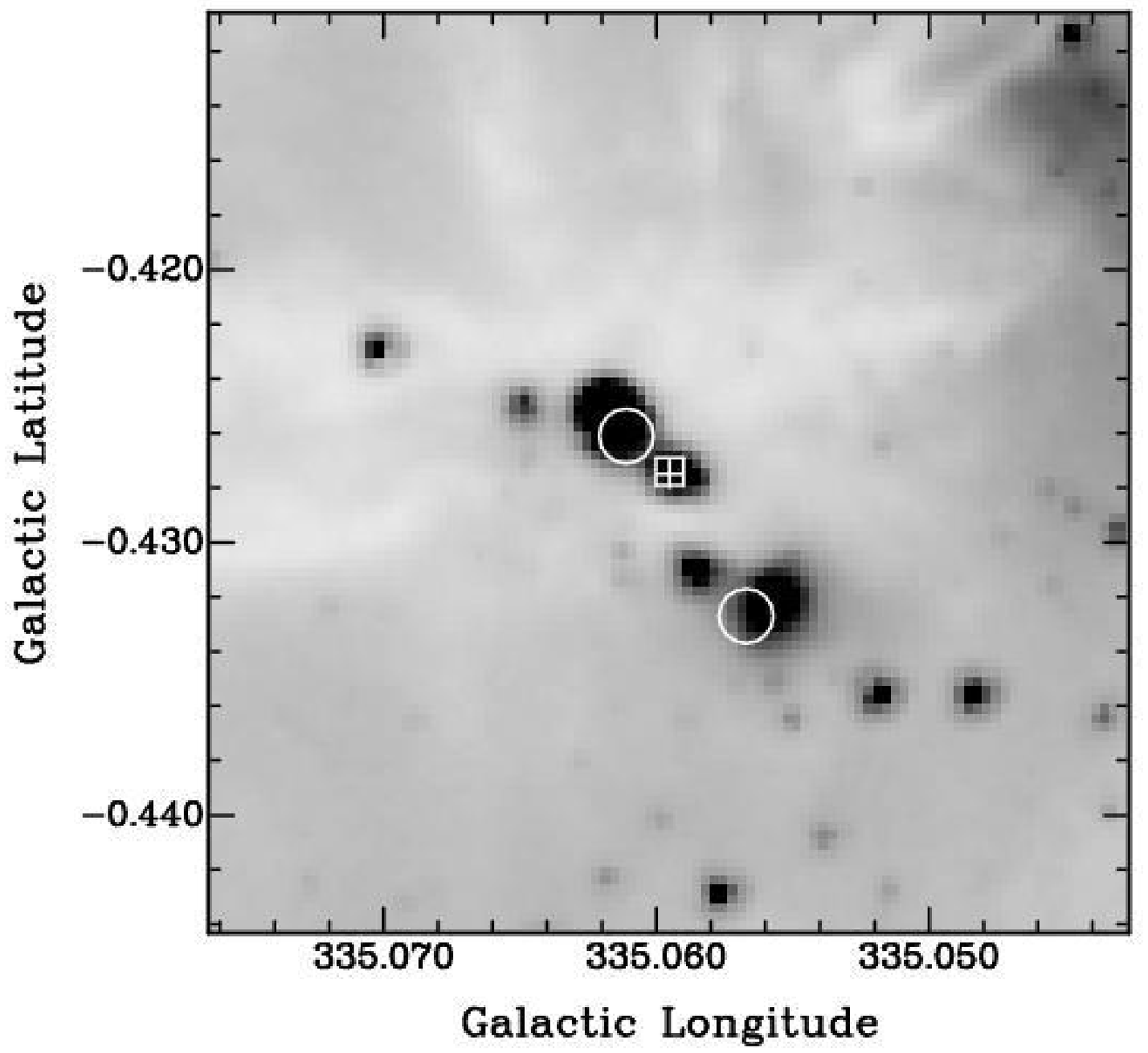}
  \caption{As for Fig.~\ref{fig:images1} for the sources  
    G\,$334.935\!-\!0.098$, G\,$335.060\!-\!0.427$}
  \label{fig:images13}
\end{figure}

G\,$332.094\!-\!0.421$ (Fig.~\ref{fig:images8}) : This is another
source where the methanol masers lie towards the centre of a region of
strong mid-infrared emission which has saturated at 8.0~\micron.
\citet{E05} showed that few class~II methanol masers have an
associated {\em MSX} point source, however, G\,$332.094\!-\!0.421$ is
one where there is a clear association.

G\,$332.295\!-\!0.094$ (Fig.~\ref{fig:images8}) : The methanol masers
are associated with a region of extended 8.0 \& 21~\micron\/ emission.
The masers are projected against a weak extension from the main region
that lies at the interface with an IRDC.

G\,$332.351\!-\!0.436$ (Fig.~\ref{fig:images8}) : The methanol masers
lie at the interface between an extended region of mid-infrared
emission and a low contrast IRDC.  No other maser species have been
detected towards this location.

G\,$332.560\!-\!0.148$ (Fig.~\ref{fig:images8}) : The methanol masers
are associated with a faint \glim\/ point source embedded in a complex
region.  No other maser species have been detected towards this
location and \citet{WBHR98} did not detect any radio continuum emission
associated with the methanol masers.

G\,$332.942\!-\!0.686$ \& G\,$332.963\!-\!0.679$
(Fig.~\ref{fig:images9}) : The G\,$332.942\!-\!0.686$ methanol masers
are associated with \glim\/ and {\em MSX} point sources that lie at
the interface between an extended mid-infrared emission region and an
IRDC.  The G\,$332.963\!-\!0.679$ masers are associated with a
point-like mid-infrared source embedded within the same IRDC as
G\,$332.942\!-\!0.686$.

G\,$333.068\!-\!0.447$ (Fig.~\ref{fig:images10}) : The methanol masers
lie near the centre of an extended region of 21~\micron\/ emission.
There is a \glim\/ archive source within 2\arcsec\/ of the methanol
maser site, but no other maser species have been detected in this region.

G\,$333.121\!-\!0.434$ \& G\,$333.128\!-\!0.440$
(Fig.~\ref{fig:images10}) : Two nearby clusters of methanol masers lie
at opposite edges of an extended region of 8.0 and 21~\micron\/
emission and are both associated with class~I methanol and water
masers.  Neither of the masers have an associated \glim\/ point
source, most likely due to the strong extended emission in the region,
but there is an archive source near G\,$333.128\!-\!0.440$.  In terms
of mid-infrared emission, this is one of the most active regions with
associated methanol masers in the $\ell$ = 325\degr--335\degr\ region.

G\,$333.315\!+\!0.105$ (Fig.~\ref{fig:images11}) : The methanol masers
are associated with both \glim\/ and {\em MSX} point sources that are
embedded within an IRDC that contains few other nearby emission
sources.

G\,$333.466\!-\!0.164$ (Fig.~\ref{fig:images11}) : The methanol masers
lie at the interface between mid-infrared emission and an IRDC in a
complex, clumpy region.  The radio continuum image of \citet{WBHR98}
shows the masers are offset from nearby \ionhy\/ regions that appear
to be associated with the stronger mid-infrared emission in the region
and are distributed roughly radially from the nearest infrared
emission.

G\,$333.562\!-\!0.025$ (Fig.~\ref{fig:images11}) : The methanol masers
lie within a small IRDC and although they are associated with a
\glim\/ point source detected in the 3.6 \& 4.5~\micron\/ bands there
is no sign of any emission at 8.0~\micron.

\section{Discussion}

\subsection{The mid-infrared colors of 6.7~GHz methanol masers} \label{sec:mircolors}

\begin{figure}
\plotone{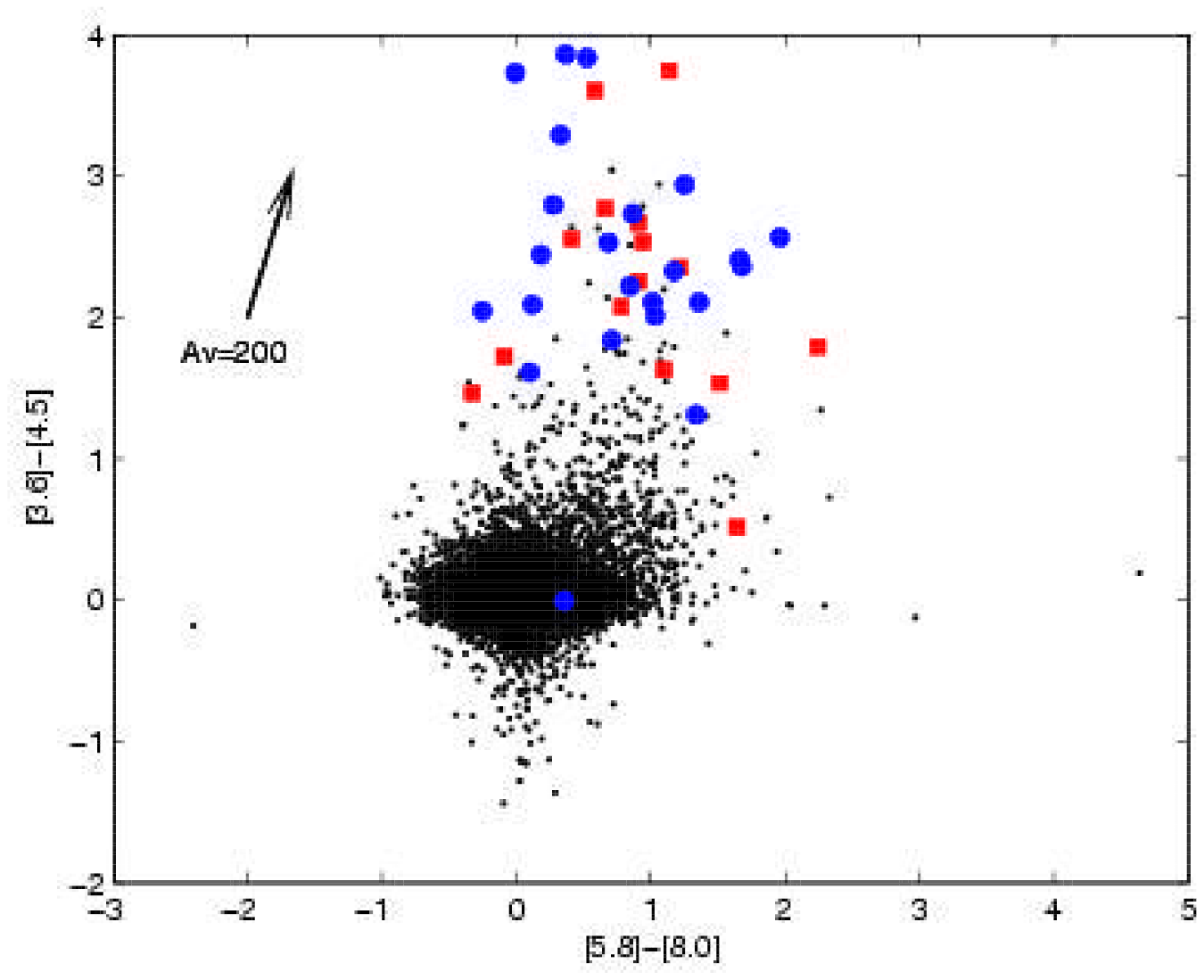}
\caption{A color-color diagram constructed from \glim\/ PSC data.
  The methanol masers in the region $\ell$=325\degr--335\degr\ are
  represented with squares, other methanol masers for which ATCA
  positions are available are represented with circles.  Sources
  within 30\arcmin\ of $\ell$=326.5\degr, $b$=0.0\degr\ are
  represented with dots.  Only sources for which there is flux density
  information for all the {\em IRAC}\ bands have been included in the
  plots, this is 15 of 29 methanol masers with a \glim\/ point source
  within 2\arcsec, 23 of 53 methanol masers for which ATCA positions
  are available and 20696 of 108918 of the general sample.  The arrow 
  represents the reddening produced by an extinction of $A_v = 200$ using 
  the \citet{IMB+05} model (see section~\ref{sec:mircolors}).}
\label{fig:colcol3412}
\end{figure}

I have constructed color-color and color-magnitude diagrams for the
\glim\/ point sources associated with methanol masers in the Galactic
Longitude range $\ell$ = 325\degr -- 326\degr\ and compared their
distribution to those within a 30\arcmin\ radius of $\ell$ =
326.5\degr, $b$ = 0.0\degr. The comparison field contains 108,918
sources.  The majority of sources from the \glim\/ archive contain
information from only two bands and few have sufficient information to
be plotted in color-color or color-magnitude diagrams.  So the color
analysis has been restricted to just the masers associated with
sources from the point source catalog.  The locations of many other
methanol maser sites within the \glim\/ survey region have also been
determined to sub-arcsecond precision from ATCA observations.  To
extend the maser sample size the properties of \glim\/ point sources
associated with 133 methanol masers from \citet{C96} and
\citet{WBHR98} that lie outside $\ell$=325\degr -- 335\degr\ and
within the \glim\/ survey region were determined.  Of these additional
6.7~GHz methanol masers 53 had associated \glim\/ point sources within
2\arcsec.  The maser associated sources are clearly offset from the
majority of field sources in most of the color-color and
color-magnitude diagrams, having colors that are much redder.  This
can be clearly be seen in Fig.~\ref{fig:colcol3412} which shows a plot
of the [5.8] - [8.0]~\micron\ versus [3.6] - [4.5]~\micron\ colors,
where the masers in general lie above the vast majority of the
comparison field sources.  Table~\ref{glimprop} shows that most of the
maser sources have a [3.6] - [8.0] $>$ 4.0 mag (corresponding to a
flux density ratio of 10), in contrast for the comparison field only
34 of the 22517 sources (0.15\%) for which there is 3.6 and
8.0~\micron\ data meet this criterion.

That the methanol masers lie in a distinctive region of \glim\/
color-color diagrams is interesting, but what can be inferred from it?
In principle it is possible to fit a single temperature grey-body
spectrum to the \glim\/ observations, however, we have no data at
longer or shorter wavelengths (with comparable resolution) to provide
meaningful constraints.  The situation is further complicated through
the influence of factors such as PAH emission and silicate absorption
lines that affect some of the {\em IRAC} bands \citep[see][for an
detailed discussion]{D03}.  PAH features at 3.3, 6.2, 7.7 and
8.6~\micron\/ lie within the [3.6], [5.8], [8.0] and [8.0] {\em IRAC}
bands respectively.  The diffuse PAH emission that extends throughout
much of the inner Galactic Plane is generally of very low intensity,
and as such is likely to have a minimal impact on the intensity
measured for point sources, except where they are near the sensitivity
limit of the observations.  Some large star forming complexes, such as
RCW49 show significant PAH emission from the region itself
\citep{WIB+04} which is much stronger than the larger scale diffuse
emission.  The majority of the methanol masers appear to lie within
star formation regions that are less evolved than RCW49 and so
probably with correspondingly lesser PAH emission.  However,
mid-infrared spectroscopy (with an instrument such as the {\em
Spitzer} Infrared Spectrograph) of a sample of sources with associated
methanol masers is clearly required to accurately quantify the
influence of PAH emission on the observed {\em IRAC} colors.  The
[8.0] {\em IRAC} band may also be affected by the silicate absorption
feature at 9.7~\micron.  Recent mid-infrared spectroscopic
observations of a number of high-mass young stellar objects (YSO) by
\citet{DOC05} show significant silicate absorption, the wings of
which extend well into the [8.0] {\em IRAC} band.  In
addition, three-dimensional radiative transfer modeling \citep{WWBC03}
has shown that for low-mass (YSO) the SEDs in the infrared are complex
and influenced by many factors including evolutionary status and
orientation.  These factors mean that until detailed mid-infrared
spectroscopy is available, interpretation of the \glim\/ colors of
sources with associated methanol and other masers (e.g. as presented
in section~\ref{sec:evol}) is to a large degree speculative.  However,
it is still possible to make meaningful comparisons of the \glim\/
point sources with and without methanol masers, as having been made
with the same instrument they all suffer the same effects and
limitations.

\begin{figure}
\plotone{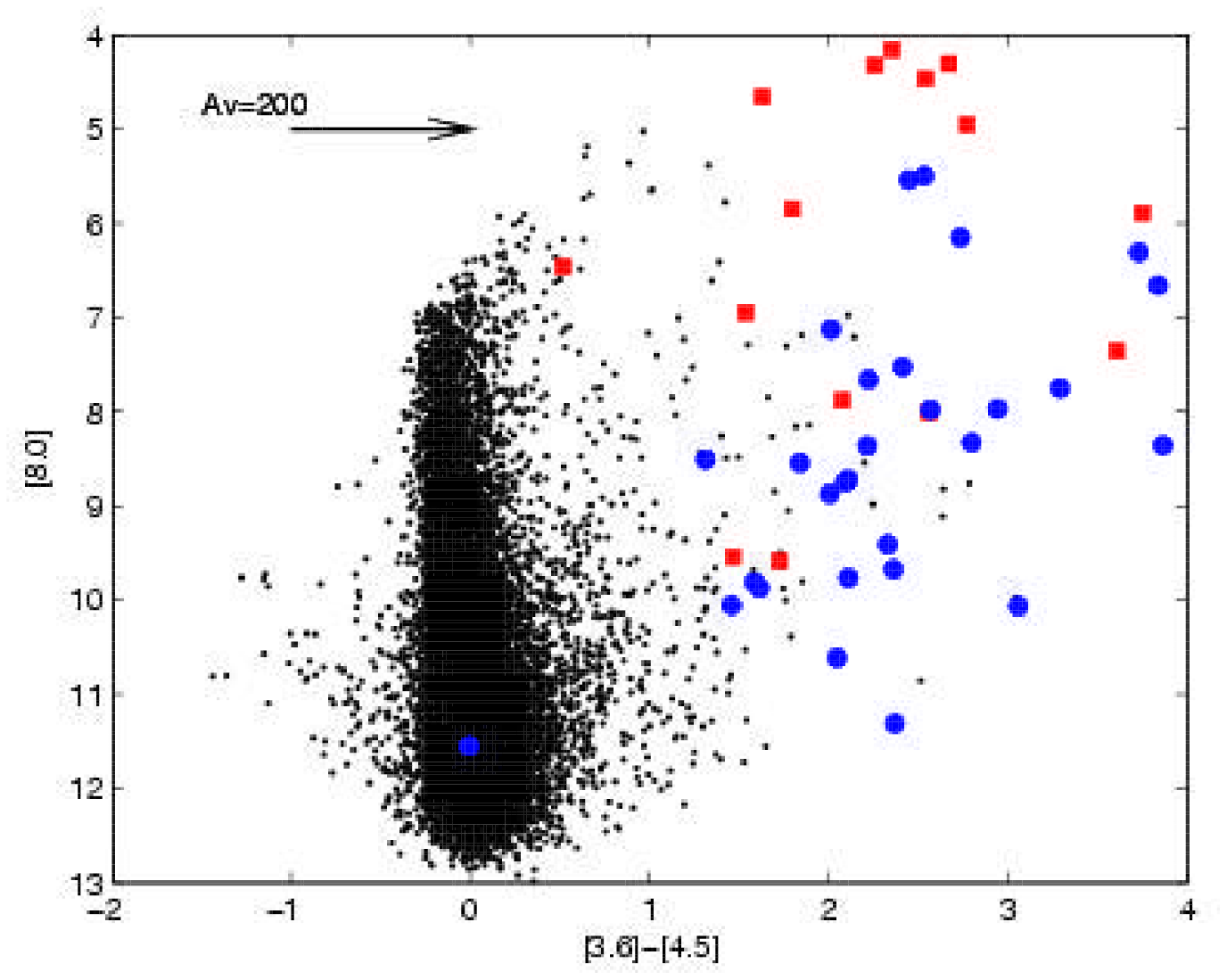}
\caption{A color-magnitude diagram constructed from \glim\/ PSC data.
  The methanol masers in the region $\ell$=325\degr--335\degr\ are
  represented with squares, other methanol masers for which ATCA
  positions are avaliable are represented with circles.  Sources
  within 30\arcmin\ of $\ell$=326.5\degr, $b$=0.0\degr\ are
  represented with dots.  Only sources for which there is flux density
  information for the 3.6, 4.5 \& 8.0~\micron\ \glim\/ bands have been
  included in the plots, this is 15 of 29 methanol masers with a
  \glim\/ point source within 2\arcsec, 29 of 53 methanol maser for
  which ATCA positions are available and 21988 of 1080918 of the
  general sample.  The arrow represents the reddening produced by an
  extinction of $A_v = 200$ using the \citet{IMB+05} model
  (see section~\ref{sec:mircolors}).}
\label{fig:colmag124}
\end{figure}

The modeling of \citet{WWBC03} shows that class~0 objects lie in a
distinctive region of the \glim\/ [5.8]-[8.0] vs [3.6]-[4.5]
color-color diagram and it is the same area as that occupied by the
methanol masers (as can be seen by comparing Fig.~\ref{fig:colcol3412}
with Fig.~7a \& 7b of \citeauthor{WWBC03}).  The extent to which the
class~0-III, low-mass evolutionary scheme can be transferred to
high-mass/clustered star formation is very much an open question.
However, the similarity of the location in the color-color diagram of
the methanol masers to predictions for class~0 objects is striking.
The fact that methanol masers are found in IRDC complicates the issue
to some degree.  \citet{MBH+05} modeled the SED of a number of
methanol masers without associated radio continuum emission and
determined the $A_v$ to range from 50-240 for the millimetre clumps with
associated methanol masers.
Figures~\ref{fig:colcol3412} \& \ref{fig:colmag124} each show that the
reddening vector of \citet{IMB+05} determined from {\em Spitzer} data
for the 1.25-8.0~\micron\ range points from the location of the bulk
of sources towards the methanol maser region.  However, although
reddening doubtless contributes to the observed colors, even an $A_v$
of 200 (as shown in Fig.~\ref{fig:colcol3412} \& \ref{fig:colmag124}) is
insufficient to fully explain them.  In Fig.~\ref{fig:colcol3412} a
reddening vector for $A_v = 200$, can be seen to be responsible for
shifting the colors by approximately 1 magnitude in [3.6]-[4.5] and a
third of a magnitude in [5.8]-[8.0].  Clearly as such promising
candidates for the early stages of high-mass star formation it is
important to determine the SED of these sources over a wider
wavelength regime.  This must be done at as high a resolution as can
be achieved to avoid the effects of confusion.  \citet{WWBC03} find
class~0 objects are well offset from all other low-mass YSO
evolutionary phases in [24]-[70] vs [8.0]-[24] color-color diagrams.
It would be very interesting to see if the high-mass protostellar
objects associated with methanol masers have similar colours to
class~0 objects over this spectral range as well.

\subsection{Targeting maser searches using mid-infrared colors}

The distinctive properties of the \glim\/ sources associated with
methanol masers also provide an opportunity to use them to target
future maser searches.  Although there have been a number of
untargetted searches covering moderately large areas
\citep{C96,EVM+96,SKH+02}, there is much of the region covered by
\glim\/ which has yet to be searched.  In developing targeting
criteria for methanol masers it is a matter of compromising between
detecting the highest possible percentage of the sources, while
keeping a managable number of target locations to be searched.  The
\glim\/ point source catalog contains more than 30 million objects and
so to produce a reasonable list of targets requires criteria which
matches of the order of 0.001-0.01\% of them.  The distinctive
characteristics of the methanol maser sources are their very steep
mid-infrared spectrum, and that they are bright in the 8.0~\micron\/
band (Fig.~\ref{fig:colcol3412} \& \ref{fig:colmag124}).
Figure~\ref{fig:colmag124} shows that the majority of methanol masers
have [3.6]-[4.5] $>$ 1.3 and an 8.0~\micron\ magnitude less than 10,
while relatively few sources in the comparison field share these
properties.  A search of the \glim\/ point source catalog matches 5675
sources on these criteria ($<$0.02\% of all \glim\/ sources).

It is possible to estimate the detection rate and efficiency of a
methanol maser search targeted towards \glim\ point sources with the
criteria outlined above through comparison with one of the untargeted
searches.  Untargeted searches by \citet{EVM+96} and \citet{C96} have
detected forty-six 6.7~GHz methanol masers in the region
$\ell$=325\degr -- 335\degr, $b$=$-$0\fdg53 -- +0\fdg53 with a peak flux
density greater than 2 Jy.  A total of 572 \glim\ point sources within
this region satisfy the criteria [3.6]-[4.5] $>$ 1.3 and an
8.0~\micron\ magnitude $<$ 10.  For a search with a single dish radio
telescope such as the University of Tasmania Mt Pleasant 26m, any
targets that lie within half the full width half maximum (FWHM) of the
telescope beam of each other could be searched in a single pointing.
The Mt Pleasant 26m telescope has a FWHM of approximately 7\arcmin\ at
6.7~GHz and this reduces the total number of target positions within
the selected region of the Galactic Plane to 352.  This number of
pointings is approximately one third the number required to completely
search the region with a telescope of this size.  If we assume that
any 6.7~GHz methanol maser within the half-power radius of a targeted
position will be detected in the search then we would find 38 of the
46 methanol maser sites in this region (it would also detected two of
the sites that lie just outside the boundaries).  This corresponds to
a detection rate of 83\% and an efficiency (percentage of target
positions that yield a detection) of 11\%.  \citet{EVM+96} performed a
similar analysis for various {\em IRAS}-based search criteria and
found typical efficiencies in the range 20-30\% (with a much smaller
number of target sources), but in general detection rates were less
than 50\% with the best being 59\%.  So maser searches targeted
towards \glim\/ point sources which meet certain criteria will detect
many sources that similar {\em IRAS}-based searches cannot, however,
at present 20\% or so of sources can only be detected through
untargeted searches.  The clustering of high-mass star formation
regions artificially enhances the detection rate of the targeting
criteria.  I have shown that approximately 50\% of 6.7~GHz methanol
masers have an associated \glim\/ point source, yet targeting criteria
based on \glim\/ data is able to detect more than 80\% of the masers.
So approximately 30\% of the detections will be made towards \glim\/
point sources that are not directly associated with the masers, but
are nearby within the same star formation complex.

Applying the same search strategy as outlined above (consolidating
targets within half the FWHM to a single pointing) to the entire
\glim\ catalog reduces the number of target positions from 5675 to
3723.  Extrapolating the findings for the $\ell$=325\degr -- 335\degr\/
region to the entire \glim\/ region suggests that 402 methanol maser
sources would be found if all sources matching the above criteria were
searched.  If the detection rate observed in the $\ell$=325\degr --
335\degr\ region holds throughout the \glim\ survey region then it
predicts that there are approximately 487 methanol masers within this
area with peak flux densities of 2 Jy (the sensitivity limit of the Mt
Pleasant survey) or more.  The catalog of \citet{PMB05} lists 391
6.7~GHz methanol masers in the \glim\ region, of which 347 have a peak
flux density of greater than 2 Jy.  This suggests that the majority of
the 6.7~GHz methanol masers in this region have been detected by
previous searches targeted towards OH masers and {\em IRAS} sources,
but that a search based on the criteria listed above would likely
yield of the order of 50 new detections.

A large number \glim\ sources that meet the criteria outlined above
are very close to either the color limit or the magnitude limit and
the detection efficiency calculated above could be raised relatively
easily by making a small increase to the color limit, or a small
decrease to the magnitude limit, however, this would naturally come at
the cost of a reduced detection rate.  The total number of \glim\/
sources that satisfy the criteria means that it must include many
objects which are not high-mass young stellar objects (since it would
imply a rate of high-mass star formation approximately an order of
magnitude in excess of that predicted from the IMF).  This
contamination is most likely from lower-mass young stellar objects and
perhaps some evolved low-mass stars.  For example the models of
\citet{WWBC03} show that the colors of the \glim\/ sources associated
with methanol masers are very similar to those predicted for class~0
YSO.  It may be possible to use cross-correlation with other
catalogues such as 2MASS to reduce this level of contamination and
develop more refined criteria for targeting high-mass young stellar
objects.  However, although this will increase the efficiency of any
surveys, it is clear from the analysis above that the majority of
stronger ($>$ 2~Jy) 6.7~GHz methanol masers from the \glim\/ region
have already been detected.

\subsection{Methanol masers and IRDC}

A striking feature of the 8.0~\micron\ \glim\/ mosaics is the diffuse
emission present throughout much of the Galactic Plane, this is due to
the PAH emission bands at 7.7 \& 8.6~\micron\ that fall within this
{\em IRAC} band. Superimposed upon this diffuse background are dark
clouds (IRDC), and when examining the \glim\/ images on large scales
it is clear that the methanol masers are preferentially associated
with these.  A brief subjective examination of \glim\ 8~\micron\
images suggests that approximately 10-20\% percent of IRDC have
associated methanol masers.  A more robust, objective determination of
this percentage is beyond the scope of this paper as the level of
extinction, foreground emission, sizes and morphologies of the IRDC
cover a large range and development and application of consistent
criteria to locate IRDC in the \glim\ observations is required.  At
the resolution and sensitivity of the \glim\/ observations, many IRDC
are seen to have one or more point or extended sources embedded within
them, typically towards the center.  The masers are typically
coincident with one of these faint embedded emission sources within
the IRDC.  The images in Figs.~\ref{fig:images1}-\ref{fig:images13}
have had the contrast increased to highlight this low-level diffuse
emission.  Only a small number of the methanol masers are associated
with prominent, well developed star formation regions such as
G\,$326.641\!+\!0.613$ and G\,$333.121\!-\!0.434$.

Infra-red dark clouds were discovered in {\em ISO} \citep{POS+96} and
{\em MSX} \citep{ESP+98} observations and are associated with dense
($N_{H_2} > 10^{23}$ cm$^{-2}$, $n > 10^5$ cm$^{-3}$), cold ($T<20$ K)
gas \citep{ESP+98,CCE+98}.  These cold clouds are sufficiently dense
to be detected in sub-millimeter emission with instruments such as the
JCMT, and show similar overall morphology in sub-millimeter emission
to that seen in infrared absorption \citep{CFR+00}.  The
sub-millimeter reveals more details of the structure of the clouds, in
particular the location of warmer clumps within the cold dust.  The
density, temperature and scale of the IRDC have lead to speculation
that they represent a very early stage in the star formation process,
perhaps prior to the existence of young stellar objects.  Recent
observations have detected ``warm'' (40-60K) cores in a number of IRDC
\citep{CFR+00,OSOH05,MPW05} and masers or young stellar objects
associated with some of these cores \citep{RFW+03,MPW05}.  The warm
cores and presence of masers suggests that some IRDC harbor high-mass
star formation in its very early stages.  The presence of class~II
methanol masers towards faint emission sources in many IRDC shows that
in a sizeable fraction, high-mass star formation has commenced.  It
also suggests that those IRDC which do not show maser emission from
either water or methanol may be the best sites to study still earlier
evolutionary phases of high-mass star formation.

The comments on individual sources (section~\ref{sec:indiv}) show that
sometimes the methanol masers are associated with mid-infrared
emission from the central regions of the dark cloud (e.g.
G\,$327.120\!+\!0.511$), while in other cases they are offset from the
strongest emission (e.g.  G\,$326.475\!+\!0.703$,
G\,$326.859\!-\!0.677$).  Where they are associated with a region that
has moderately bright, extended 8.0~\micron\ emission surrounded by a
dark cloud, the masers are typically located at the interface between
the two.  This suggests that the methanol masers may be marking
regions where the interaction of an earlier epoch of star formation
with the parent molecular cloud has triggered further high-mass star
formation.  That multiple epochs of high-mass star formation occur
within an individual giant molecular cloud is well established for
regions such as Orion \citep{L99}.  The location of the methanol
masers within the \glim\/ images suggests that in many cases star
formation commences in the central regions of a molecular core and
proceeds outwards.

\subsection{The dark filament G\,$329.05\!-\!0.30$} \label{sec:filament}

\begin{figure}
\plotone{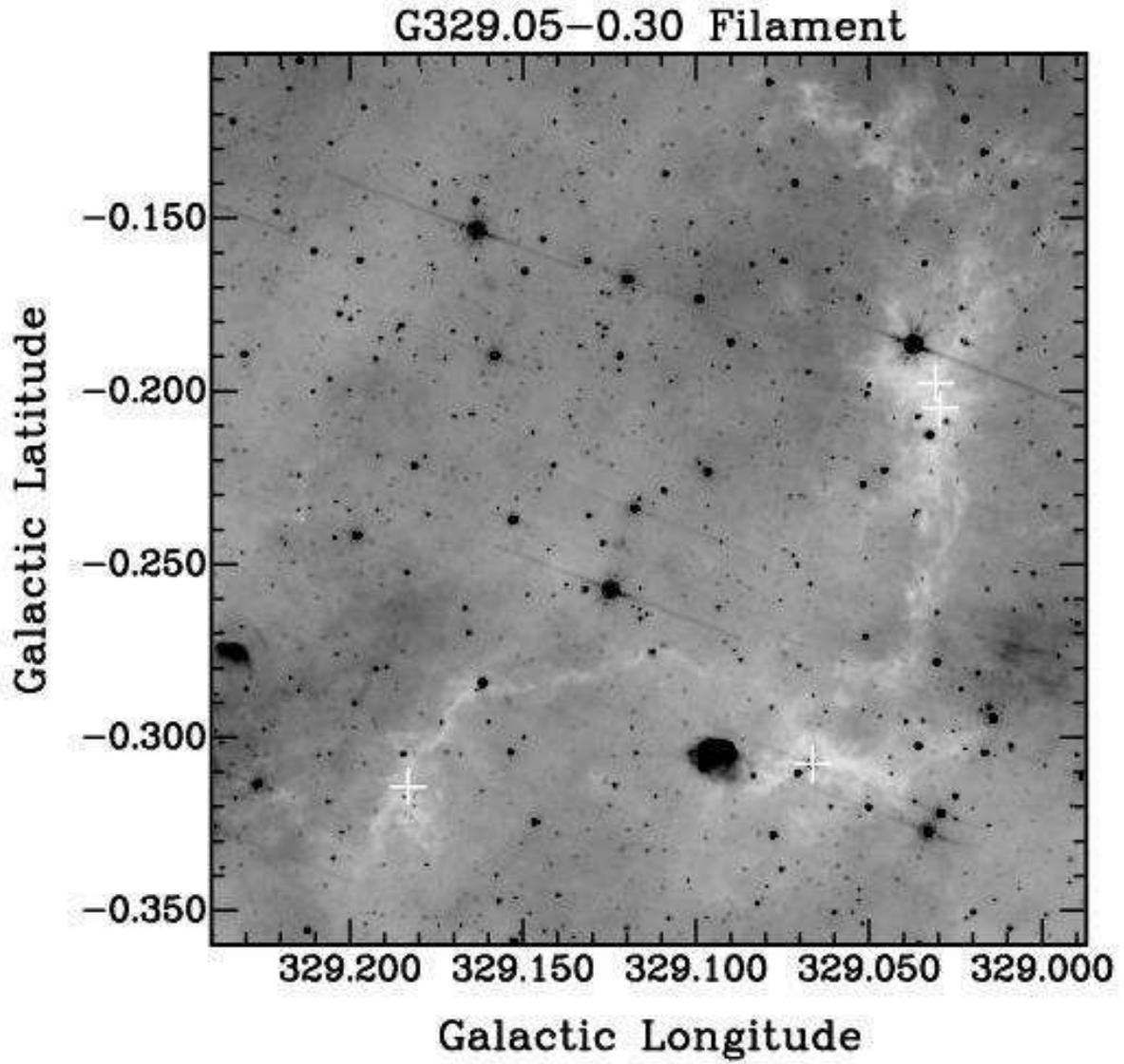}
\caption{The grey scale is the GLIMPSE band 4 emission (8.0~\micron)
  the crosses mark the positions of 6.7~GHz methanol masers (from top
  to bottom G\,$329.031\!-\!0.198$, G\,$329.029\!-\!0.205$,
  G\,$329.066\!-\!0.308$, G\,$329.183\!-\!0.314$).}
  \label{fig:filament}
\end{figure}

The majority of the methanol masers are associated with IRDC which
show relatively little elongation.  Notable exceptions are the four
methanol masers (G\,$329.031\!-\!0.198$, G\,$329.029\!-\!0.205$,
G\,$329.066\!-\!0.308$, G\,$329.183\!-\!0.314$) associated with the
dark filament centered on Galactic coordinates $\ell$=329.05,
$b$=$-$0.30 shown in Figure~\ref{fig:filament}.  More detailed images
of the regions surrounding each of the masers are shown in
Fig.~\ref{fig:images4}.  The G\,$329.05\!-\!0.30$ filament shows a
number of similarities to one of the best studied IRDC
G\,$11.11\!-\!0.12$ \citep{MPW05}, which has been modeled as a
nonmagnetic isothermal filament \citep{JFR+03}.  VLBI images of the
G\,$329.029\!-\!0.205$ 6.7~GHz methanol masers shows two sites, the
largest of which is aligned north-south, the same as the dark filament
\citep{DOE04}.  Class~I methanol and OH masers are found towards each
of the four class~II sites \citep{E05,C98} and water masers are
associated with three of the four, the exception being
G\,$329.066\!-\!0.308$ \citep{H97}.  There are three regions along the
filament where the 8~\micron\ absorption peaks and width of the
filament broadens.  At each of these locations there are one or more
methanol maser sites.  The near kinematic distance determined from the
velocity of the peak of the 6.7~GHz methanol maser emission in each of
the sources varies from 2.6-3.6~kpc.  Using a distance of 3~kpc for
the filament, the three clumps of methanol masers are separated by a
linear distance of about 6~pc from their nearest neighbouring maser.
Assuming the 6.7~GHz methanol masers have a lifetime of $10^5$ years
it is improbable that any localised event triggered the fragmentation
and subsequent star formation occurring along the filament as it would
require any disturbance to propagate at speeds in excess of
60~kms$^{-1}$.

\subsection{Identifying an evolutionary sequence for masers} \label{sec:evol}
An important question relating to the common maser species (OH,
methanol and water) is whether or not we can use them to help identify
an evolutionary sequence in high-mass star formation?  Previous
attempts \citep[e.g.][]{E05} have been hampered by the limitations of
earlier generations of infrared catalogs.  Many star forming regions
show emission in two or more of these maser species, so clearly there
is a large degree of overlap in the evolutionary phases, however,
there are also many sources where only one of the common maser species
is observed.  \citet{FC89} surveyed a large number of star forming
regions with the VLA, observing main-line OH masers, water masers and
radio continuum emission.  They concluded that the water masers are
associated with younger regions, with those which show both species
being of intermediate age, and those with only OH masers the oldest.
Early searches for class II methanol masers towards OH maser sites
were very successful.  However, untargetted searches, and those
targetted towards {\em IRAS} sources detected many in regions with no
OH masers and as outlined in the introduction there is good evidence
from millimeter and submillimeter observations that some methanol
masers trace an early stage of the high-mass star formation process
\citep{PHB02,WMA+03,MBH+05}.  This is interpretted similarly to the
water/OH relationship with the methanol masers arising prior to the OH
phase, but with a significant degree of overlap.  The question then
naturally arises, which species, methanol or water, traces the
earliest phase?

Information on associated water, OH and class~I methanol masers
\citep{C98,E05,H97} is available for the 6.7~GHz methanol masers from
the Mt Pleasant sample in the region $\ell$=325-335\degr\ region, as
is outlined in section~\ref{sec:results}.  The water and class~I
methanol maser observations were both made with single dish radio
telescopes and were targeted towards the 6.7~GHz maser position, so a
definite association between the two different maser species has not
yet been made for these transitions.  Future high resolution
observations are likely to show that in some cases the two maser
species are associated with independent objects within a larger
star-forming complex.  These ``false'' associations will add confusion
to attempts to find any evolutionary trend, however, they are unlikely
to mask it completely.

To investigate whether there is any evidence for the various maser
species tracing different evolutionary phases the 29 \glim\/ point
sources associated with 6.7~GHz methanol masers were split into two
sub-samples on the basis of whether or not they also had an associated
OH maser.  A number of color-color and color-magnitude diagrams,
including those shown in Fig.~\ref{fig:colcol3412} \&
\ref{fig:colmag124} were then constructed to compare the two samples.
This procedure was repeated for sub-samples constructed on the basis
of water and class~I methanol maser association. The total number of
6.7~GHz methanol masers with associated \glim\/ point sources which
had sufficient data to be plotted in Fig.~\ref{fig:colcol3412} \&
\ref{fig:colmag124} was around 15 in each case, so typically the
sub-samples had less than 10 members in each case.  The small numbers
in each sample and the possibility of contamination from false
associations for the water and class~I methanol masers mean that it is
not possible to draw any firm conclusions, however, a number of
interesting trends worthy of further investigation are evident.

\begin{figure}
\plotone{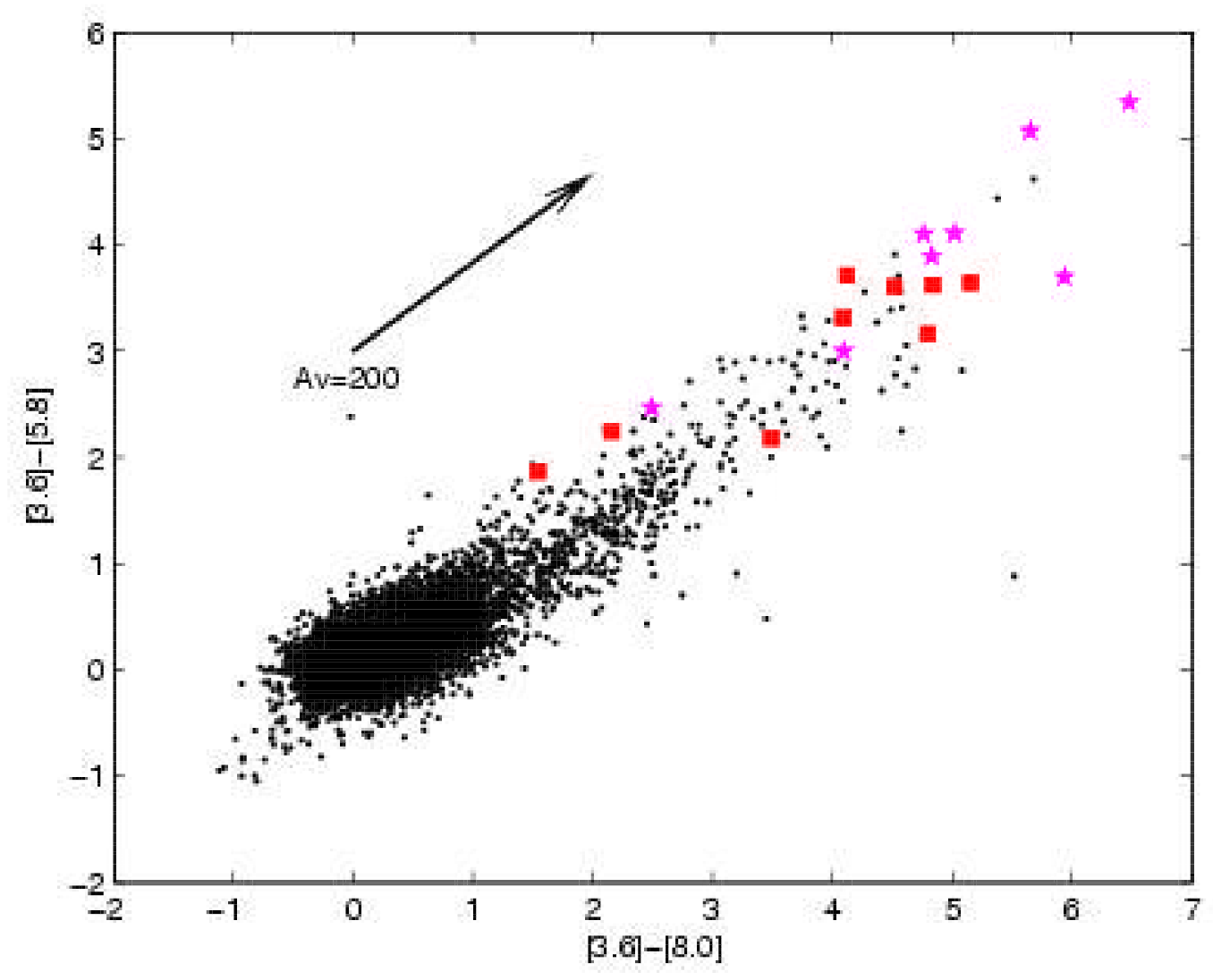}
\caption{A \glim\/ [3.6]-[8.0] versus [3.6]-[5.8] color-color diagram.
  The methanol masers with associated \glim\ point sources in the region
  $\ell$=325\degr--335\degr\ are represented by stars if they also have
  an associated class~I methanol maser and as squares if not.  Sources
  within 30\arcmin\ of $\ell$=326.5\degr, $b$=0.0\degr\ are represented
  with dots.  The arrow represents the reddening produced by an
  extinction of $A_v = 200$ using the \citet{IMB+05} model (see
  section~\ref{sec:mircolors}).}
\label{fig:classI}
\end{figure}

\begin{figure}
\plotone{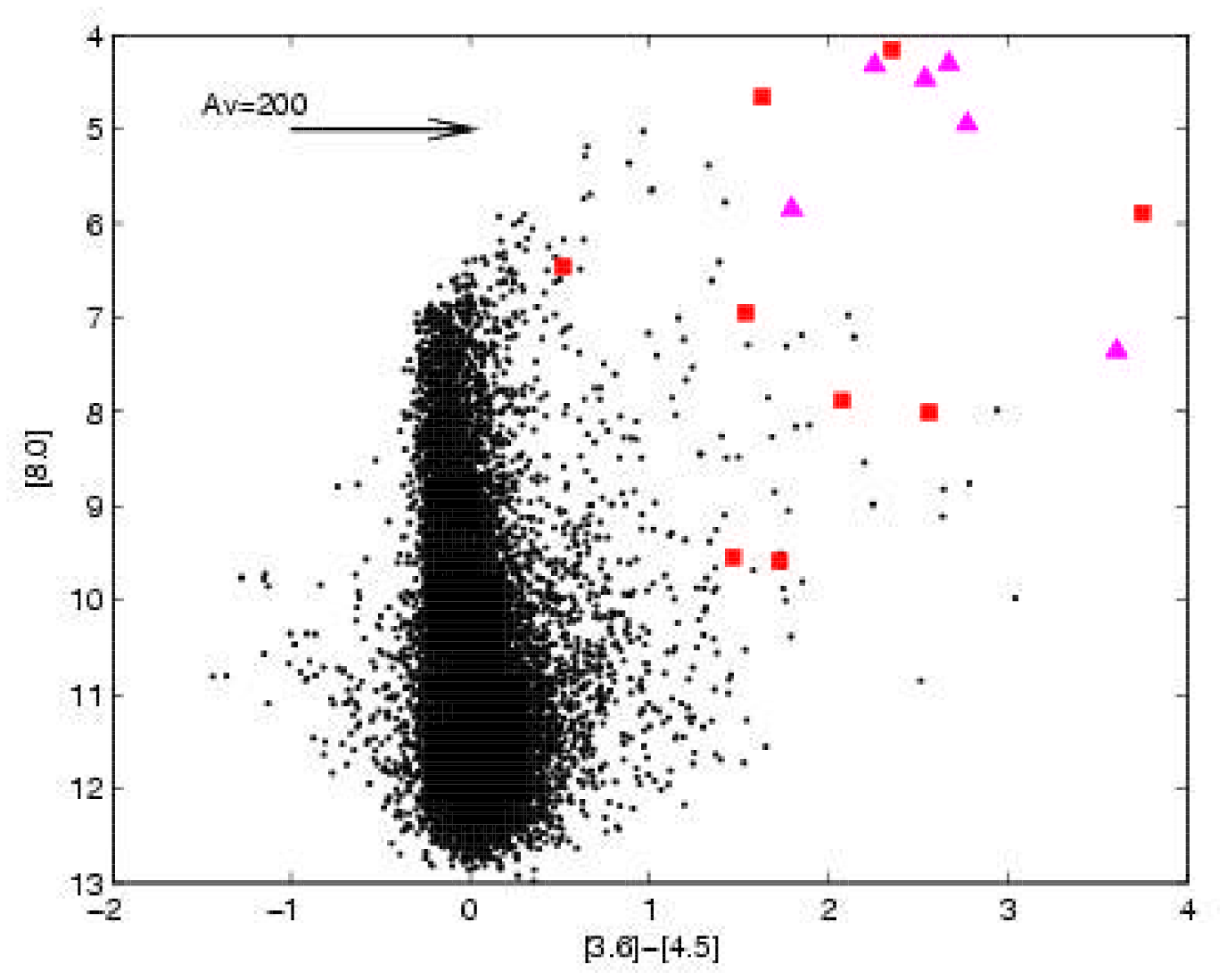}
\caption{A \glim\/ [3.6]-[4.5] versus 8.0~\micron\ color-magnitude diagram,
  similar to Fig.~\ref{fig:colmag124}.  The methanol masers with
  associated \glim\ point sources in the region
  $\ell$=325\degr--335\degr\ are represented with triangles if they
  also have an associated OH maser and as squares if not.  Sources
  within 30\arcmin\ of $\ell$=326.5\degr, $b$=0.0\degr\ are
  represented with dots.  The arrow represents the reddening produced by an
  extinction of $A_v = 200$ using the \citet{IMB+05} model (see
  section~\ref{sec:mircolors}).}
\label{fig:OH}
\end{figure}

Figure~\ref{fig:classI} plots a [3.6]-[8.0] versus [3.6-[5.8]
color-color diagrams using different symbols for the two sub-samples
of \glim\/ associated methanol masers, it also shows sources from the
comparison field.  Figure~\ref{fig:classI} shows that for class~II
methanol masers with an associated class~I methanol maser, there is a
trend for the \glim\/ point sources to be redder than those without an
associated class~I methanol maser.  Figure~\ref{fig:OH} shows that for
class~II methanol masers with an associated OH maser, there is a trend
for the 8.0~\micron\/ magnitude of the \glim\/ point sources to be
less than for those without an associated OH maser.  As stated above,
these trends could be artifacts of small number statistics, however,
they clearly warrent further investigation.

What might these apparent trends mean for a maser evolutionary
sequence?  Class~I methanol masers are created at the interface
between molecular outflows and the parent cloud \citep{PM90,JGS+92}.
Such outflows are expected to be most energetic and prevalent during
the infall/accretion phase of star formation and hence it has been
speculated that sources with associated class~I methanol masers may
signpost an earlier phase of high-mass star formation than those
without.  In general the youngest stellar objects will be the most
deeply embedded and coldest and this implies that they will have the
steepest mid-infrared spectrum.  The trend shown in
Fig.~\ref{fig:classI} appears consistent with this picture, with the
youngest maser sources with associated class~I methanol masers
generally having the redest \glim\ colors.

As outlined above, methanol masers with associated OH masers are
thought to be typically associated with a more advanced evolutionary
phase than those without.  Figure~\ref{fig:OH} shows that methanol
masers with associated OH masers are generally brighter at 8.0~\micron\ 
than those without.  All of these sources lie in the same region of
the Galactic Plane and so the range of distances for the two
sub-samples should be very similar, implying that the sources with
associated OH masers really are more luminous at 8.0~\micron\ and not
simply closer.  There are two possible explanations for this :
\begin{itemize}
\item The stellar luminosity range traced by OH masers may cutoff at a
  higher mass than is the case for methanol masers.  
\item The sources with associated OH masers have less silicate
  absorption and/or more PAH emission in the 8.0~\micron\ band than
  those without.
\end{itemize}
Both of the above possibilities are consistent with other observed
properties of OH and methanol maser sources.  A greater percentage of
OH masers are associated with ultra-compact \ionhy\/ regions than are
methanol masers \cite{C97}.  The more massive the star, the brighter
the \ionhy\ region and the easier it will be to detect at centimeter
wavelengths.  Further, if the methanol masers extend to a lower mass
range then we would expect there to be more methanol maser sources
than OH maser sources, as is observed \citep{CVE+95}.  However, the
6.7~GHz methanol masers typically have a peak flux density an order of
magnitude greater than main-line OH masers from the same source
\citep{CVE+95} and so this may partly be due to the sensitivity of
searches for the two maser species.  Alternatively, if OH and methanol
masers trace similar stellar mass ranges then the observation that OH
masers are more often associated with \ionhy\/ regions implies that
they are generally at a later evolutionary phase.  As an \ionhy\/
region forms and evolves it will gradually destroy or clear the
surrounding dust reducing silicate absorption and the UV radiation
will stimulate PAH emission.  Both of these effects will increase the
8.0~\micron\ emission.  These two hypotheses can be tested through
mid-infrared spectroscopy of the \glim\/ sources.

\section{Conclusions}

Data from the \glim\/ legacy science program of the {\em Spitzer Space
  Telescope} has been used to investigate the properties of the
infrared sources associated with class~II methanol masers.
Approximately 70\% of class~II methanol masers have a \glim\ point
source or archive source within 2\arcsec, a much higher rate of
association than found in previous infrared catalogs such as {\em
  IRAS} or {\em MSX}.  The colors of the \glim\/ point sources
associated with methanol masers are very red, similar to low-mass
class~0 YSO, and consistent with other recent observations
\citep[e.g.][]{PHB02,WMA+03,MBH+05} suggesting that they are
associated with a very early phase of high-mass star formation.
\glim\/ colors have been used to develop criteria for targeting
methanol maser searches which will detect more than 80\% of all
sources with an efficiency of greater than 10\%.

IRDC are considered prime candidates to be the location for the
earliest stages of high-mass star formation.  \glim\/ 8.0~\micron\/
images show that the majority of class~II methanol masers are
associated with sources embedded within IRDC, and that a significant
fraction of IRDC contain methanol masers.  This is consistent with
both IRDC and class~II methanol masers signposting regions where
high-mass star formation is in its early stages and confirms that
star-formation has commenced in many IRDC.  IRDC without associated
methanol and/or water masers may represent regions at a still earlier
stage of the high-mass star formation evolutionary sequence (although
some may simply be regions where the beaming angle or some other
factor is unfavourable for our line of sight).

It appears likely that the properties of \glim\/ sources associated
with masers will allow a qualitative evolutionary sequence for the
early stages of high-mass star formation to be developed and tested
against models.  Comparison of the colors of class~II methanol maser
sources with and without associated class~I methanol masers shows that
those with an associated class~I methanol maser have redder \glim\ 
colors.  This is consistent with class~I methanol masers being
generally associated with younger objects.  While a comparison of the
properties of \glim\ sources related to class~II methanol masers with
and without associated OH masers shows that those with an associated
OH maser are generally stronger in the 8.0~\micron\/ band.  This may
indicate that the stellar mass range associated with methanol maser
extends to lower masses than for OH masers, or alternatively it may
show that OH masers are generally associated with a later evolutionary
phase.

\section*{Acknowledgements}
I would like to thank the referee for many useful comments that have
improved the paper.  I would like to thank John Dickey for stimulating
discussions on the \glim\ program and the nature of IRDC.  Financial
support for this work was provided by the Australian Research Council.
This research has made use of NASA's Astrophysics Data System Abstract
Service.  This research has made use of data products from the \glim\
survey, which is a legacy science program of the {\em Spitzer Space
Telescope}, funded by the National Aeronautics and Space
Administration.  This research has made use of data products from the
{\em Midcourse Space Experiment}.  Processing of the data was funded
by the Ballistic Missile Defence Organization with additional support
from the NASA Office of Space Science.  The research has made use of
the NASA/IPAC Infrared Science Archive, which is operated by the Jet
Propulsion Laboratory, California Institute of Technology, under
contract with the National Aeronautics and Space Administration.

Facilities: \facility{Sa.Sptizer(IRAC)}.

\clearpage

\begin{deluxetable}{llcccrrrr}
\tabletypesize{\scriptsize}
\tablecaption{6.7~GHz methanol masers in the Galactic longitude region
$l$ = 325\degr -- 335\degr.  For discovery and other references for
each maser see \citet{PMB05}.  The kinematic distances have been
calculated from the velocity of the strongest methanol maser emission
using the model of \citet{BB93}.  For the associations column, I =
Class~I methanol maser \citep{E05}, o = OH maser \citep{C98}, w =
water maser \citep{H97}.  Sources marked $^{a}$ lie outside the statistically 
complete sample of 6.7~GHz methanol masers.\label{results}}
\tablehead{
\colhead{Source} & \colhead{Methanol} & \colhead{Right}     & 
  \colhead{Declination} & \colhead{\glim\/} & \colhead{Sep.} & 
  \multicolumn{2}{c}{Kinematic Distance} & \colhead{Assoc.} \\
\colhead{\#} & \colhead{Maser}    & \colhead{Ascension} & 
                        & \colhead{Source}           & \colhead{(\arcsec)}   &
  \colhead{Near} & \colhead{Far} & \\
                 &                    & \colhead{(J2000)}   & 
  \colhead{(J2000)}     & \colhead{Name}          &                      &
  \colhead{(kpc)} & \colhead{(kpc)} & \\
}
\startdata
 1 & G\,$326.475\!+\!0.703$     & 15:43:16.648 & -54:07:12.71 & {\em GLMA}$\mathit{326.4746\!+\!0.7030}$ &  1.8  & 2.6  & 11.6    & I,w \\
 2 & G\,$326.641\!+\!0.613^{a}$ & 15:44:32.938 & -54:05:28.55 & GLMC$326.6405\!+\!0.6125$                &  0.6  & 2.9  & 11.3    & I,w \\
 3 & G\,$326.662\!+\!0.521$     & 15:45:02.854 & -54:09:03.20 & {\em GLMA}$\mathit{326.6620\!+0\!.5209}$ &  0.2  & 2.8  & 11.4    & w \\
 4 & G\,$326.859\!-\!0.677^{a}$ & 15:51:14.190 & -54:58:04.94 &                          &        & 3.7  & 10.5 & I \\
 5 & G\,$327.120\!+\!0.511$     & 15:47:32.729 & -53:52:38.90 & GLMC$327.1196\!+\!0.5105$                &  0.9   & 5.5  & 8.7  & o,w \\
 6 & G\,$327.402\!+\!0.444$     & 15:49:19.523 & -53:45:14.21 & GLMC$327.4019\!+\!0.4447$                &  1.7   & 5.2  & 9.1  & o,w \\
 7 & G\,$327.392\!+\!0.199$     & 15:50:18.491 & -53:57:06.36 & {\em GLMA}$\mathit{327.3915\!+\!0.1991}$ &  1.2   & 5.3  & 9.0     & I \\
 8 & G\,$327.590\!-\!0.094$     & 15:52:36.824 & -54:03:18.97 & GLMC$327.5900\!-\!0.0944$                &  0.7  & 5.4  & 8.9   \\
 9 & G\,$327.618\!-\!0.111$     & 15:52:50.241 & -54:03:00.71 & GLMC$327.6186\!-\!0.1110$                &  0.6   & 6.5  & 7.9  & I \\
10 & G\,$327.945\!-\!0.115$     & 15:54:33.917 & -53:50:44.58 & GLMC$327.9447\!-\!0.1148$                &  0.2   & 3.4  & 11.0 &  \\
11 & G\,$328.237\!-\!0.548^{a}$ & 15:57:58.381 & -53:59:23.14 &                          &        & 3.0  & 11.4    & I,o,w \\
12 & G\,$328.254\!-\!0.532$     & 15:57:59.790 & -53:58:00.85 & GLMC$328.2542\!-\!0.5323$                &  0.5   & 2.6  & 11.9    & o,w \\
13 & G\,$328.809\!+\!0.633^{a}$ & 15:55:48.608 & -52:43:06.20 &                          &        & 3.0  & 11.5 & I,o \\
14 & G\,$329.031\!-\!0.198$     & 16:00:30.326 & -53:12:27.35 & GLMC$329.0307\!-\!0.1982$                &  1.6   & 2.9  & 11.7 & I,o,w \\
15 & G\,$329.029\!-\!0.205$     & 16:00:31.799 & -53:12:49.66 & {\em GLMA}$\mathit{329.0296\!-\!0.2049}$ &  0.7   & 2.6  & 12.0    & I,o,w \\
16 & G\,$329.066\!-\!0.308$     & 16:01:09.940 & -53:16:02.65 & GLMC$329.0663\!-\!0.3076$                &  0.3   & 3.0  & 11.6 & I,o \\
17 & G\,$329.183\!-\!0.314$     & 16:01:47.034 & -53:11:44.19 & GLMC$329.1833\!-\!0.3143$                &  0.7   & 3.6  & 11.0 & I,o,w \\
18 & G\,$329.339\!+\!0.148$     & 16:00:33.154 & -52:44:40.00 &                          &        & 7.3  & 7.3     & \\
19 & G\,$329.407\!-\!0.459$     & 16:03:32.662 & -53:09:26.98 & {\em GLMA}$\mathit{329.4066\!-\!0.4592}$ &  0.8   & 4.2  & 10.4    & o,w \\
20 & G\,$329.469\!+\!0.502$     & 15:59:40.727 & -52:23:27.70 &                          &        & 4.5  & 10.1 & I \\
21 & G\,$329.622\!+\!0.138$     & 16:02:00.292 & -52:33:59.16 &                          &        & 3.9  & 10.8 & w \\
22 & G\,$329.610\!+\!0.114$     & 16:02:03.103 & -52:35:33.53 & GLMC$329.6101\!+\!0.1137$                &  0.4   & 3.9  & 10.8 & \\
23 & G\,$330.952\!-\!0.182^{a}$ & 16:09:52.372 & -51:54:57.89 &                          &        & 5.3  & 9.5     & o,w \\
24 & G\,$331.120\!-\!0.118^{a}$ & 16:10:23.050 & -51:45:20.10 & GLMC$331.1195\!-\!0.1180$                &  0.3   & 5.6  & 9.2  &  \\
25 & G\,$331.132\!-\!0.244$     & 16:10:59.743 & -51:50:22.70 & GLMC$331.1314\!-\!0.2439$                &  1.4   & 5.2  & 9.7  & I,o,w \\
26 & G\,$331.278\!-\!0.188$     & 16:11:26.596 & -51:41:56.67 & GLMC$331.2780\!-\!0.1883$                &  1.1   & 4.8  & 10.1 & o,w \\
27 & G\,$331.342\!-\!0.346$     & 16:12:26.456 & -51:46:16.86 & GLMC$331.3417\!-\!0.3465$                &  0.1   & 4.3  & 10.6 & I,o \\
28 & G\,$331.425\!+\!0.264^{a}$ & 16:10:09.354 & -51:16:04.60 & GLMC$331.4249\!+\!0.2643$                &  1.1  & 5.4  & 9.6  &  \\
29 & G\,$331.442\!-\!0.187^{a}$ & 16:12:12.478 & -51:35:10.32 & GLMC$331.4420\!-\!0.1866$                &  0.3   & 5.4  & 9.6  & I,o,w \\
30 & G\,$331.542\!-\!0.066$     & 16:12:09.020 & -51:25:47.60 &                          &        & 5.1  & 9.8  & o \\
31 & G\,$331.556\!-\!0.121^{a}$ & 16:12:27.210 & -51:27:38.20 &                          &        & 6.3  & 8.7  & o,w \\
32 & G\,$332.094\!-\!0.421$     & 16:16:16.487 & -51:18:25.46 &                          &        & 4.0  & 11.0    & w \\
33 & G\,$332.295\!-\!0.094$     & 16:15:45.381 & -50:55:53.85 &                          &        & 3.2  & 11.8    & I,w \\
34 & G\,$332.351\!-\!0.436$     & 16:17:31.560 & -51:08:21.55 & GLMC$332.3506\!-\!0.4362$                &  1.8   & 3.6  & 11.5 & \\
35 & G\,$332.560\!-\!0.148$     & 16:17:12.123 & -50:47:12.27 & GLMC$332.5599\!-\!0.1479$                &  0.5   & 3.5  & 11.6 &  \\
36 & G\,$332.604\!-\!0.167$     & 16:17:29.320 & -50:46:12.51 & GLMC$332.6039\!-\!0.1677$                &  1.2   & 3.5  & 11.6 & I \\
37 & G\,$332.942\!-\!0.686^{a}$ & 16:21:19.018 & -50:54:10.41 & GLMC$332.9418\!-\!0.6857$                &  1.0   & 3.6  & 11.5 & I,w \\
38 & G\,$332.963\!-\!0.679^{a}$ & 16:21:22.926 & -50:52:58.71 & GLMC$332.9633\!-\!0.6791$                &  0.3   & 3.2  & 11.9 & I \\
39 & G\,$333.029\!-\!0.015$     & 16:18:44.167 & -50:21:50.77 & GLMC$333.0295\!-\!0.0147$                &  1.4   & 3.6  & 11.5 &  \\
40 & G\,$333.029\!-\!0.063^{a}$ & 16:18:56.735 & -50:23:54.17 &                          &        & 3.6  & 11.5    & I \\
41 & G\,$333.068\!-\!0.447$     & 16:20:48.995 & -50:38:40.72 & {\em GLMA}$\mathit{333.0679\!-\!0.4469}$ &  0.6   & 3.7  & 11.5    &  \\
42 & G\,$333.121\!-\!0.434$     & 16:20:59.704 & -50:35:52.32 &                          &        & 3.4  & 11.8    & I,w \\
43 & G\,$333.128\!-\!0.440$     & 16:21:03.300 & -50:35:49.75 & {\em GLMA}$\mathit{333.1284\!-\!0.4405}$ & 1.9    & 3.1  & 12.0    & I,w \\
44 & G\,$333.130\!-\!0.560^{a}$ & 16:21:35.392 & -50:40:56.97 &                          &        & 3.8  & 11.4    & I,w \\
45 & G\,$333.163\!-\!0.101^{a}$ & 16:19:42.670 & -50:19:53.20 &                          &        & 5.7  & 9.5     & I \\
46 & G\,$333.184\!-\!0.091$     & 16:19:45.620 & -50:18:35.00 & GLMC$333.1841\!-\!0.0908$                &  0.1   & 5.0  & 10.1 & I \\
47 & G\,$333.234\!-\!0.062$     & 16:19:51.250 & -50:15:14.10 &                          &        & 5.2  & 10.0    & I,o,w \\
48 & G\,$333.315\!+\!0.105$     & 16:19:29.016 & -50:04:41.45 & GLMC$333.3150\!+\!0.1052$                &  0.1   & 3.1  & 12.1 & I,o \\
49 & G\,$333.466\!-\!0.164$     & 16:21:20.180 & -50:09:48.60 &                          &        & 3.0  & 12.2    & I,o,w \\
50 & G\,$333.562\!-\!0.025$     & 16:21:08.797 & -49:59:48.26 & GLMC$333.5622\!-\!0.0248$                &  0.6   & 2.7  & 12.6 & I \\
51 & G\,$333.646\!+\!0.058^{a}$ & 16:21:09.140 & -49:52:45.90 & {\em GLMA}$\mathit{333.6454\!+\!0.0578}$ &  0.6   & 5.3  & 10.0 & \\
52 & G\,$333.683\!-\!0.437$     & 16:23:29.794 & -50:12:08.69 & GLMC$333.6824\!-\!0.4367$                &  0.5   & 0.4  & 14.9 & \\
53 & G\,$333.931\!-\!0.135$     & 16:23:14.831 & -49:48:48.87 & {\em GLMA}$\mathit{333.9307\!-\!0.1345}$ &  0.7   & 2.7  & 12.5    &\\
54 & G\,$334.635\!-\!0.015$     & 16:25:45.729 & -49:13:37.51 & GLMC$334.6350\!-\!0.0146$ &  0.5   & 2.3  & 13.0 & \\
55 & G\,$334.935\!-\!0.098$     & 16:27:24.250 & -49:04:11.30 &                          &        & 1.6  & 13.8 & \\
56 & G\,$335.060\!-\!0.427$     & 16:29:23.146 & -49:12:27.34 & GLMC$335.0595\!-\!0.4274$                &  0.3   & 3.4  & 12.1 & I,o,w \\
\enddata
\end{deluxetable}

\clearpage

\begin{deluxetable}{rlrrrrrrrr}
\tabletypesize{\scriptsize}
\tablecaption{Properties of the \glim\/ sources associated
  with 6.7~GHz methanol masers in the Galactic Longitude range
$l$ = 325\degr -- 335\degr.\label{glimprop}}
\tablehead{
\colhead{Src} & \colhead{\glim\/} & \multicolumn{4}{c}{Magnitudes} &
  \multicolumn{4}{c}{Flux Densities} \\
\colhead{\#} & \colhead{Source}           & \colhead{3.6~\micron} & 
  \colhead{4.5~\micron} & \colhead{5.8~\micron} & \colhead{8.0~\micron} &
  \colhead{3.6~\micron} & \colhead{4.5~\micron} & \colhead{5.8~\micron} & 
  \colhead{8.0~\micron} \\
                 & \colhead{Name}          & \colhead{(mag)}       & 
  \colhead{(mag)} & \colhead{(mag)} & \colhead{(mag)} & 
  \colhead{(mJy)} & \colhead{(mJy)} & \colhead{(mJy)} & \colhead{(mJy)}\\
}
\startdata
 1 &  {\em GLMA}$\mathit{326.4746\!+\!0.7030}$ & $14.15\pm0.18$ & $10.50\pm0.24$ & $9.68\pm0.10$  & $8.46\pm0.05$ & 0.61 & 11.3  & 15.7  & 26.0   \\
 2 &  GLMC$326.6405\!+\!0.6125$ & $13.60\pm0.15$ & $13.82\pm0.34$ &                &              & 1.01   & 0.53 &               &                \\
 3 & {\em GLMA}$\mathit{326.6620\!+0\!.5209}$ & $6.92\pm0.11$  &                & $2.75\pm0.08$  &               & 474  &       & 9290  &        \\ 
 5 &  GLMC$327.1196\!+\!0.5105$ & $7.51\pm0.07$  &                & $4.54\pm0.05$  &              & 275      &                & 1787   &                \\
 6 &  GLMC$327.4019\!+\!0.4447$ &                &                & $4.77\pm0.06$  & $4.11\pm0.16$&                 &                & 1444   & 1435   \\
 7 & {\em GLMA}$\mathit{327.3915\!+\!0.1991}$ &                &                & $7.32\pm0.08$  & $6.47\pm0.04$ &      &       & 137   & 163.00 \\ 
 8 &  GLMC$327.5900\!-\!0.0944$ & $11.26\pm0.07$ & $10.74\pm0.16$ & $8.10\pm0.05$  & $6.46\pm0.04$& 8.69  & 9.0    & 67.0 & 164.1 \\
 9 &  GLMC$327.6186\!-\!0.1110$ & $8.76\pm0.07$  & $7.12\pm0.11$  & $5.76\pm0.04$  & $4.66\pm0.03$& 87.3    & 254     & 581   & 863    \\
10 &  GLMC$327.9447\!-\!0.1148$ & $8.10\pm0.13$  &                & $5.91\pm0.07$  & $4.60\pm0.26$& 160      &        & 503    & 910    \\
12 &  GLMC$328.2542\!-\!0.5323$ & $8.84\pm0.11$  & $6.58\pm0.15$  & $5.24\pm0.06$  & $4.32\pm0.05$& 80.9     & 418    & 938   & 1179 \\
14 &  GLMC$329.0307\!-\!0.1982$ &                & $12.04\pm0.18$ & $11.20\pm0.14$ & $12.27\pm0.35$ &                 & 2.75  & 3.87 & 0.77  \\
15 & {\em GLMA}$\mathit{329.0296\!-\!0.2049}$ &                &                &                & $11.99\pm0.13$&      &       &       & 1.01   \\     
16 &  GLMC$329.0663\!-\!0.3076$ & $9.30\pm0.05$  & $6.76\pm0.06$  & $5.41\pm0.03$  & $4.47\pm0.03$& 52.9    & 354     & 801    & 1030    \\
17 &  GLMC$329.1827\!-\!0.3137$ & $11.54\pm0.20$ &                & $9.07\pm0.17$  & $9.04\pm0.12$ & 6.7    &                & 27.4  & 15.2  \\
19 & {\em GLMA}$\mathit{329.4066\!-\!0.4592}$ &                &                & $7.85\pm0.17$  &               &      &       & 84.5  &        \\        
22 &  GLMC$329.6101\!+\!0.1137$ & $8.99\pm0.15$  & $6.64\pm0.08$  & $5.37\pm0.03$  & $4.15\pm0.03$& 70.4    & 397     & 830    & 1377    \\
24 &  GLMC$331.1195\!-\!0.1180$ & $12.13\pm0.14$ & $9.57\pm0.09$  & $8.42\pm0.11$  & $8.01\pm0.18$& 3.90   & 26.7   & 50.1  & 39.6   \\
25 &  GLMC$331.1314\!-\!0.2439$ & $11.79\pm0.14$ & $9.99\pm0.34$  & $8.09\pm0.20$  & $5.85\pm0.27$& 5.35   & 18.0   & 68     & 289     \\
26 &  GLMC$331.2780\!-\!0.1883$ &                & $10.66\pm0.34$ & $9.25\pm0.19$  &              &                 & 9.8    & 23.3  &                \\
27 &  GLMC$331.3417\!-\!0.3465$ & $9.33\pm0.10$  & $6.66\pm0.15$  & $5.22\pm0.04$  & $4.31\pm0.03$& 51.6    & 391     & 954    & 1196    \\
28 &  GLMC$331.4249\!+\!0.2643$ & $12.10\pm0.15$ & $10.56\pm0.21$ & $8.46\pm0.10$  & $6.95\pm0.17$& 4.01  & 10.7   & 48.2  & 105     \\
29 &  GLMC$331.4420\!-\!0.1866$ & $13.01\pm0.10$ & $9.40\pm0.05$  & $7.94\pm0.06$  & $7.36\pm0.04$& 1.74   & 31.2   & 77.7  & 72.1   \\
34 &  GLMC$332.3506\!-\!0.4362$ & $12.76\pm0.17$ & $11.11\pm0.30$ & $10.32\pm0.18$ &              & 2.19   & 6.44   & 8.7   &                \\
35 &  GLMC$332.5599\!-\!0.1479$ & $11.08\pm0.28$ & $9.62\pm0.18$  & $9.21\pm0.12$  & $9.54\pm0.15$& 10.3   & 25.6   & 24.0  & 9.6    \\
36 &  GLMC$332.6043\!-\!0.1666$ & $13.70\pm0.10$ & $11.90\pm0.15$ &                &              & 0.92   & 3.12   &       &        \\
37 &  GLMC$332.9418\!-\!0.6857$ &                &                & $6.93\pm0.09$  & $6.55\pm0.11$&                 &                & 197    & 151     \\
38 &  GLMC$332.9633\!-\!0.6791$ & $12.37\pm0.07$ & $8.62\pm0.10$  & $7.03\pm0.04$  & $5.89\pm0.03$& 3.13   & 64.0   & 180.1 & 277.8 \\
39 &  GLMC$333.0295\!-\!0.0147$ & $11.74\pm0.15$ & $10.02\pm0.21$ & $9.49\pm0.08$  & $9.58\pm0.08$& 5.59   & 17.7   & 18.5  & 9.26  \\
41 & {\em GLMA}$\mathit{333.0679\!-\!0.4469}$ & $8.66\pm0.10$  & $5.83\pm0.09$  & $4.22\pm0.04$  &               & 95.2 & 834   & 2390  &        \\   
43 & {\em GLMA}$\mathit{333.1284\!-\!0.4405}$ & $8.47\pm0.05$  & $8.47\pm0.14$  & $8.42\pm0.22$  &               & 114  & 73.5  & 50.0  &        \\   
46 &  GLMC$333.1841\!-\!0.0908$ &                &                & $7.03\pm0.14$  & $6.67\pm0.07$&                 &                & 180    & 135.2  \\
48 &  GLMC$333.3150\!+\!0.1052$ & $9.71\pm0.15$  & $6.94\pm0.10$  & $5.61\pm0.05$  & $4.95\pm0.08$& 36.2    & 300     & 664    & 661     \\
50 &  GLMC$333.5622\!-\!0.0248$ & $13.72\pm0.12$ & $12.66\pm0.15$ &                &              & 0.899 & 1.55  &               &                \\
51 & {\em GLMA}$\mathit{333.6454\!+\!0.0578}$ &                & $12.96\pm0.22$ & $11.32\pm0.28$ &               &      & 1.17  & 3.47  &        \\     
52 &  GLMC$333.6824\!-\!0.4367$ & $13.05\pm0.18$ & $11.69\pm0.13$ &                &              & 1.69   & 3.77  &               &                \\
53 & {\em GLMA}$\mathit{333.9307\!-\!0.1345}$ & $12.28\pm0.12$  & $9.55\pm0.12$  & $7.97\pm0.09$  &               & 3.39 & 27.1  & 75.4  &        \\     
54 &  GLMC$334.6350\!-\!0.0146$ & $11.98\pm0.06$ & $9.90\pm0.06$  & $8.67\pm0.09$  & $7.88\pm0.13$& 4.50   & 19.6   & 39.8  & 44.4   \\
56 &  GLMC$335.0595\!-\!0.4274$ &                &                & $7.13\pm0.07$  & $6.38\pm0.06$&                 &                & 164    & 176.4  \\
\enddata
\end{deluxetable}

\clearpage

\clearpage

\clearpage

\clearpage

\clearpage

\clearpage

\clearpage

\clearpage

\clearpage

\clearpage

\clearpage

\clearpage

\clearpage

\clearpage

\clearpage

\clearpage

\clearpage

\clearpage

\clearpage

\end{document}